\documentclass[a4paper,fleqn,usenatbib]{mnras}

\usepackage[T1]{fontenc}
\usepackage{times}

\usepackage{color,soul}
\usepackage{graphicx}	
\usepackage{amsmath}	
\usepackage{amssymb}	
\usepackage{etoolbox}
\makeatletter
\patchcmd\@combinedblfloats{\box\@outputbox}{\unvbox\@outputbox}{}{%
   \errmessage{\noexpand\@combinedblfloats could not be patched}%
}%
 \makeatother

\newcommand{\md}{\mathrm{d}}

\newcommand{\Rg}{\mathbf{R}_\mathrm{g}}
\newcommand{\rp}{r_\mathrm{p}}

\newcommand{\nn}{\nonumber}

\def\ba{\begin{eqnarray}}
\def\ea{\end{eqnarray}}

 \title[Secular dynamics of binaries II: dynamical evolution]{Secular dynamics of binaries in stellar clusters II: dynamical evolution}

\author[C. Hamilton \& R. R. Rafikov]{
  Chris Hamilton$^{1}$\thanks{E-mail: ch783@cam.ac.uk} and Roman R. Rafikov$^{1,2}$\\
$^1$Department of Applied Mathematics and Theoretical Physics, University of
Cambridge, Wilberforce Road, Cambridge CB3 0WA, UK\\
$^2$Institute for Advanced Study, Einstein Drive, Princeton, NJ 08540, USA}

\date{Accepted XXX. Received YYY; in original form ZZZ}

\pubyear{2018}

\begin{document}
\label{firstpage}
\pagerange{\pageref{firstpage}--\pageref{lastpage}}
\maketitle

\begin{abstract}
Dense stellar clusters are natural sites for the origin and evolution of exotic objects such as relativistic binaries (potential gravitational wave sources), blue stragglers, etc. We investigate the secular dynamics of a binary system driven by the global tidal field of an axisymmetric stellar cluster in which the binary orbits. In a companion paper (Hamilton \& Rafikov 2019a) we developed a general Hamiltonian framework describing such systems. The effective (doubly-averaged) Hamiltonian derived there encapsulates all information about the tidal potential experienced by the binary in its orbit around the cluster in a single parameter $\Gamma$. Here we provide a thorough exploration of the phase-space of the corresponding secular problem as $\Gamma$ is varied. We find that for $\Gamma > 1/5$ the phase-space structure and the evolution of binary orbital elements are qualitatively similar to the Lidov-Kozai problem.  However, this is only one of four possible regimes, because the dynamics are qualitatively changed by bifurcations at $\Gamma = 1/5,0,-1/5$. We show how the dynamics are altered in each regime and calculate characteristics such as secular evolution timescale, maximum possible eccentricity, etc. We verify the predictions of our doubly-averaged formalism numerically and find it to be very accurate when its underlying assumptions are fulfilled, typically meaning that the secular timescale should exceed the period of the binary around the cluster by $\gtrsim 10-10^2$ (depending on the cluster potential and binary orbit). Our results may be relevant for understanding the nature of a variety of exotic systems harboured by stellar clusters.
\end{abstract}

\begin{keywords}
gravitation -- celestial mechanics -- stars: kinematics and
dynamics -- galaxies: star clusters: general -- binaries: general
\end{keywords}




\section{Introduction}


Orbital evolution of binary systems in dense stellar clusters is a very rich problem. Historically, the focus in this area has been on the cumulative effect of multiple encounters of the binary with the other stellar members of the cluster \citep{Heggie1975}. In particular, the excitation of binary eccentricity via this mechanism was explored \citep{Heggie1996} as a means of understanding the properties of binary pulsars in globular clusters \citep{Phinney92,Phinney94}. At the same time, the effect of the smooth component of the cluster gravitational field on the internal dynamics of the binary was largely overlooked in the past (although it was routinely accounted for in the studies of the Oort Cloud dynamics and wide binaries driven by the Galactic tide, see e.g.  \citealt{Heisler1986,Brasser2001,Veras2013}).

To fill this gap, in \citet{Hamilton2019a} --- hereafter `Paper I' --- we explored secular evolution of a binary system driven by the smooth tidal field of a cluster in which the binary orbits. We developed a general Hamiltonian framework which describes the evolution of the orbital elements of a binary fully accounting for the details of the (axisymmetric or spherical) cluster potential and the binary orbit in it. This was done by expanding the cluster potential near the barycentre\footnote{To sufficient accuracy, the binary's barycentre follows the trajectory of a test particle, $\md^2 \Rg/\md t^2 = -\nabla \Phi$ where $\Phi$ is the smooth cluster potential.} $\Rg$ of the binary consisting of two point masses $m_1$ and $m_2$ and performing two averages of the resultant tidal perturbation: first over the binary's `inner' orbital period (i.e. over its mean anomaly), and second over many `outer' orbits of the binary itself around the cluster.  As a result, defining the binary orbital elements --- semi-major axis $a$, eccentricity $e$, inclination $i$, longitude of the ascending node $\Omega$, argument of pericentre $\omega$ and mean anomaly $M$ --- with respect to a frame tied to the equatorial plane of the axisymmetric cluster (or the plane of the outer orbit of the binary in the case of a spherical cluster), Paper I arrived at a secular (`doubly-averaged') perturbing Hamiltonian
\begin{align} 
\overline{\langle H_1 \rangle}_{M} = C H_1^*, \,\,\,\,\,\,\,\,\,
\mathrm{where}\,\,\,\,\,\,\,\, 
C =\frac{Aa^2}{8}
\label{H1Mt} 
\end{align} 
is a constant with dimensions of energy per unit mass and $H_1^*$ is the `dimensionless Hamiltonian'
\begin{align} 
H_1^* &= (2+3e^2)\left( 1-3\Gamma \cos^2 i\right) - 15\Gamma e^2\sin^2i\cos2\omega. 
\label{H1starorb} 
\end{align} 
The quantities $A$ (with units of $($frequency$)^2$) and $\Gamma$ (dimensionless) emerging in these expressions are constants, which encode all of the information about the time-averaged tidal potential experienced by the binary as it moves around the cluster. Paper I explored in detail the dependence of $A$ and $\Gamma$ upon the shape of the background potential and the binary's orbit within it. 

The quantity $A$ is an overall proportionality constant and hence sets the timescale for the secular eccentricity oscillations. The dimensionless part of the Hamiltonian, $H_1^*$, depends on the potential and the binary's outer orbit only through the constant $\Gamma$ (see equation \ref{H1starorb}). The value of $\Gamma$ determines the phase-space morphology for the dynamics of the inner orbit of the binary, as we will see in this work. In particular, in Paper I we found that $\Gamma\to 1$ is reached when the cluster potential is Keplerian, in which case $H_1^*$ reduces to the test particle quadrupole Lidov-Kozai (hereafter `LK') Hamiltonian \citep{Lidov1962, Kozai1962, Antognini2015,Naoz2016}.  In addition, the Hamiltonian for binaries orbiting in a thin Galactic disk (first derived by \citealt{Heisler1986}) was recovered from equation (\ref{H1starorb}) when $\Gamma=1/3$. It was also shown that for binaries in realistic spherical potentials $0 \leq \Gamma \leq 1$, while in non-spherical potentials $\Gamma$ can be (but is not necessarily) negative.  

The goal of the present paper is to systematically explore the dynamics that result from the general secular theory based on the Hamiltonian \eqref{H1Mt}-\eqref{H1starorb}. It turns out that the dynamics in the range $\Gamma > 1/5$ are qualitatively very similar to those arising in the LK case, but that bifurcations occur when $\Gamma=1/5$, $0$ and $-1/5$, which change the picture significantly. As a result, in this work we separately treat four distinct dynamical regimes: 
\ba
&\Gamma & > 1/5, 
\label{eq:1}\\
0 < &\Gamma & \leq 1/5, 
\label{eq:2}\\
-1/5 < & \Gamma & \leq 0, 
\label{eq:3}\\
&\Gamma &\leq -1/5.
\label{eq:4}
\ea   

We recognise that our `clean', idealised scenario in which binaries are torqued by an entirely smooth, time-independent potential ignores various effects which must be included if our theory is to be of astrophysical relevance. The most obvious of these is perturbations of the binary by stellar flybys.  However for clarity we choose to focus exclusively on the idealised problem for most of the paper, before discussing non-ideal effects at the end, including flybys (\S\ref{scattering}).

In \S\ref{sec_General} we derive general results that hold for all $\Gamma$: conditions for the existence of fixed points (\S\ref{sec_fixedpoints}), the criteria for phase-space trajectories to librate or circulate (\S\ref{sect:gen-lib-circ}), the values of maximum and minimum eccentricities (\S\ref{sect:gen-min-max}), the timescale of eccentricity oscillations (\S\ref{toeo}), etc.  
Then, in \S\S\ref{sec_GammaRegime1}-\ref{sec_GammaRegime4}, we explore the details of each of the $\Gamma$ regimes (\ref{eq:1})-(\ref{eq:4}) separately. The validity of the doubly-averaged (secular) theory is verified numerically in \S\ref{sec_NumericalVerification}. The impact of general relativistic (GR) pericentre precession on the cluster tide-driven secular evolution is explored in \S \ref{GReffect}.
We collect our results in \S\ref{discussion}, discuss them in light of the existing literature and comment on the applicability of our formalism. We summarise our findings and discuss future applications in \S\ref{conclusions}.


\section{General aspects of secular dynamics}
\label{sec_General}


Our goal is to understand evolution of the orbital elements of the inner orbit of the binary --- eccentricity, inclination, etc. --- as it moves in the cluster potential. We do this by carefully investigating the phase-space of the dimensionless Hamiltonian $H_1^*$ (equation \eqref{H1starorb}).  

It will prove useful to express the Hamiltonian in terms of Delaunay variables.  We therefore define the actions $L=\sqrt{G(m_1+m_2) a}, J=L\sqrt{1-e^2}$, and $J_z = J\cos i$, their corresponding angles being the mean anomaly $M$, the argument of pericentre $\omega$ and the longitude of ascending node $\Omega$ respectively. (As in Paper I, the symmetry axis of the axisymmetric cluster is chosen to be the $Z$ axis.  In the case of a spherical cluster, the $Z$ axis is chosen perpendicular to the plane of the binary's outer orbit).

The Hamiltonian \eqref{H1Mt} does not depend on the mean anomaly $M$, so the action $L$ is conserved.  Hence we can choose to work with dimensionless versions of our variables $(J,J_z)$; following the notation of \cite{Antognini2015} we define
\begin{align} 
\Theta \equiv J_z^2/L^2 = (1-e^2)\cos^2 i, \,\,\,\,\,\,\,\,\,\,\,\,\,\,\, 
j\equiv J/L = \sqrt{1-e^2}, 
\label{eq:AMdefs}
\end{align} 
and we clearly have $0 \leq \Theta \leq j^2 \leq 1$. Obviously $j$ is just the dimensionless angular momentum.  Then 
\begin{align}  
\nn H_1^* =& \frac{1}{j^2}\left[ (j^2 - 3\Gamma \Theta)( 5-3j^2) \right. \\ &\left. - 15\Gamma(j^2-\Theta)(1-j^2) \cos 2\omega \right]. 
\label{H1star} 
\end{align} 

Since $H_1^*$ is independent of the longitude of the ascending node $\Omega$, the dimensionless quantity $\Theta$ is also an integral of motion (the analog of the `Kozai constant').  This simply reflects conservation of the $Z$-component of angular momentum since the doubly-averaged potential is axisymmetric. Since $\Theta$ is conserved, we can always infer the time evolution of binary inclination from the behavior of its eccentricity via $\cos^2i=(1-e^2)^{-1}\Theta$. Finally, definitions (\ref{eq:AMdefs}) imply that $e$ and $j$ must obey
\begin{align}   
0\leq e\leq \sqrt{1-\Theta},~~~~~\Theta^{1/2}\leq j\leq 1,
\label{eq:AMconstr}
\end{align}
to be physically meaningful for a fixed $\Theta$.

Given these considerations, the secular Hamiltonian is a function of the dimensionless angle-action variables $\omega$ and $j$. The equations of motion fully describing their evolution are
\begin{align}
\frac{\md \omega}{\md t} &= \frac{C}{L} \frac{\partial H_1^*}{\partial j}= \frac{6C}{L}\frac{[5\Gamma\Theta - j^4 +5\Gamma(j^4-\Theta)\cos2\omega]}{j^3}, 
\label{eom1} \\ 
\frac{\md j}{\md t} &=  -\frac{C}{L} \frac{\partial H_1^*}{\partial \omega}
 = -\frac{30\Gamma C}{L} \frac{(j^2-\Theta)(1-j^2)}{j^2}\sin 2\omega.
\label{eom2} 
\end{align} 
Our subsequent analysis of binary dynamics is largely based on these evolution equations.


\subsection{Phase portrait} 
\label{phase}


Since the dimensionless doubly-averaged Hamiltonian \eqref{H1star} ends up being a function of $j=\sqrt{1-e^2}$ and $\omega$, one can get a good perception of the secular dynamics by plotting contours of $H_1^*$ in the $(\omega,e)$ plane. We do this in Figures \ref{EccOmegaPlots}, \ref{EccOmegaPlots2}, \ref{EccOmegaPlots3}, \& \ref{EccOmegaPlots4} for the $\Gamma$ ranges \eqref{eq:1}-\eqref{eq:4} respectively,
and for varying $\Theta$.
In each panel the limiting eccentricity $e_\mathrm{lim}=\sqrt{1-\Theta}$ is represented by a dashed black line.  The direction in which orbits traverse their trajectories is indicated with green arrows in Figures \ref{EccOmegaPlots}a, \ref{EccOmegaPlots2}d, \ref{EccOmegaPlots3}a \& \ref{EccOmegaPlots4}d.

We will explain the features of these phase portraits as we discuss each of them individually in \S\S \ref{sec_GammaRegime1}-\ref{sec_GammaRegime4}. An observation that we would like to make now is that all phase-space trajectories are split into two families: librating and circulating. The librating orbits are closed contours of $H_1^*$ which loop around a fixed point (always located at $\omega=\pm \pi/2$ as explained in \S \ref{sec_fixedpoints}), whereas circulating orbits run over all $\omega \in (-\pi,\pi)$. The separatrices between families of librating and circulating orbits are indicated with red dashed lines. Depending on the values of $\Gamma$ and $\Theta$, one could have either only circulating orbits (typically the case for large $\Theta$, and always true for $-1/5<\Gamma\leq 0$), only librating orbits (a rare case realized for $\Gamma=1/5$, see Figure \ref{EccOmegaPlots2}), or a mix of both (for low enough $\Theta$ in most dynamical regimes). 


\subsection{Fixed points and orbit families} 
\label{sec_fixedpoints}


We start by exploring characteristics of the fixed points of the system, around which phase-space orbits librate. As these points are extrema of $H_1^*$ in $(\omega,j)$ space, they must be solutions of $\md \omega/\md t = \md j/\md t= 0$.  With a small amount of algebra, we find from equations (\ref{eom1})-(\ref{eom2}) two possible formal solutions for the non-trivial fixed points in our phase-space\footnote{Fixed points also exist for all $\omega$ along the lines $j=1$ and $j=\sqrt{\Theta}$, but these are trivial in the sense that they can never be reached by orbits which do not start on those lines. However, they are still important because they bound the phase-space and are often the locations of maximum/minimum $H_1^*$, see \S\ref{sec_ParameterRanges}.}, namely $(\omega,j) = (\pm\pi/2, j_\mathrm{f})$ where 
\begin{align} 
j_\mathrm{f}(\Theta,\Gamma) = \left(\frac{\Theta}{\Lambda
}\right)^{1/4},
~~~~~~~~~~~~
\Lambda(\Gamma)=\frac{5\Gamma+1}{10\Gamma}. 
\label{jfeqn} 
\end{align}  
The value of $\omega$ corresponding to fixed points agrees with the phase portraits discussed in \S \ref{phase}.

\begin{figure}
\centering
\includegraphics[width=0.98\linewidth]{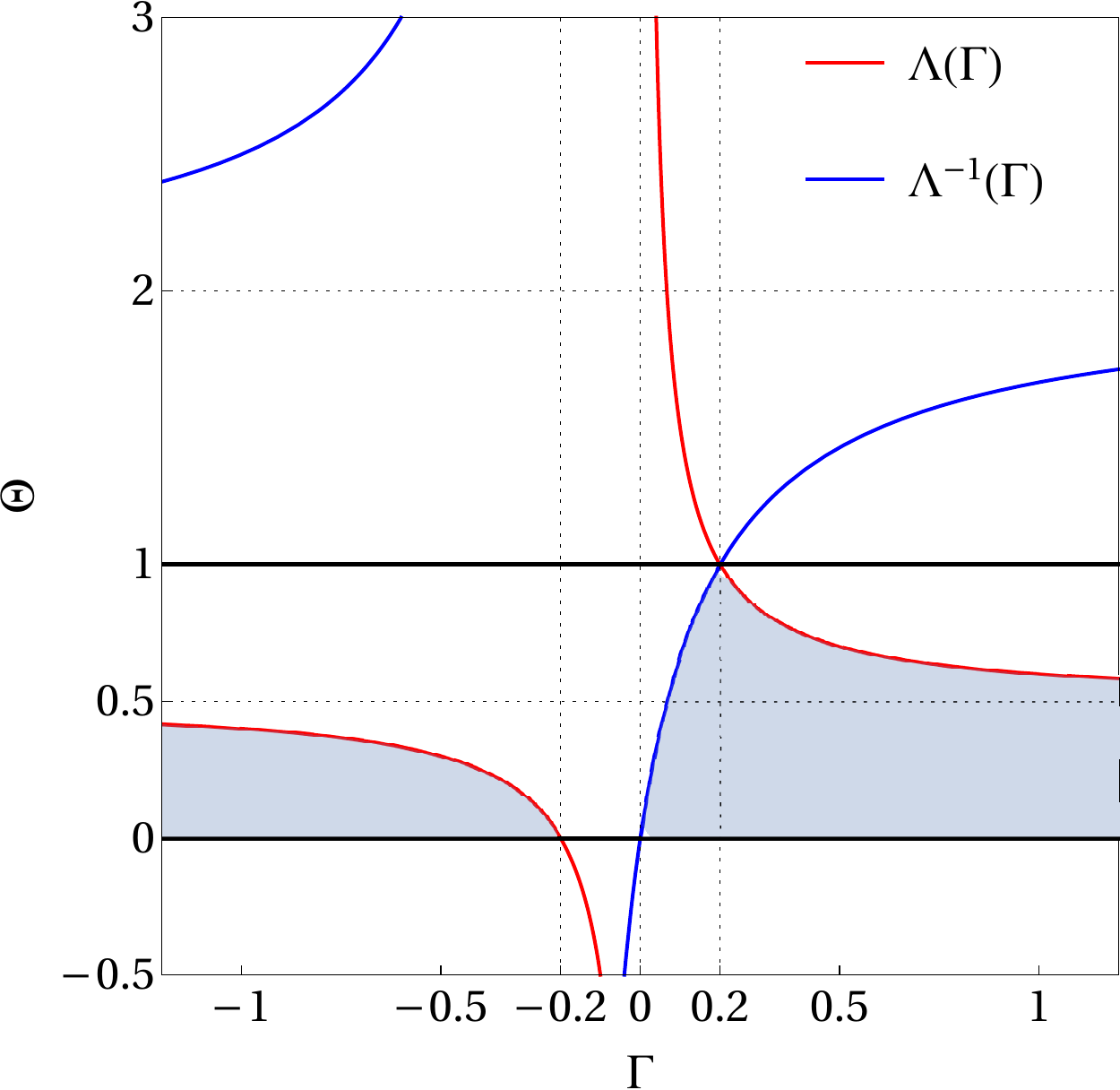}
\caption{Ranges of $\Theta$ for which fixed points and librating orbits can exist (shaded regions), shown as a function of $\Gamma$. Red and blue curves show $\Lambda$ (defined by equation \eqref{jfeqn}) and $\Lambda^{-1}$ respectively. Shaded regions correspond to the constraints (\ref{eq:Theta_constr}) and $0\leq\Theta\leq 1$.}
\label{fig_LambdaOfGamma}
\end{figure}

For a given $\Gamma$, fixed points can exist in the $(\omega,j)$ phase-space as long as $j_\mathrm{f}(\Theta,\Gamma)$ satisfies the condition (\ref{eq:AMconstr}). Solving the inequality $\sqrt{\Theta}<j_\mathrm{f}<1$ gives the following constraint on $\Theta$ for the existence of fixed points:
\begin{align}   
\Theta<\min\left(\Lambda,\,\Lambda^{-1}\right).
\label{eq:Theta_constr}
\end{align}   
Depending on the value of $\Gamma$, this constraint may or may not have meaningful solutions.  The functions $\Lambda(\Gamma)$ and $\Lambda^{-1}(\Gamma)$ are plotted in Figure \ref{fig_LambdaOfGamma}.  
We can see from this plot that there are four distinct $\Gamma$ regimes, given by the ranges (\ref{eq:1})-(\ref{eq:4}).    We will return to Figure \ref{fig_LambdaOfGamma} when discussing the existence of fixed points in \S\S\ref{sec_GammaRegime1}-\ref{sec_GammaRegime4}.

\subsection{Does a given orbit librate or circulate?}
\label{sect:gen-lib-circ}
 
  
Next we work out whether a given phase-space orbit with specified $H_1^*$, $\Theta$, $\Gamma$ librates or circulates. We do this by considering the behaviour at $\omega = 0$: from the morphology of the phase portraits (Figures \ref{EccOmegaPlots}, \ref{EccOmegaPlots2}, \ref{EccOmegaPlots3}, \& \ref{EccOmegaPlots4}), it is clear that if constant $H_1^*=H_1^*(\Theta,\Gamma,\omega,j)$ has a physical solution for $j$ at $\omega=0$ then the orbit circulates; if not, it has to librate about one of the fixed points. 

We find from equation \eqref{H1star} a formal solution
\begin{align} 
\label{j3} 
j^2(\omega=0) = \frac{5/3-2\Gamma\Theta -5\Gamma -H_1^*/3}{1-5\Gamma}. 
\end{align} 
The trajectory is circulating whenever the condition (\ref{eq:AMconstr}) is satisfied for $j(\omega=0)$. If $j(\omega=0)$ does not obey this inequality, then it does not represent a physically meaningful solution. As a result, the orbit must librate around one of the fixed points and never reach $\omega=0$.  

\begin{figure*}
\centering
\includegraphics[width=0.355\linewidth]{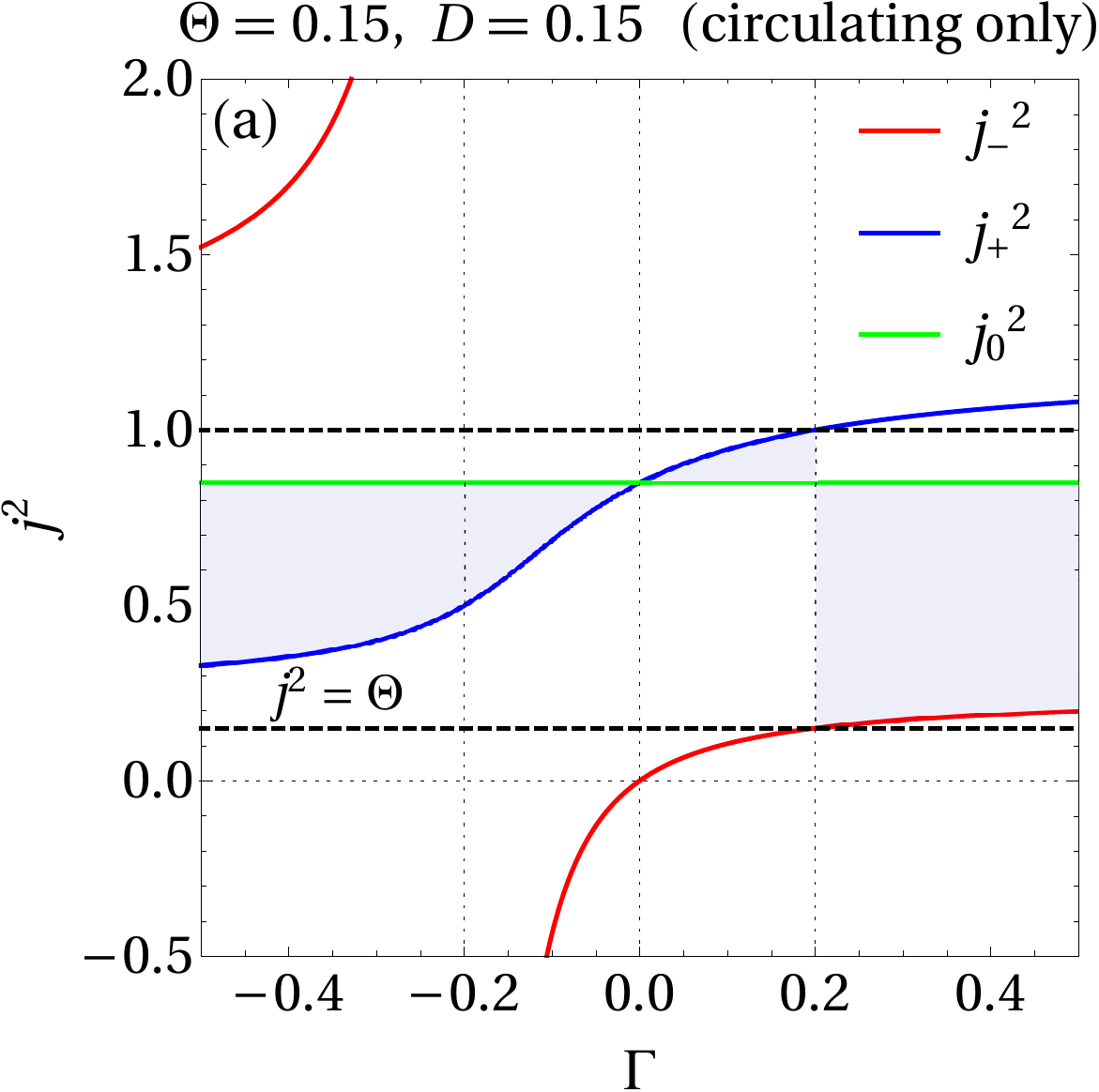}
\includegraphics[width=0.31\linewidth]{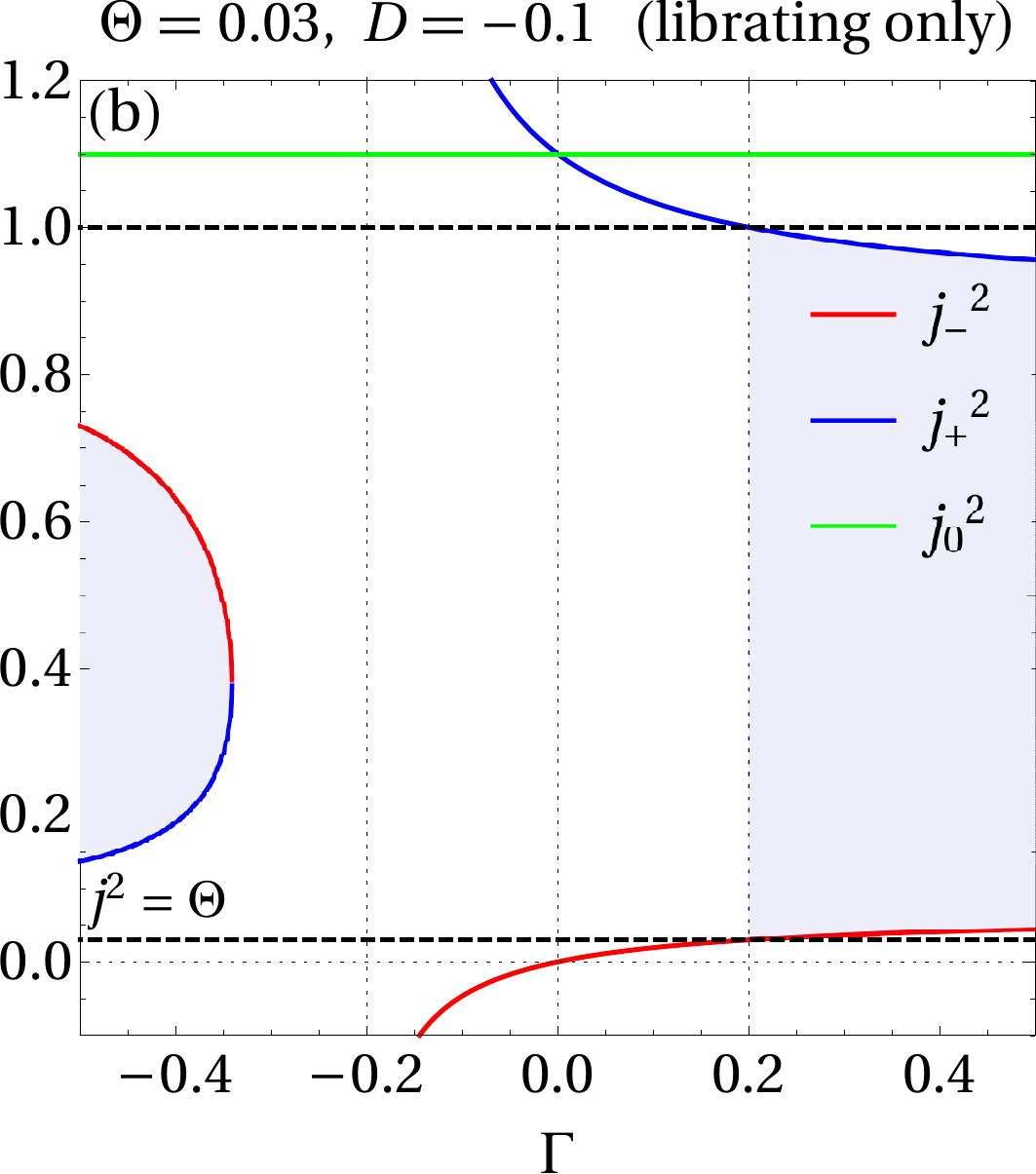}
\includegraphics[width=0.31\linewidth]{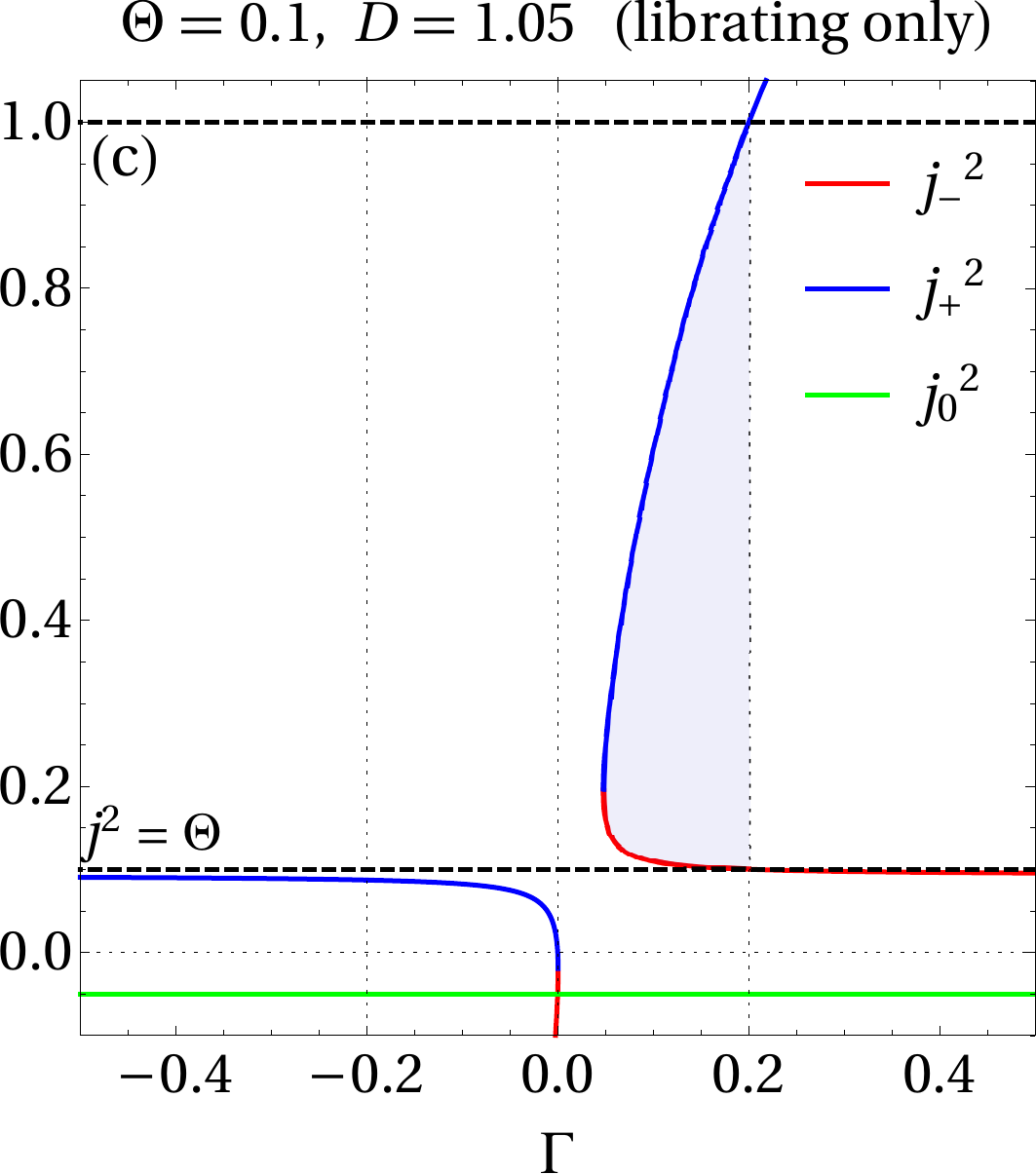}
\caption{Plot showing $j_0^2$ and $j_\pm^2$ defined by equations (\ref{eq:j3}) and (\ref{jpm}) as a function of $\Gamma$ for fixed $(D,\Theta)$. Vertical dotted lines indicate $\Gamma=\pm 1/5$, $0$. Thick horizontal dashed lines correspond to $j^2=\Theta$ and $j^2=1$.  Note that $j_\pm^2, j_0^2$ can take values outside the allowed $\Theta < j^2 < 1$ range and can even be negative, because they are simply formal solutions (see the text after equation (\ref{eq:Sigma})).  In each $\Gamma$ range, either two or none of $j_\pm^2, j_0^2$ are physically relevant --- physical solutions $j^2$ must always lie in the shaded regions.  Panel (a) corresponds to circulating orbits (which exist for all $\Gamma \neq 1/5$), while panels (b) and (c) correspond to librating orbits. From these plots one can read off the relative amplitudes of $j_\pm^2,j_0^2$ (and hence the values of $j_\mathrm{min/max}$ and $\Delta$) in each $\Gamma$ regime.}
\label{jpmplots}
\end{figure*}

Let us define the quantity 
\begin{align} 
\label{libconst} 
D\equiv 1-j^2(\omega=0) = \frac{H_1^*/3-2/3+2\Gamma\Theta}{1-5\Gamma}.
\end{align} 
It represents $e^2(\omega=0)$ and is an integral of motion since it depends on $H_1^*$ and $\Theta$. It will prove useful to eliminate $\Theta$ and $H_1^*$ in the above expression in favour of $e,i,\omega$.  We find after some algebra 
\begin{align} 
\label{Deqn} 
D= e^2 \left(1+ \frac{10\Gamma}{1-5\Gamma}\sin^2i\sin^2\omega \right).
\end{align} 

Equations (\ref{eq:AMconstr}) and (\ref{libconst}) imply that for $0<D< 1-\Theta$ the trajectory is circulating, whereas for $D<0$ or $D>1-\Theta$ it librates around a fixed point. If both families exist then the separatrix between them corresponds to either $D=0$ or $D=1-\Theta$ (see the upcoming Figure \ref{fig_Timescales}). 


\subsection{Maximum and minimum eccentricities}
\label{sect:gen-min-max}


We now find the minimum and maximum eccentricities that a binary reaches as it evolves along its phase-space trajectory. We do this by finding the extrema of $j$, which we call $j_\mathrm{min/max}$. 

From the phase portraits in Figures \ref{EccOmegaPlots}, \ref{EccOmegaPlots2}, \ref{EccOmegaPlots3}, \& \ref{EccOmegaPlots4} it is clear that librating orbits, whenever they exist, have both their minimum and maximum eccentricities at $\omega=\pm\pi/2$. On the other hand, for circulating orbits the maximum and minimum eccentricities can be at either $\omega=0$ or $\omega=\pm\pi/2$ depending on the value of $\Gamma$; see Figures \ref{EccOmegaPlots} and \ref{EccOmegaPlots2}. To find $j_\mathrm{min/max}$, we therefore separately plug $\omega=0, \pi/2$ into $H_1^*$ (equation \eqref{H1star}) and solve for $j$.  

The solution for $\omega=0$ is simply the square root of equation \eqref{j3}, and we denote it $j_0$: 
\begin{align}
j_0\equiv j(\omega=0)= \sqrt{1-D}.
\label{eq:j3}
\end{align} 
For $\omega = \pi/2$ there are two solutions, which we call $j_\pm$: 
\begin{align} 
j_\pm \equiv j(\omega=\pi/2) = \sqrt{\frac{\Sigma \pm \sqrt{\Sigma^2-10\Gamma\Theta\left(1+ 5\Gamma\right)}}{1+5\Gamma}}, 
\label{jpm} 
\end{align} 
where 
\begin{align} 
\Sigma &= \frac{1+5\Gamma}{2} +5\Gamma\Theta + \left(\frac{5\Gamma-1}{2} \right)D \\ &=\frac{1+5\Gamma}{2}(1-e^2) + 5\Gamma \left( \cos^2 i + e^2 \sin^2 i \cos^2 \omega \right).
\label{eq:Sigma}
\end{align}
We would like to stress here that although we use the notation $j_\pm^2, j_0^2$, these quantities are nothing more than possible solutions to the equations $\md j/\md t = \md \omega / \md t = 0$, and should not therefore be interpreted as always positive.  Indeed, depending on the $\Gamma$ regime either two or none of $j_\pm^2, j_0^2$ will lie in the allowed physical range $(\Theta,1)$ (equation \eqref{eq:AMconstr}) --- the remaining one or three will lie outside this range and can even be negative, but are physically irrelevant.  

This is demonstrated in Figure \ref{jpmplots}, which shows $j_\pm^2, j_0^2$ as functions of $\Gamma$ for various points in $(D,\Theta)$ space.  Depending on the dynamical regime (determined by the value of $\Gamma$) each of these $j_\pm$, $j_0$ solutions, if they exist, can correspond to either the minimum or maximum of $j$, as we describe in further detail in \S\S \ref{sec_GammaRegime1}-\ref{sec_GammaRegime4}.


\subsection{Range of parameter values} 
\label{sec_ParameterRanges}

We would first like to determine the range of values that the integral of motion $H_1^*$ can take. To do this we need to find the extrema of our Hamiltonian $H_1^*$ in the $(\omega,e)$ (or equivalently $(\omega,j)$) phase-space.  It is clear from the phase portraits (Figures \ref{EccOmegaPlots}, \ref{EccOmegaPlots2}, \ref{EccOmegaPlots3} \& \ref{EccOmegaPlots4}) that extrema of $H_1^*$ can only occur in three distinct locations: fixed points\footnote{Evaluating at $\omega=-\pi/2$ gives the same answers as evaluating at $\omega=\pi/2$.  Fixed points exist in both locations, but to avoid confusion with other upcoming $\pm$ signs we prefer just to consider $\omega=\pi/2$ from here on.} $(\omega=\pi/2, j=j_\mathrm{f})$, the line $j=1$, and the line $j=\sqrt{\Theta}$.  Evaluating $H_1^*$ in these locations, we find \begin{align} 
&H_1^*(j=1)=2(1-3\Gamma\Theta), 
\label{eq:Hj1}\\  
&H_1^*(j=\Theta)=(5-3\Theta)(1-3\Gamma), 
\label{eq:HjTheta}\\ 
& H_1^*(\omega=\pi/2, j=j_\mathrm{f})= 
\begin{cases} H_-, \,\,\,\,\,\Gamma > 0, \\ 
H_+, \,\,\,\,\,\Gamma \leq -1/5, 
\end{cases} 
\label{eq:Hjf}
\end{align}
where\footnote{Note that the functions $H_\pm$, unlike $j_\pm$, are \textit{not} two independent solutions.  The $\pm$ sign is due merely to an algebraic peculiarity that arises when evaluating the single-valued function $H_1^*$ at $(\pi/2,j_\mathrm{f})$, depending on the $\Gamma$ range. For a given $\Gamma$, only one of $H_\pm$ is correct. The same consideration holds for $D_\pm$ in the next paragraph.}
\begin{align} 
H_\pm = 5(1+3\Gamma)+24\Gamma\Theta \pm 6\sqrt{10\Gamma\Theta(1+5\Gamma)}.
\label{eq:Hpm}
\end{align} 
(No fixed points exist in the range $-1/5 < \Gamma \leq 0$). One must then investigate each $\Gamma$ regime independently to work out which of the above corresponds to a maximum or minimum.  We will not pursue the details here but we state the results for each $\Gamma$ regime in \S\S\ref{sec_GammaRegime1}-\ref{sec_GammaRegime4}, and summarise them in Table \ref{table_Hamiltonian}.  

\begin{figure*}
\centering
\includegraphics[width=0.85\linewidth]{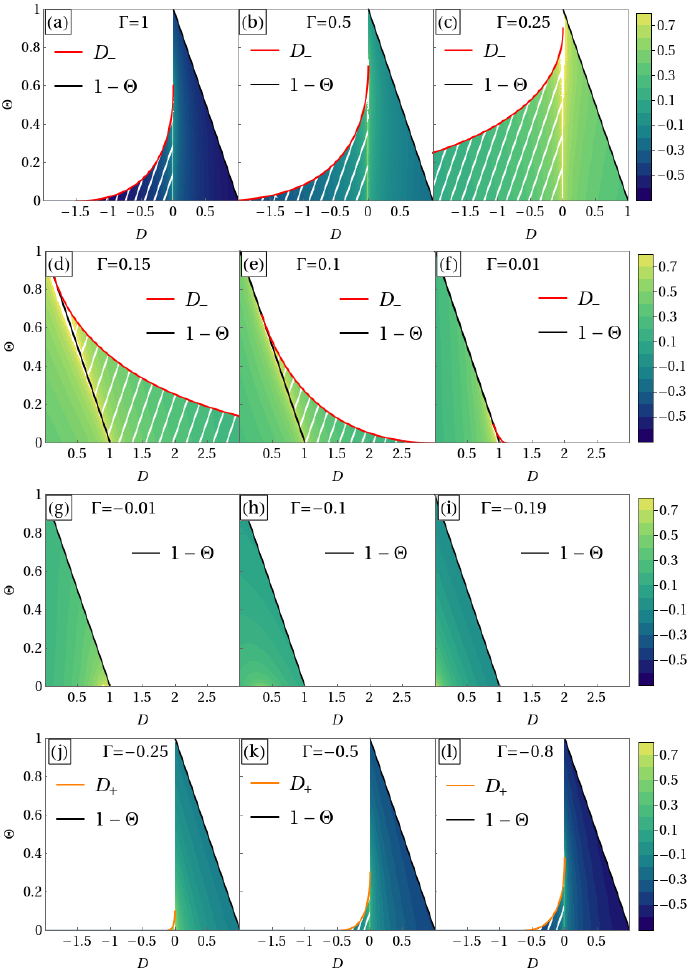}
\caption{Contour plots of $\log_{10}(t_\mathrm{sec}/t_1)$, where $t_\mathrm{sec}$ is the period of secular eccentricity oscillations (equation \eqref{timescaleeqn}), in the $(D,\Theta)$ plane for various values of $\Gamma$. Each row corresponds to one of the dynamical regimes (\ref{eq:1})-(\ref{eq:4}). In each panel the circulating orbits fill the triangle $0< D<1-\Theta$, while regions of librating orbits are indicated with white hashing (one of their boundaries is given by $D_\pm(\Theta)$ defined by equation (\ref{Dpm})).  The secular period diverges at the separatrix between the regions of librating and circulating orbits. It also diverges everywhere in $(D,\Theta)$ space in the special cases of $\Gamma=\pm 1/5$.  Bifurcations at $\Gamma = 1/5, 0, -1/5$ change the morphology of the $(D,\Theta)$ plane. }
\label{fig_Timescales}
\end{figure*}

Rather than $H_1^*, \Theta$, it is often convenient to take $D$ and $\Theta$ to be our primary integrals of motion, so that for fixed $\Gamma$ each phase-space trajectory of a binary corresponds to a point in the $(D,\Theta)$ plane (see the upcoming Figure \ref{fig_Timescales}).  It is obvious that $\Theta$ can always run between $0$ and $1$ for circulating orbits, and is bounded by equation \eqref{eq:Theta_constr} for librating orbits.  We would like to know which values $D$ can take for a given $\Theta, \Gamma$. From equation \eqref{libconst} we see that extrema of $D$ are also extrema of $H_1^*$, so we must evaluate $D$ at the same three locations as $H_1^*$, see equations (\ref{eq:Hj1})-(\ref{eq:Hjf}).  We find 
\begin{align} 
&D(j=1)=0, 
\label{eq:Dj1}\\  
&D(j=\Theta)=1-\Theta, 
\label{eq:DjTheta}\\ 
& D(\omega= \pi/2, j=j_\mathrm{f})= \begin{cases} D_-, \,\,\,\,\,\Gamma > 0, \\ D_+, \,\,\,\,\,\Gamma \leq -1/5, \end{cases} 
\label{eq:Djf}
\end{align}
where
\begin{align} 
\label{Dpm}
D_\pm =\frac{1}{(1-5\Gamma)} \left[1+5\Gamma +10\Gamma\Theta \pm2\sqrt{10\Gamma\Theta\left(1+5\Gamma\right)} \right]. 
\end{align}  
In each $\Gamma$ regime and for each type of orbit the maximum and minimum $D$ will correspond to some combination of $D_\pm$, $D=0$ and $D=1-\Theta$. These limits give rise to distinctive morphologies of the physically allowed regions in $(D,\Theta)$ plane --- see Figure \ref{fig_Timescales}. They are summarised in Table \ref{table_Ranges} and discussed in \S\S\ref{sec_GammaRegime1}-\ref{sec_GammaRegime4}.


\subsection{Timescales of eccentricity oscillations} 
\label{toeo}


We can also derive a general expression for the timescale of secular eccentricity oscillations, for any binary and for any given $\Gamma$. 

Since $H_1^*$ is a conserved quantity we can rearrange \eqref{H1star} to get $\omega$ explicitly in terms of $j$: 
\begin{align}
\cos 2\omega = \frac{(j^2-3\Gamma \Theta)(5-3j^2)-j^2H_1^*}{15\Gamma(j^2-\Theta)(1-j^2)}.
\end{align}
Plugging this into equation \eqref{eom2} allows us to eliminate $\omega$ from $\md j/\md t$, turning it into an equation for $j$ only. Factorizing the result and multiplying both sides by $j$ we arrive at a simple equation for the rate of change of $j^2$:
\begin{align} 
\frac{\md j^2}{\md t} = \pm \frac{12C}{L} \sqrt{(25\Gamma^2-1)(j_0^2-j^2)(j_+^2-j^2)(j^2-j_-^2)}. 
\label{djsquareddt} 
\end{align} 
The square root here is well defined because for $|\Gamma|>1/5$ and $<1/5$ the signs of bracketed terms change in such a way that the whole expression under the square root is positive\footnote{For $\Gamma=\pm 1/5$ we have $\md j^2/\md t = 0$.  This reflects the fact that secular evolution stalls and $t_{\rm sec}\to \infty$ for $\Gamma=\pm 1/5$ --- see equation \eqref{timescaleeqn} and Figure \ref{fig_Timescales}.}, as can be checked using the results collected in Table \ref{table_Ranges} (also see Figure \ref{jpmplots}).  
 
\begin{figure*}
\centering
\includegraphics[width=0.96\linewidth]{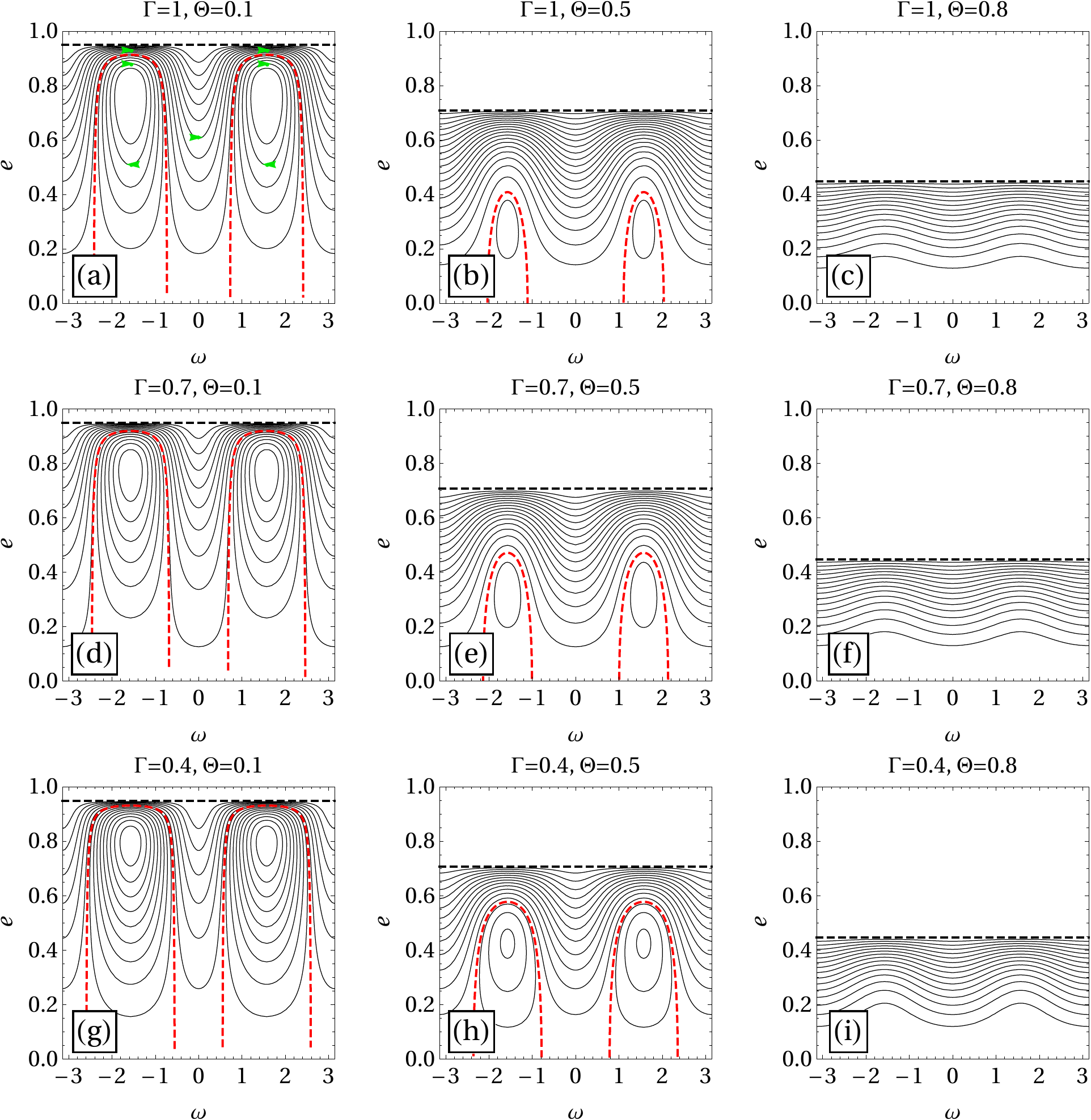}\quad
\caption{Contour plots of constant $H_1^*$ in the $(\omega,e)$ phase-space for $\Gamma>1/5$. Phase portraits are shown for $\Gamma =1, 0.7, 0.4$ and $\Theta =0.1,0.5,0.8$.  Contours are spaced linearly from $H_{1,\mathrm{min}}^*$ to $H_{1,\mathrm{max}}^*$.  The black dashed line shows the limiting eccentricity $e_\mathrm{lim}=\sqrt{1-\Theta}$ and the red dashed lines show separatrices between regions of librating and circulating orbits.  Green arrows in panel (a) show the direction of phase-space trajectories.}
\label{EccOmegaPlots}
\end{figure*}	

The maximum and minimum $j$ reached by a given phase-space orbit and satisfying the constraint (\ref{eq:AMconstr}) are denoted $j_\mathrm{min/max}$.  They correspond to two of the three possible roots $j_\pm, j_0$ depending on the orbit type and the value of $\Gamma$, as we will see in \S\S\ref{sec_GammaRegime1}-\ref{sec_GammaRegime4}. Regardless of their precise values, an entire oscillation runs from $j_\mathrm{min}$ to $j_\mathrm{max}$ and back again, so an entire secular oscillation takes 
\begin{align} 
\label{takes} 
t_\mathrm{sec} = 2 \int_{j^2_\mathrm{min}}^{j^2_\mathrm{max}} \left( \frac{\md j^2}{\md t} \right)^{-1} \md j^2, 
\end{align}  
which is expressible in terms of complete elliptic integrals of the first kind $K(k)=\int_0^{\pi/2} \md \alpha /\sqrt{1-k^2\sin^2\alpha}$ \citep{Gradshteyn2014}. Defining\footnote{Note that, in general, $\mathrm{min}[j_+^2,j_-^2,j_0^2] \neq j_\mathrm{min}^2$. This is because $\mathrm{min}[j_+^2,j_-^2,j_0^2]$ can take any value (including negative values, see Figure \ref{jpmplots}), whereas $j_\mathrm{min}$ is the physical minimum angular momentum reached by a given binary, which must lie between $\sqrt{\Theta}$ and $1$.  An analogous statement holds for $j_\mathrm{max}$.}
\begin{align} 
\Delta \equiv \mathrm{max}[j_+^2,j_-^2,j_0^2] - \mathrm{min}[j_+^2,j_-^2,j_0^2], 
\end{align} 
we find, in general, that 
\begin{align} 
t_\mathrm{sec} = \frac{t_1}{\sqrt{\vert 1-25\Gamma^2\vert  \Delta}}K\left( \sqrt{\frac{j_\mathrm{max}^2-j_\mathrm{min}^2}{\Delta}}\right), 
\label{timescaleeqn} 
\end{align}  
where 
\begin{align} 
t_1 &\equiv \frac{8}{3A}\sqrt{\frac{G(m_1+m_2)}{a^3}} 
\label{eq:t1}\\ 
&= 1.7 \, \mathrm{Gyr} \times \left( \frac{A^*}{0.5} \right)^{-1}\left( \frac{M}{10^5 M_\odot} \right)^{-1}\left( \frac{b}{\mathrm{pc}} \right)^{3} \nn \\ 
& \times \left( \frac{m_1+m_2}{M_\odot} \right)^{1/2}\left( \frac{a}{10\mathrm{AU}} \right)^{-3/2}
\end{align} 
is the characteristic secular timescale. In the numerical estimate we assumed the binary orbits a spherical cluster with scale radius $b$ and total mass $M$, and defined $A^* \equiv A/(GM/b^3)$. Maps of $A^*$ for different cluster potentials and binary orbits are presented in Paper I. 

In Paper I we also noted that $A\sim 4\pi^2 A^* T_\phi^{-2}$, where $T_\phi=2\pi(GM/b^3)^{-1/2}$ is the characteristic azimuthal period of the outer orbit of the binary around the cluster. Introducing the period of the inner orbit of the binary $T_b=2\pi[G(m_1+m_2)/a^3]^{-1/2}$, we can then use equation (\ref{eq:t1}) to estimate $t_1$ as
\begin{align}
t_1\sim \frac{4}{3\pi}\frac{T_\phi^2}{T_b}. \label{eq:roughtime}
\end{align}
A similar estimate of the characteristic secular timescale --- ratio of the square of the outer orbital period to the inner orbital period --- is known to hold for the LK problem \citep{Naoz2016}. Although equation \eqref{eq:roughtime} was derived for a spherical cluster potential, we also expect the same scaling to hold in (non-spherical) axisymmetric potentials.

The ratio $t_\mathrm{sec}/t_1$ is plotted in the $(D,\Theta)$ plane for various fixed $\Gamma$ in Figure \ref{fig_Timescales}.  In each panel, circulating orbits fill the triangle $0<D<1-\Theta$ while regions of librating orbits are shown with white hashing. In \S\ref{sec_fixedpoints} we noted that the four distinct $\Gamma$ regimes (equations (\ref{eq:1})-(\ref{eq:4})) would give rise to different phase-space behaviours.  We see in Figure \ref{fig_Timescales} that they also give rise to different morphologies of the allowed regions in the $(D,\Theta)$ plane, to be discussed in \S\S\ref{sec_GammaRegime1}-\ref{sec_GammaRegime4}.

The analytic derivation of eccentricity oscillation period (equation \eqref{timescaleeqn}) was previously done in the LK limit ($\Gamma=1$) by \citet{Vash1999} and \citet{Kinoshita2007}, and in the $\Gamma=1/3$ limit by \citet{Brasser2001}. Our expression (\ref{timescaleeqn}) generalizes their results for arbitrary external tidal potentials of the type explored in Paper I.
\\

 Much of \S\S\ref{sec_GammaRegime1}-\ref{sec_GammaRegime4} will be focused on deriving the values of $j_\mathrm{min/max}$, $\Delta$, and the bounds on $\Theta$, $D$ and $H_1^*$ appropriate to each of the distinct $\Gamma$ regimes (\ref{eq:1})-(\ref{eq:4}). For ease of reference all of the results are collected in Tables \ref{table_Ranges} and \ref{table_Hamiltonian}, which will be discussed in more detail in \S \ref{discussion}.


\section{Secular dynamics in the case \texorpdfstring{$\Gamma > 1/5$}{}} 
\label{sec_GammaRegime1}


In this section we focus exclusively on the case $\Gamma > 1/5$. In short, we find the dynamics in this regime to be qualitatively similar to (but quantitatively different from) the `test particle quadrupole' LK problem. A similar investigation in the LK limit ($\Gamma=1$) was carried out by \cite{Antognini2015}. This regime also covers the case of $\Gamma=1/3$ relevant for Oort Cloud comets perturbed by the Galactic tide \citep{Heisler1986}.

In Figure \ref{EccOmegaPlots} we plot contours of the dimensionless doubly-averaged perturbing Hamiltonian $H_1^*$ (equation \eqref{H1starorb}) in the $(\omega,e)$ plane, with $\Gamma=1, \,0.7, \,0.4$ from top to bottom and $\Theta=0.1, \,0.5, \,0.8$ from left to right.  

Fixed points exist in the left and centre columns, but not in the rightmost column (large $\Theta$) where there are only circulating orbits.  Circulating orbits show prograde pericentre precession $\dot{\omega}>0$, while librating orbits traverse clockwise loops in the $(\omega,e)$ plane.  As we increase $\Theta$ (i.e. move from left to right), the maximum eccentricity reached by the average orbit sharply decreases.  

We see that whenever a fixed point is present, the circulating orbits run `over the top' of the librating orbits.  As a result, \textit{the eccentricity of a fixed point provides a lower bound on the maximum eccentricity reached by any orbit in the phase-space}\footnote{This is not true for general $\Gamma$ --- c.f. \S\ref{sec_GammaRegime2}.}. Furthermore, the eccentricity of the fixed point increases slightly as we decrease $\Gamma$ (move down the page)  --- see equation \eqref{jfeqn}. Thus for $\Gamma$ close to (but greater than) $0.2$ and $\Theta \to 0$ more binaries may reach very high eccentricities than for $\Gamma \sim 1$.

We now proceed to explain these features quantitatively.


\subsection{Fixed points and orbit families}


Looking at Figure \ref{fig_LambdaOfGamma}, we see that whenever $\Gamma>1/5$, for fixed points to exist (shaded regions) $\Theta$ must be less than $\Lambda$.  However, $\Lambda$ has a minimum value of $1/2$ in this $\Gamma$ range (namely as $\Gamma \to \infty$); hence fixed points \textit{always} exist for $\Gamma > 1/5$ provided we choose $\Theta$ small enough, see equation (\ref{eq:Theta_constr}).  The precise requirement is 
\begin{align} 
\Theta \in \left(0, \Lambda \right),\,\,\,\,\,\,\,\, \Gamma > 1/5.
\label{thetarange} 
\end{align} 
The range of $\Theta$ for which fixed points exist increases as $\Gamma$ decreases.

This result allows us to understand the lack of librating orbits in panels (c), (f) and (i) in Figure \ref{EccOmegaPlots}: their $\Theta$ value is too large for the range \eqref{thetarange}.  Physically, at large $\Theta$ the inclination of the binary with respect to the symmetry plane of the cluster is too low for its tidal field to efficiently torque the binary to high eccentricities.  In the LK case $\Gamma=1$ we recover the classic result that the critical $\Theta$ range for fixed points to exist is $\Theta\in(0,3/5)$.  For initially circular binaries the resulting minimum inclination in then $\cos^{-1}(\sqrt{3/5})\approx 39.2^\circ$ (see also \S\ref{sec:crit_inc_circular}).

We can convert \eqref{jfeqn} into an eccentricity via $e=\sqrt{1-j^2}$: 
\begin{align} 
\label{fixedecc} 
e_\mathrm{f} = \left[1- \sqrt{ \frac{\Theta}{\Lambda}}\right]^{1/2}. \end{align} 
This helps us to understand why in Figure \ref{EccOmegaPlots}, the maximum eccentricity is largest for small $\Theta$, and only weakly dependent on $\Gamma$: since each trajectory reaches a maximum eccentricity which is \textit{at least} $e_\mathrm{f}$,  decreasing $\Theta$ will increase that maximum. And since $\Lambda$ is a weak function of $\Gamma$ in this range (taking values $\Lambda \in (1/2,1)$ --- see Figure \ref{fig_LambdaOfGamma}), dependence on $\Gamma$ is not very strong.

	
\subsection{Range of parameter values} 

In \S\ref{sec_ParameterRanges} we mentioned that $\Theta$ is bounded by equation \eqref{eq:Theta_constr} for librating orbits and runs between $0$ and $1$ for circulating orbits.  From equations \eqref{eq:Dj1}-\eqref{eq:Djf} we know that in the regime $\Gamma > 1/5$, the extrema of $D$ correspond to some combination of $D=0$, $D=1-\Theta$ and $D_-$. It can be shown that $D_-$ is negative all for $\Theta$ when $\Gamma > 1/5$, while $D=1-\Theta$ is obviously positive. Hence the bounds on $D$ are: 
\begin{align} 
D\in \begin{cases} \left(D_-, 0\right), \,\,\,\,\,\,\, &\Gamma>1/5, \,\,\mathrm{librating \,\, orbits,} \\ \left(0, 1-\Theta\right), \,\,\,\,\,\,\, &\Gamma>1/5,\,\, \mathrm{circulating \,\, orbits.} \end{cases} 
\label{Drange} 
\end{align} 
These ranges dictate the morphology of the $(D,\Theta)$ plane in the top row of Figure \ref{fig_Timescales}.

It is easy to verify that for $\Gamma>1/5$ the minimum of $H_1^*$ is attained at $j^2=\Theta$ (i.e. along the black dashed lines of limiting eccentricity in Figure \ref{EccOmegaPlots}), so $H^*_{1,\mathrm{min}}$ is given by equation \eqref{eq:HjTheta}. As for $H^*_{1,\mathrm{max}}$, if fixed points exist for a given $\Gamma, \Theta$, then the Hamiltonian is maximised at the fixed point and $H^*_{1,\mathrm{max}}=H_-$, see equation \eqref{eq:Hjf}. If fixed points do not exist, then $H^*_{1,\mathrm{max}}$ is attained at $j^2=1$ (i.e. along the line of zero eccentricity), and hence is given by equation \eqref{eq:Hj1}.


\subsection{Maximum and minimum eccentricities}
\label{sec_maxecc1}

For $\Gamma>1/5$, librating orbits, if they exist, will have minimum angular momentum $j_\mathrm{min}=j_-$ and maximum angular momentum $j_\mathrm{max}=j_+$, since these are the two solutions at $\omega=\pi/2$, and $j_+ > j_-$ (see Figure \ref{EccOmegaPlots} and equation \eqref{jpm}). 

Meanwhile, circulating orbits also reach minimum angular momentum at $\omega=\pi/2$ (Figure \ref{EccOmegaPlots}), so that their $j_\mathrm{min}=j_-$. At the same time, $j_\mathrm{max}=j_0$ for circulating orbits since the maximum value of their angular momentum (lowest eccentricity) is reached at $\omega=0$. 

Maximum and minimum eccentricites are then given by $e_\mathrm{max/min} = (1-j_\mathrm{min/max}^2)^{1/2}$ for both types of phase-space trajectories.


\subsection{Timescales of eccentricity oscillations} 
\label{toeo_a}


To find the timescale $t_\mathrm{sec}$ using equation (\ref{takes}) we need to know the values of $j_\mathrm{min/max}$ (\S\ref{sec_maxecc1}) and $\Delta$ for each orbit family.  First of all, in the $\Gamma>1/5$ regime it is clear (equation \eqref{jpm}) that we always have $j_+^2 > j_-^2 \geq 0$ (see Figure \ref{jpmplots} for an illustration).  

Librating orbits in this regime have $D<0$ (see Figure \ref{fig_Timescales}), so we see from equation \eqref{j3} that $j_0^2 > 1$.  Since $j_\pm^2$ provide upper and lower bounds on the true angular momentum $j^2$, it must be the case that $j_-^2 < j^2 < j_+^2 < j_0^2$, from which we read off that librating orbits have $\Delta = j_0^2-j_-^2$.  

Meanwhile for circulating orbits $D>0$ and we know that $j^2$ is bounded from above by $j_0^2<1$, so we must have $j_-^2 < j^2 < j_0^2 < j_+^2$, see Figure \ref{jpmplots}a.  Hence $\Delta = j_+^2-j_-^2$ for circulating orbits.

Using these results we plot the ratio $\log_{10}(t_\mathrm{sec}/t_1)$ in $(D,\Theta)$ space for $\Gamma=1,0.5,0.25$ in the top row of Figure \ref{fig_Timescales}.  The triangle $0 < D <1-\Theta$ contains the circulating orbits in each case; the orbits outside of this triangle librate.  The bounds on $\Theta$ and $D$ are given by equations \eqref{thetarange} and \eqref{Drange} respectively. The timescale for oscillations is seen to depend primarily on the proximity to the separatrix between librating and circulating orbits at $D=0$.  Along the separatrix the timescale for secular oscillations diverges\footnote{To see this note that as $D\to 0$, we have $j_+ \to j_0=1$, and the function $K(k)$ diverges logarithmically as $k\to 1$.}. As we decrease $\Gamma$ from $1$, the region containing librating orbits gets larger, though of course the triangle of circulating orbits is unchanged.  The value $t_\mathrm{sec}/t_1$ is amplified when we decrease $\Gamma$, so that the timescale for oscillations increases as we decrease $\Gamma$ (at fixed $A$). 

For any $\Gamma$ in the approximate range $0.25 \leq \Gamma \lesssim 0.5$, and sufficiently far from the separatrix, $t_1$ provides a good estimate of $t_\mathrm{sec}$. For $\Gamma \gtrsim 0.5$, large portions of the $(D,\Theta)$ space have secular timescales that are shorter than $t_1$ by  a factor of a few. As $\Gamma \to 0.2$, the timescale diverges everywhere in $(D,\Theta)$ space (see equation \eqref{timescaleeqn}).
\\

All of the results arrived at in this section will change when we leave the regime $\Gamma > 1/5$.  


\section{The case \texorpdfstring{$0<\Gamma \leq 1/5$}{}} 
\label{sec_GammaRegime2}


We now turn to the second regime, $0<\Gamma \leq 1/5$, which is realised quite naturally for example by binaries orbiting close to the core of a spherical cluster (see Paper I).

\begin{figure*}
\centering
\includegraphics[width=0.96\linewidth]{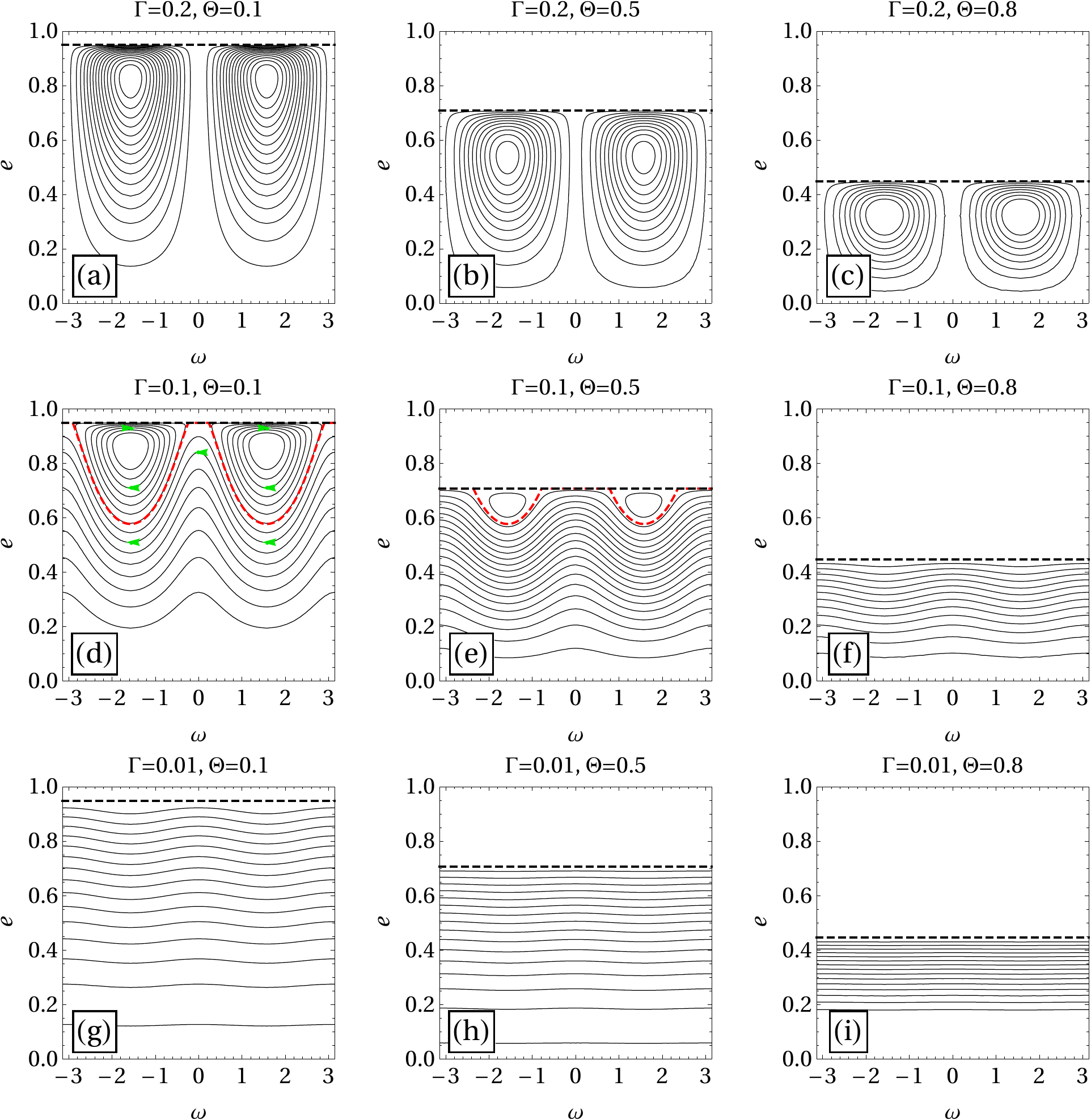}\quad
\caption{Contour plots of constant $H_1^*$ in the $(\omega,e)$ plane as in Figure \ref{EccOmegaPlots}, but now for the regime $0<\Gamma \leq 1/5$. Phase portraits are shown for $\Gamma=0.2,0.1,0.01$. Note the different sign of precession of circulating orbits compared to Figure \ref{EccOmegaPlots} as well as the change of morphology: circulating orbits (when they exist, i.e. for $\Gamma\neq 1/5$) now run {\it below} the islands of libration.}
\label{EccOmegaPlots2}
\end{figure*}

We begin as in \S\ref{sec_GammaRegime1} by showing the phase portraits as one varies $\Gamma$ and $\Theta$. In Figure \ref{EccOmegaPlots2} we plot contours of $H_1^*$, with $\Gamma=0.2,0.1,0.01$ from top to bottom and $\Theta=0.1, \,0.5, \,0.8$ from left to right.  The green arrows in panel (d) show the sense in which orbits traverse their trajectories. We immediately note qualitative differences between the plots with $0<\Gamma \leq 1/5$ (Figure \ref{EccOmegaPlots2}) and those for $\Gamma>1/5$ (Figure \ref{EccOmegaPlots}).  For $\Gamma=1/5=0.2$ only librating orbits exist, as we can see in the top row of Figure \ref{EccOmegaPlots2}.  In panels (d) and (e) we again have fixed points at $\omega = \pm \pi/2$ and librating orbits surrounding them, but note that the circulating orbits which exist for $\Gamma=0.1, 0.01$ now have eccentricity \textit{minima} at $\omega=\pm\pi/2$ and \textit{maxima} at $\omega=0$, which is the opposite of the $\Gamma >1/5$ case.  In the $(\omega,e)$ phase plane, circulating orbits now run `underneath' the librating orbits, whereas for $\Gamma > 1/5$ they ran `over the top'.  As a result, \textit{fixed points no longer provide a lower bound on the maximum eccentricity}. The librating orbits still run clockwise but the circulating orbits now display retrograde precession, $\dot{\omega}<0$, whereas in the $\Gamma > 1/5$ case we had $\dot{\omega}>0$.

\subsection{Fixed points and orbit families}


Figure \ref{fig_LambdaOfGamma} shows that for $\Gamma=1/5$ we have $\Lambda = \Lambda^{-1} = 1$. This means that all $\Theta \in (0,1)$ accommodate librating trajectories, and nothing circulates. Moving to lower $\Gamma$ in Figure \ref{fig_LambdaOfGamma} one finds $\Lambda^{-1}(\Gamma)<\Lambda(\Gamma)$, so that for fixed points and librating trajectories to exist $\Theta$ must now now be less than $\Lambda^{-1}$ (i.e. the shaded region is now bounded by the blue curve):
\begin{align} 
\Theta \in \left(0, \Lambda^{-1} \right),\,\,\,\,\,\,\,\, 0 < \Gamma < 1/5 
\label{thetarange2}. 
\end{align} 
The range of $\Theta$ for which fixed points exist diminishes as $\Gamma \to 0$.  

The fact that circulating orbits have changed their sense of pericentre precession from prograde to retrograde is easily explained by calculating $\md \omega/\md t$ (equation \eqref{eom1}) at $\omega=0$.  The result is proportional to $(5\Gamma-1)$, so that $\dot{\omega}$ is positive when $\Gamma > 1/5$ and negative when $\Gamma < 1/5$.
	

\begin{figure*}
\centering
\includegraphics[width=0.96\linewidth]{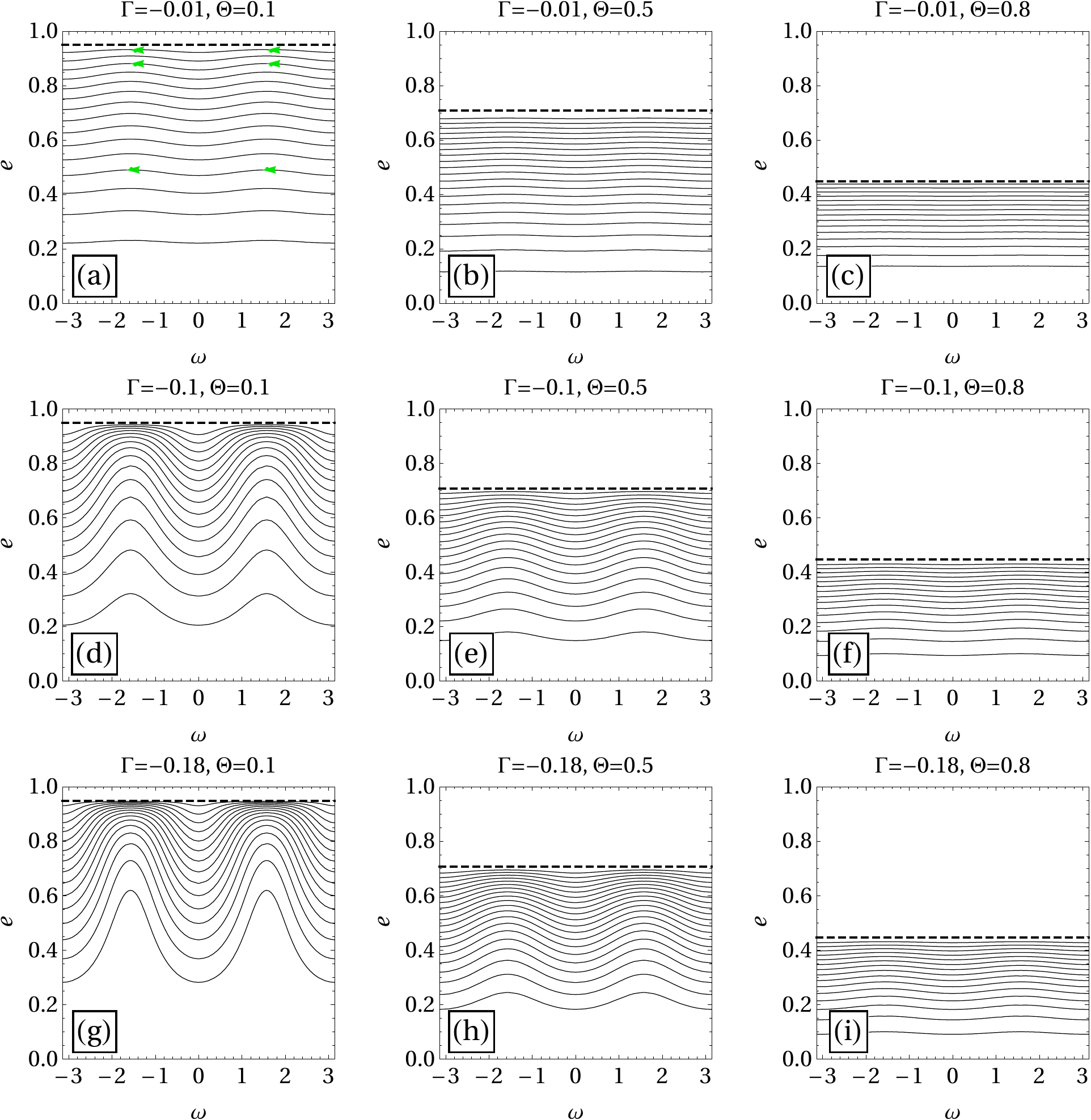}\quad
\caption{Contour plots of constant $H_1^*$ in the $(\omega,e)$ plane as in Figure \ref{EccOmegaPlots}, but now for the regime $-1/5<\Gamma \leq 0$. Phase portraits are shown for $\Gamma=-0.01,-0.1,-0.18$. Note the absence of fixed points and librating orbits for all $\Gamma$ and $\Theta$ in this regime.}
\label{EccOmegaPlots3}
\end{figure*}
 
\begin{figure*}
\centering
\includegraphics[width=0.96\linewidth]{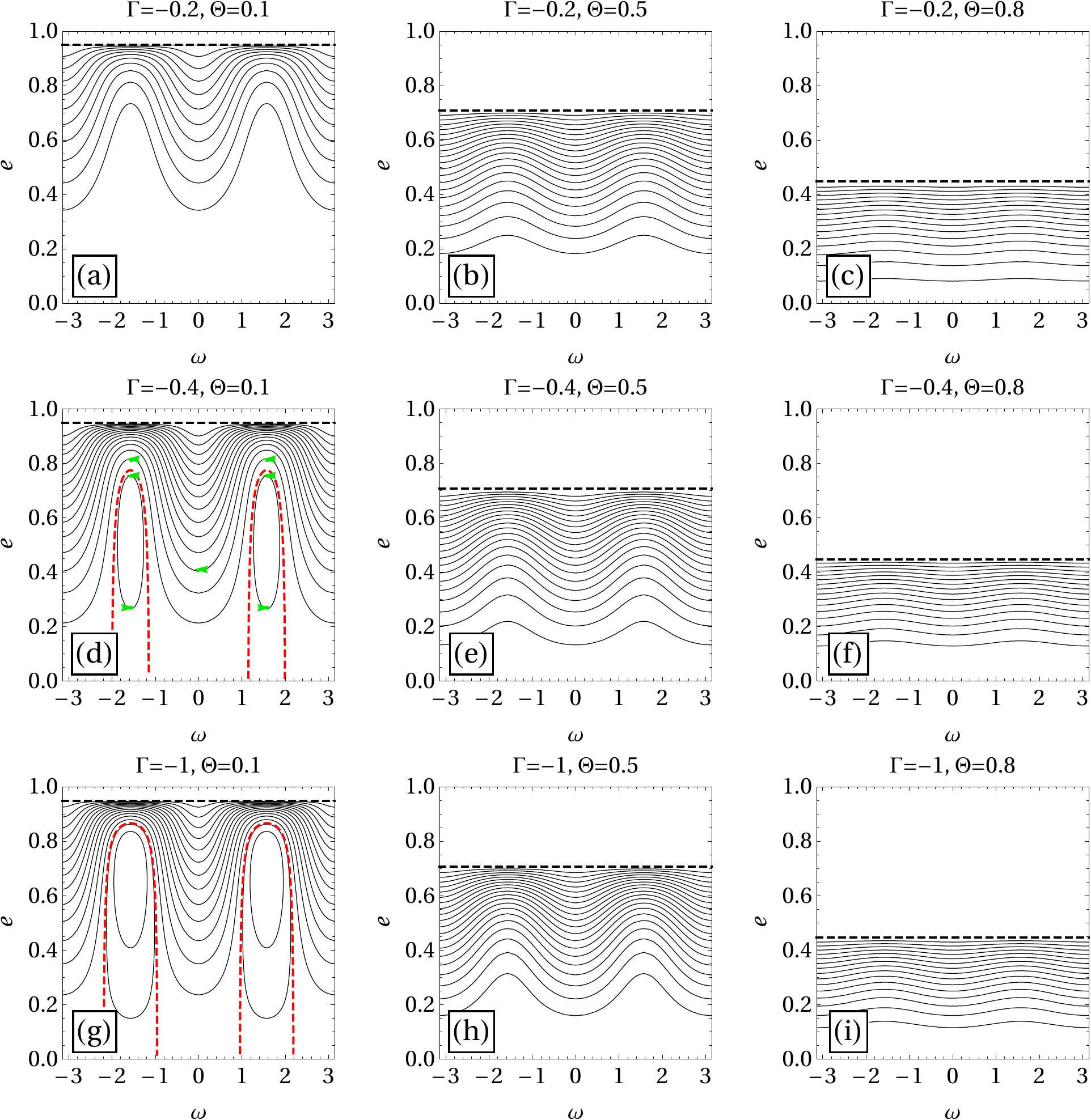}\quad
\caption{Contour plots of constant $H_1^*$ in the $(\omega,e)$ plane as in Figure \ref{EccOmegaPlots}, but now for the regime $\Gamma \leq -1/5$. Phase portraits are shown for $\Gamma=-0.2,-0.4,-1$.}
\label{EccOmegaPlots4}
\end{figure*}


\subsection{Range of parameter values} 
\label{rangeof2}


We again want to derive the bounds on the $(D,\Theta)$ plane and to find the extrema of $H_1^*$.  We begin as in \S \ref{sec_ParameterRanges} by considering the limits on $D$.

First of all, according to its definition (\ref{eq:j3}), $D$ diverges when $\Gamma=1/5$, which is in agreement with the absence of circulating orbits in the top row of Figure \ref{EccOmegaPlots2} (circulating orbits require $0<D<1-\Theta$). For $0 <\Gamma<1/5$ the quantity $10\Gamma/(1-5\Gamma)$ is positive; then it follows from equation (\ref{Deqn}) that the minimum value of $D$ for any fixed $\Theta$ is zero (attained for $e=0$). Hence, $D=0$ is the lower bound.  The other possible bounds on $D$ are $D_-$ and $1-\Theta$ (see equations \eqref{eq:Dj1}-\eqref{eq:Djf}); it turns out that $D_- \geq 1-\Theta$ for all $\Theta$ in this $\Gamma$ range, so we conclude that: 
\begin{align} 
D\in \begin{cases} (1-\Theta,D_-), \,\,\,\,\,\,\, &0 < \Gamma<1/5,\,\, \mathrm{librating \,\, orbits,} \\ (0, 1-\Theta), \,\,\,\,\,\,\, &0 < \Gamma<1/5, \,\,\mathrm{circulating \,\, orbits.} \end{cases} 
\label{Drange2} 
\end{align} 
Looking at the timescale plots in the second row of Figure \ref{fig_Timescales}, we see that the $(D,\Theta)$ plane morphology has completely changed compared to $\Gamma >1/5$ (top row).

This time the minimum of $H_1^*$ is situated at $j^2=1$ (i.e. along the line of zero eccentricity in Figure \ref{EccOmegaPlots2}), so  $H_{1,\mathrm{min}}^*$ is given by equation \eqref{eq:Hj1}. If fixed points exist for a given ($\Gamma,\Theta$), then $H_{1,\mathrm{max}}^*=H_-$ is found at the fixed point, see equation \eqref{eq:Hjf}. Otherwise $H_{1,\mathrm{max}}^*$ is found on the line $j^2=\Theta$ (the line of limiting eccentricity), and its value is given by equation \eqref{eq:HjTheta}.


\subsection{Maximum and minimum eccentricities}


For $\Gamma>1/5$ it was easy to determine for example that librating orbits have $j_-^2 < j^2 < j_+^2 < j_0^2$.  From this we were able to instantly read off $j_\mathrm{min/max} = j_\pm$ and $\Delta = j_0^2 - j_-^2$ (\S\ref{toeo_a}).  When $\Gamma\leq 1/5$ things become more complicated.  While it is possible to calculate the analagous results in each $\Gamma$ regime algebraically, it is easier and more informative to look at Figure \ref{jpmplots}, in which we plot $j_\pm^2, j_0^2$ as a function of $\Gamma$ for various fixed $(D,\Theta)$.  

Let us consider only Figure \ref{jpmplots}a to begin with, which is for $\Theta=0.15, D=0.15$ and therefore sits inside the triangle of circulating orbits ($0 < D <1-\Theta$) for any $\Gamma\neq 1/5$.  The properties of circulating orbits in each $\Gamma$ regime can be read off from Figure \ref{jpmplots}a.   In the current regime $0<\Gamma<1/5$ we see that $j_-^2 < j_0^2 < j_+^2$. Moreover, the horizontal dotted lines represent $j^2=\Theta$ and $j^2=1$, so physical solutions $j^2$ must lie between these two lines.  Hence $j^2$ must be bounded by the green and blue lines (that is by $j_0^2$ and $j_+^2$). We therefore deduce that $j_\mathrm{min}=j_0$, $j_\mathrm{max}=j_+$, and $\Delta =j_+^2- j_-^2$.

This exercise can be repeated with panels (b) and (c) to find the corresponding results for librating orbits.  The only subtlety is that as one changes $\Gamma$, the region of the $(D,\Theta)$ plane corresponding to librating orbits changes shape and moves (this is most clearly seen by comparing different panels along the same row in Figure \ref{fig_Timescales}).  The $(D,\Theta)$ points we have chosen for panels (b) and (c) in Figure \ref{jpmplots} do not \textit{quite} fall inside the region of librating orbits for some $\Gamma$ because that region becomes too small, which is why for example the red and blue curves in panel (b) are not defined in a small range near $\Gamma=-0.2$.  However these plots are good enough for anticipating the results of the algebraic calculations. When $0 < \Gamma < 1/5$ the $j^2$ ranges for librating orbits can be read off from Figure \ref{jpmplots}c: for $j^2$ to lie between the upper and lower horizontal dotted lines we must have $j_0^2<j_-^2<j^2<j_+^2$; as a result, $j_\mathrm{min}=j_-$, $j_\mathrm{max}=j_+$, and $\Delta =j_+^2- j_0^2$.   


\subsection{Timescales of eccentricity oscillations} 
\label{toeo2}


In the second row of Figure \ref{fig_Timescales} we present contour plots of $\log_{10}(t_\mathrm{sec}/t_1)$ in $(D,\Theta)$ space for $\Gamma=0.15,0.1,0.01$.  This time the bounds on $\Theta$ and $D$ are given by equations \eqref{thetarange2} and \eqref{Drange2} respectively.  The triangle $0 < D <1-\Theta$ again contains the circulating orbits, but now the librating orbits have moved to the right of this triangle, and the separatrix corresponds to $D=1-\Theta$. Along the separatrix the timescale for secular oscillations again diverges.  The timescale is also infinite everywhere in the $(D,\Theta)$ plane in the special case $\Gamma=1/5$ (see equation \eqref{timescaleeqn}).  

For $0 \leq \Gamma \lesssim 0.15$, we see that $t_1$ again provides a fair estimate of $t_\mathrm{sec}$, although it is really a lower bound on $t_\mathrm{sec}$ in much of the $(D,\Theta)$ space, whereas for $\Gamma > 1/5$ it was typically an effective upper bound.


\section{The case \texorpdfstring{$-1/5 < \Gamma \leq 0$}{}} 
\label{sec_GammaRegime3}

The two regimes $-1/5 < \Gamma \leq 0$ and $\Gamma \leq -1/5$ cannot be realised by binaries orbiting spherical potentials.  In fact, they typically require rather extreme orbits in strongly aspherical potentials.  For this reason, here and in \S\ref{sec_GammaRegime4} we simply summarise the qualitative results for each regime --- the details are given in Appendices \ref{ap:regimeIII} and \ref{ap:regimeIV} respectively.\\

The regime $-1/5 < \Gamma \leq 0$ is typically realised when the binary's outer orbit makes large excursions in the $Z$ direction in an oblate potential, i.e. is highly inclined with respect to the potential's symmetry plane (see Paper I). Orbits in prolate potentials can also result in this range of $\Gamma$.
	
In Figure \ref{EccOmegaPlots3} we plot contours of constant $H_1^*$ for $\Gamma=-0.01,-0.1,-0.18$ from top to bottom and $\Theta=0.1, \,0.5, \,0.8$ from left to right. The phase portrait has undergone another bifurcation as we passed through $\Gamma=0$. We see from Figure \ref{EccOmegaPlots3} that only circulating orbits exist in this $\Gamma$ regime, all with retrograde precession ($\dot{\omega}<0$). Fixed points and librating trajectories do not emerge at all in this regime, which is explained in Appendix \ref{ap:regimeIII} along with some more technical details.


\section{The case \texorpdfstring{$\Gamma \leq -1/5$}{}} 
\label{sec_GammaRegime4}

Situations where $\Gamma \leq -1/5$ may arise, for example, from orbits that are highly inclined with respect to the symmetry plane of a strongly flattened potential. Cylindrical potentials, i.e. ones in which $\Phi(R,Z)=\Phi(R)$ (no $Z$-dependence, extremely prolate configurations), also fall into this regime as they always have $\Gamma=-1/3$, see Paper I.  

In Figure \ref{EccOmegaPlots4} we plot contours of $H_1^*$ with $\Gamma=-0.2,-0.4,1$ from top to bottom and $\Theta=0.1, \,0.5, \,0.8$ from left to right.  A final bifurcation has occured as we moved below $\Gamma=-1/5$, such that the phase portrait morphology now looks similar to the case $\Gamma >1/5$, with fixed points at $\omega = \pm \pi/2$ and circulating orbits running `over the top' of librating orbits in the $(\omega,e)$ plane. However, the direction of $\omega$ precession is the opposite to the $\Gamma>1/5$ case: circulating orbits in the $\Gamma \leq -1/5$ regime have retrograde precession ($\dot{\omega}<0$) while librating orbits run anticlockwise. Appendix \ref{ap:regimeIV} provides additional details on this $\Gamma$ regime.\\

This completes our detailed exploration of the dynamical regimes corresponding to the different ranges of $\Gamma$ (equations (\ref{eq:1})-(\ref{eq:4})) considered in this work.


\section{Accuracy of the doubly-averaged approximation} 
\label{sec_NumericalVerification}


In Paper I we developed three successive levels of approximation for the evolution of the binary's inner orbital elements.  

\begin{figure*}
\centering
\includegraphics[width=0.31\linewidth]{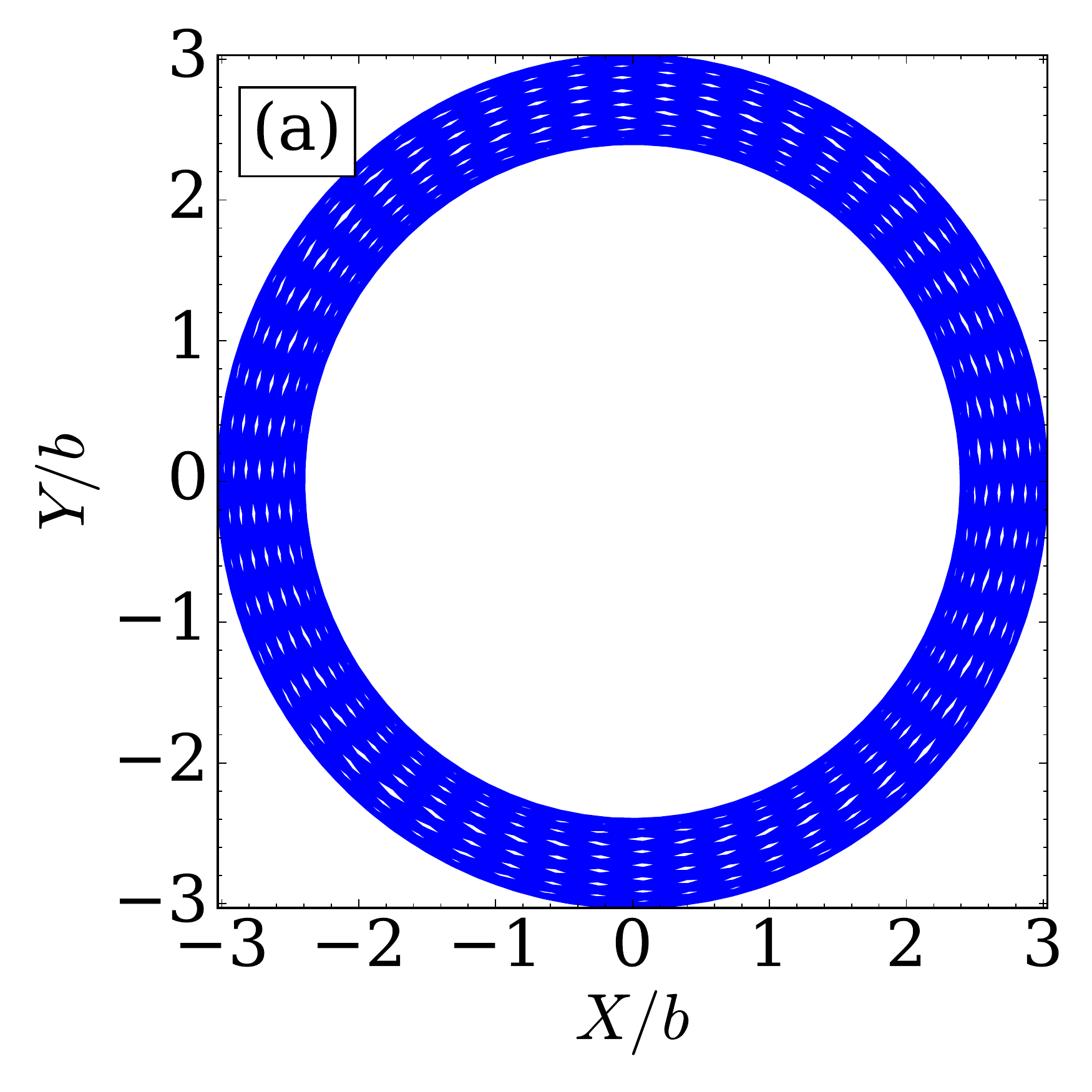}
\raisebox{0.05cm}{\includegraphics[width=0.31\linewidth]{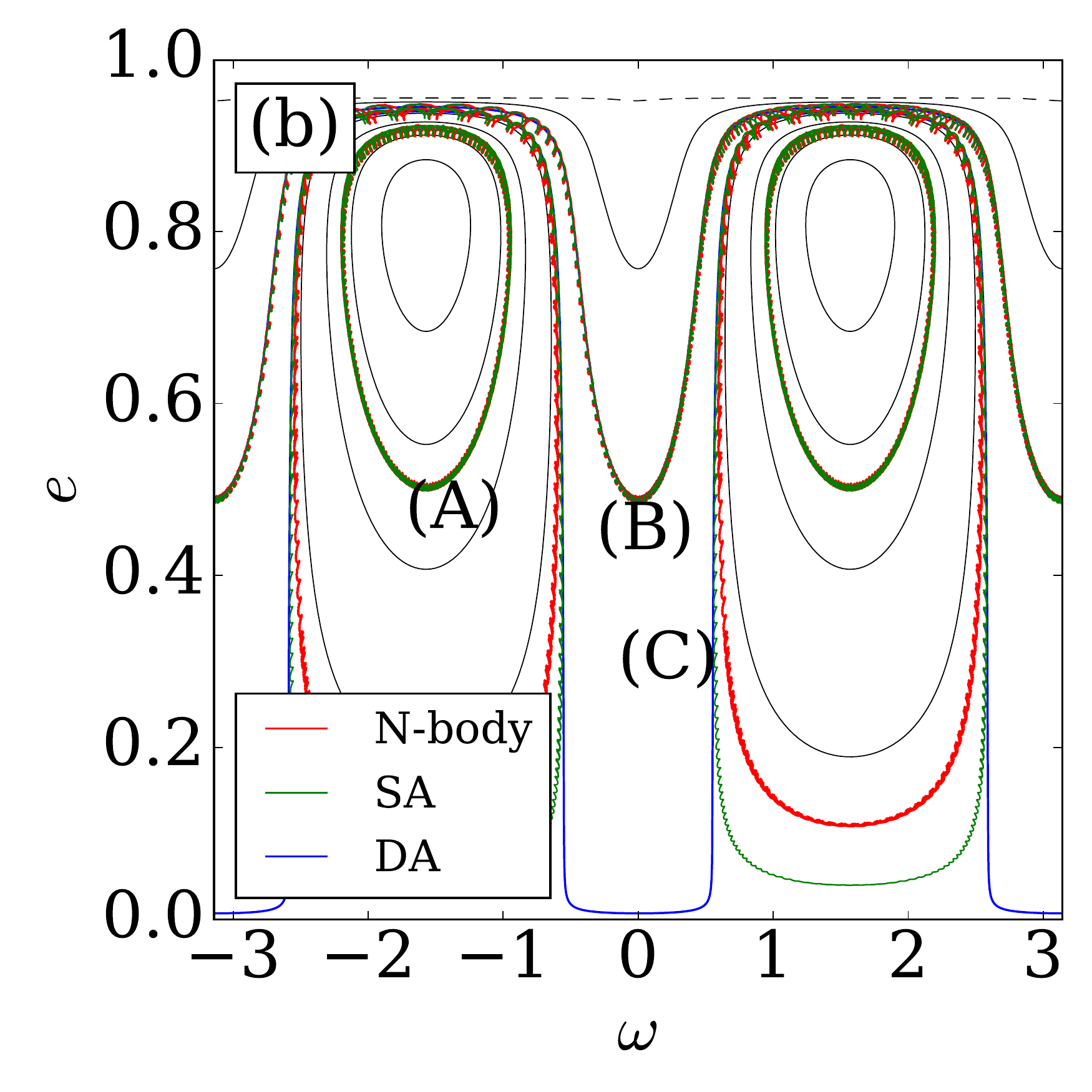}}
\raisebox{0.12cm}{\includegraphics[width=0.322\linewidth]{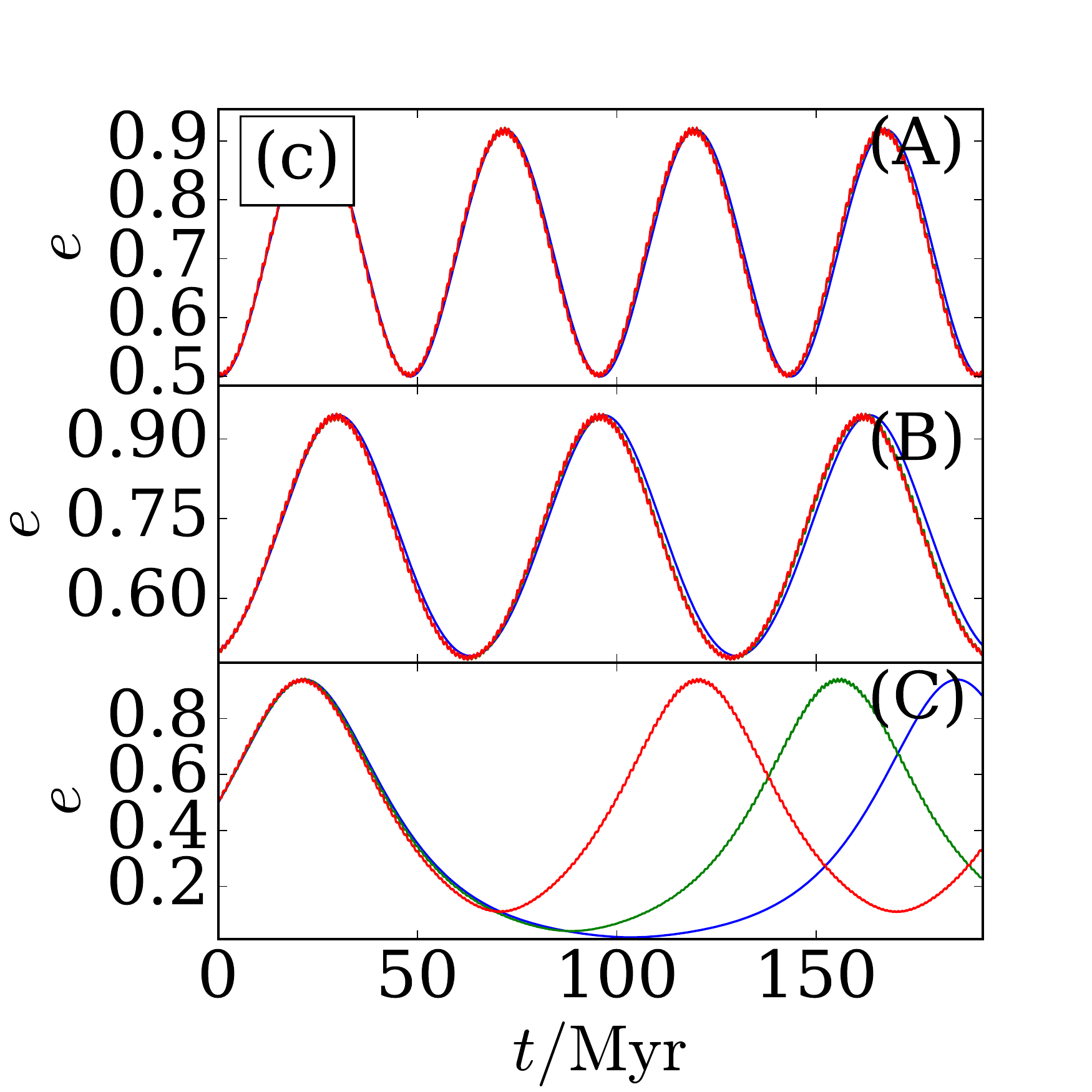}}
\caption{Comparison of the evolution of inner binary orbital elements computed via N-body, SA and DA integrations (see legend in panel (b)), over 100 azimuthal periods of $\Rg$.  The binary ($m_1=m_2=0.5M_\odot$) orbits a spherical isochrone cluster \eqref{IsoPot} with total mass $M=10^5M_\odot$ and scale radius $b=1$pc.  The outer orbit (plotted in the $(X,Y)$ plane in panel (a)) has initial conditions $(R,v_\mathrm{R},Z,v_\mathrm{Z},\phi,v_\mathrm{\phi})=(3.03b,\, 0,\,0,\,0,\,0,\,0.363\sqrt{GM/b})$, giving the theoretical values $A = 0.0206(GM/b^3)$ and $\Gamma=0.4$. Panels (b) and (c) display the $(\omega,e)$ phase-space portrait and time dependence of eccentricity, respectively, for three binaries. Their initial orbital elements are $(a,e,i,\Omega,M)=(10^3\mathrm{AU},0.5,70^{\circ}, 17.188^{\circ},  161.36^{\circ})$, with initial $\omega$ taking the values (A) $91.67^{\circ}$, (B) $5.73^{\circ}$, (C) $34.14^{\circ}$. See \S \ref{sec_NumericalVerification1} for discussion.}
\label{Iso_GamPt4_Plots}
\end{figure*}

First, we had the set of six time-dependent equations for the relative position and velocity of the binary components, subject to the gravitational force of each other and the smooth background potential of the cluster. No tidal approximation had been made at this stage. These six equations can only be solved by direct numerical (`N-body') integration.  

Second, we had a set of four `singly-averaged' (SA) equations, which were obtained by tidally expanding the six N-body equations, recasting them in Hamiltonian form, and averaging over the binary's mean anomaly (i.e. over the `inner orbit').  The singly-averaged equations are still explicitly time-dependent through the external potential $\Phi(\Rg(t))$, where $\Rg(t)$ is the barycentric position of the binary (assumed to move as a test particle in the cluster potential).

Finally, we derived a system of two `doubly-averaged' (DA) time-independent equations \eqref{eom1}-\eqref{eom2} resulting from the secular Hamiltonian \eqref{H1Mt}, which was itself obtained by averaging the singly-averaged Hamiltonian over many `outer orbits' of the binary around the cluster. Our time-averaging procedure relied on the assumption that $\Rg(t)$ densely fills an axisymmetric torus whose symmetry axis coincides with the symmetry ($Z$) axis of the potential (in the case of a spherical potential, the `torus' is a two-dimensional annulus perpendicular to $Z$).  More technically, we required that the time-averages of the derivatives $\Phi_{\alpha\beta}$ of the potential (where $\Phi_{xy}=\partial^2\Phi/\partial X\partial Y$, etc.) converge to constant values --- in particular, $\overline{\Phi}_{xy}\equiv \overline{\Phi}_{xz}=\overline{\Phi}_{yz}=0$. This condition is almost always satisfied for orbits in any axisymmetric potential after a sufficient number of outer orbital periods.  Then, the time-dependent torque on the binary can be replaced with a converged time-independent torque that arises from an axisymmetric mass distribution as seen from the binary's frame. In \S\S\ref{sec_General}-\ref{sec_GammaRegime4} of the present paper we have examined the dynamics that arise from the doubly-averaged equations \eqref{eom1}-\eqref{eom2}.

However, it is to be expected that the DA theory will break down under certain circumstances. The goal of this section is to explore the validity of the DA approximation for computing inner binary dynamics in the different $\Gamma$ regimes covered in \S\S \ref{sec_GammaRegime1}-\ref{sec_GammaRegime4}.
To do so, we present several examples comparing numerical integrations of the N-body\footnote{Note the term `N-body' is not meant to imply that we integrate an entire cluster of, say, $10^5$ particles.  Instead we integrate the exact two-body equations of motion in the presence of the time-dependent external potential calculated via numerical orbit integration of $\Rg(t)$.}, SA and DA equations.

We will see that the secular approximation is good as long as the ratio of the secular timescale to the outer orbital period, $t_\mathrm{sec}/T_\phi$, is large enough. Given that $t_\mathrm{sec} \sim T_\phi^2/T_\mathrm{b}$ (see equations \eqref{timescaleeqn} and \eqref{eq:roughtime}), this is equivalent to the statement that $T_\phi/T_\mathrm{b}$ be sufficiently large. Hence, alongside each case where the DA theory fails we present another example with identical initial conditions except for a smaller binary semi-major axis. The effect of this is to decrease the inner binary orbital period $T_\mathrm{b}$, rendering the DA theory valid.  

\subsection{Method}


We use two particular forms of the background potential $\Phi$ in our examples. The first is the spherical isochrone potential 
\begin{align} 
\label{IsoPot}
\Phi_\mathrm{iso}(r) = -\frac{GM}{b+\sqrt{b^2+r^2}}, 
\end{align} 
where $r=\sqrt{X^2+Y^2+Z^2}$ is the spherical radius.  This is a model potential for a spherical star cluster with total mass $M$ and scale radius $b$ \citep{Binney2008}. Since the potential \eqref{IsoPot} is spherical we can always choose the plane of the binary's outer orbit $\Rg$ to be the $(X,Y)$ plane. The other potential we will use is the Miyamoto-Nagai potential \citep{Miyamoto1975}: 
\begin{equation}
\Phi_\mathrm{MN}(R,Z)=-\frac{GM}{\sqrt{R^2+(b_\ell+\sqrt{Z^2+b_h^2})^2}}, 
\label{MNPot}
\end{equation} 
where $R = \sqrt{X^2+Y^2}$ is the usual cylindrical radius.  Here $M$ is the total mass of the model, $b_\ell$ is the scale length and $b_h$ is the scale height.  By changing the value of $b_h/b_\ell$, the Miyamoto-Nagai potential interpolates between the Kuzmin potential of a razor thin disk ($b_h \ll b_\ell $) and the spherical Plummer potential ($b_h \gg b_\ell$) frequently used to model globular clusters \citep{Binney2008}.

\begin{figure*}
\centering
\hspace{-0.7cm}  
\includegraphics[width=0.34\linewidth,]{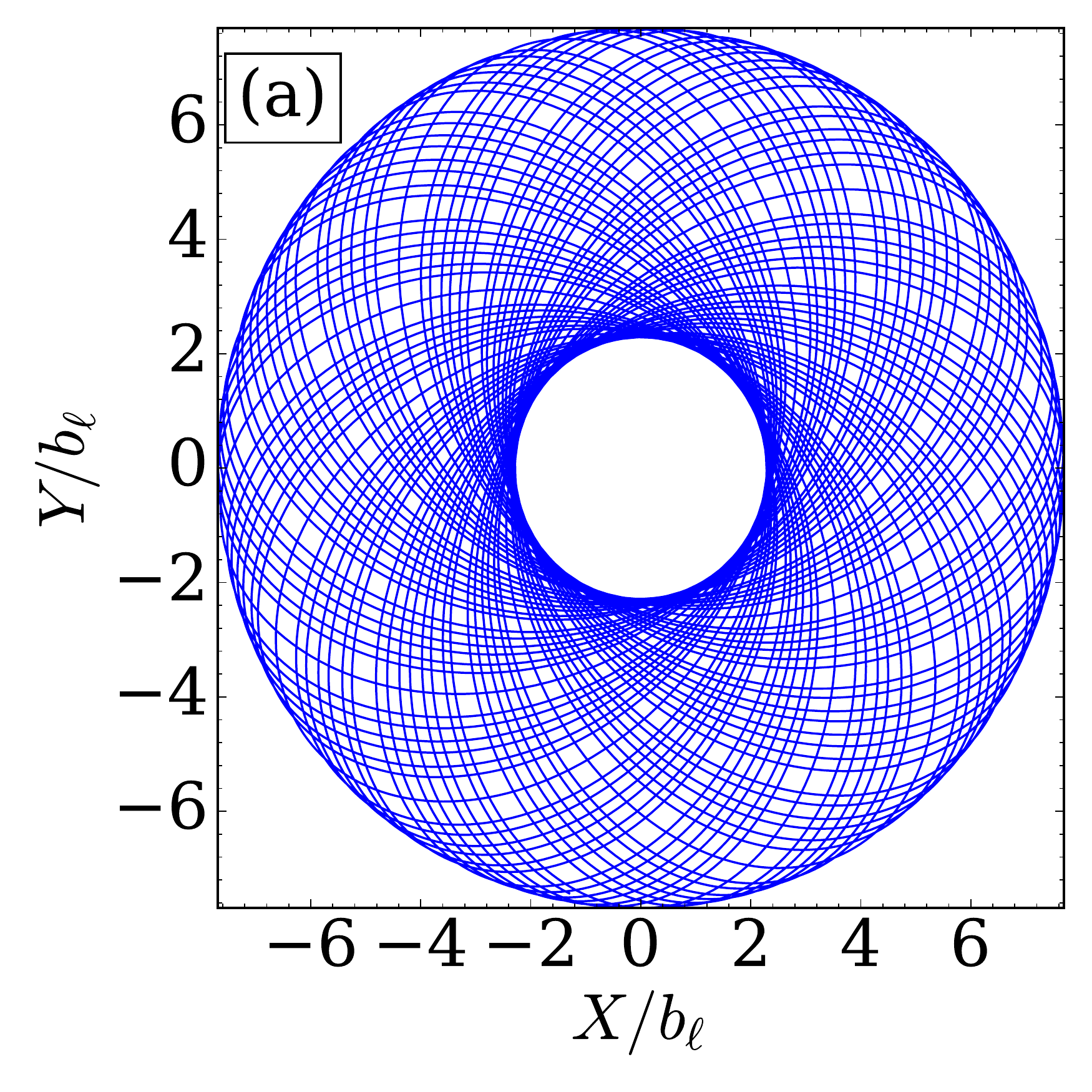}
\includegraphics[width=0.34\linewidth]{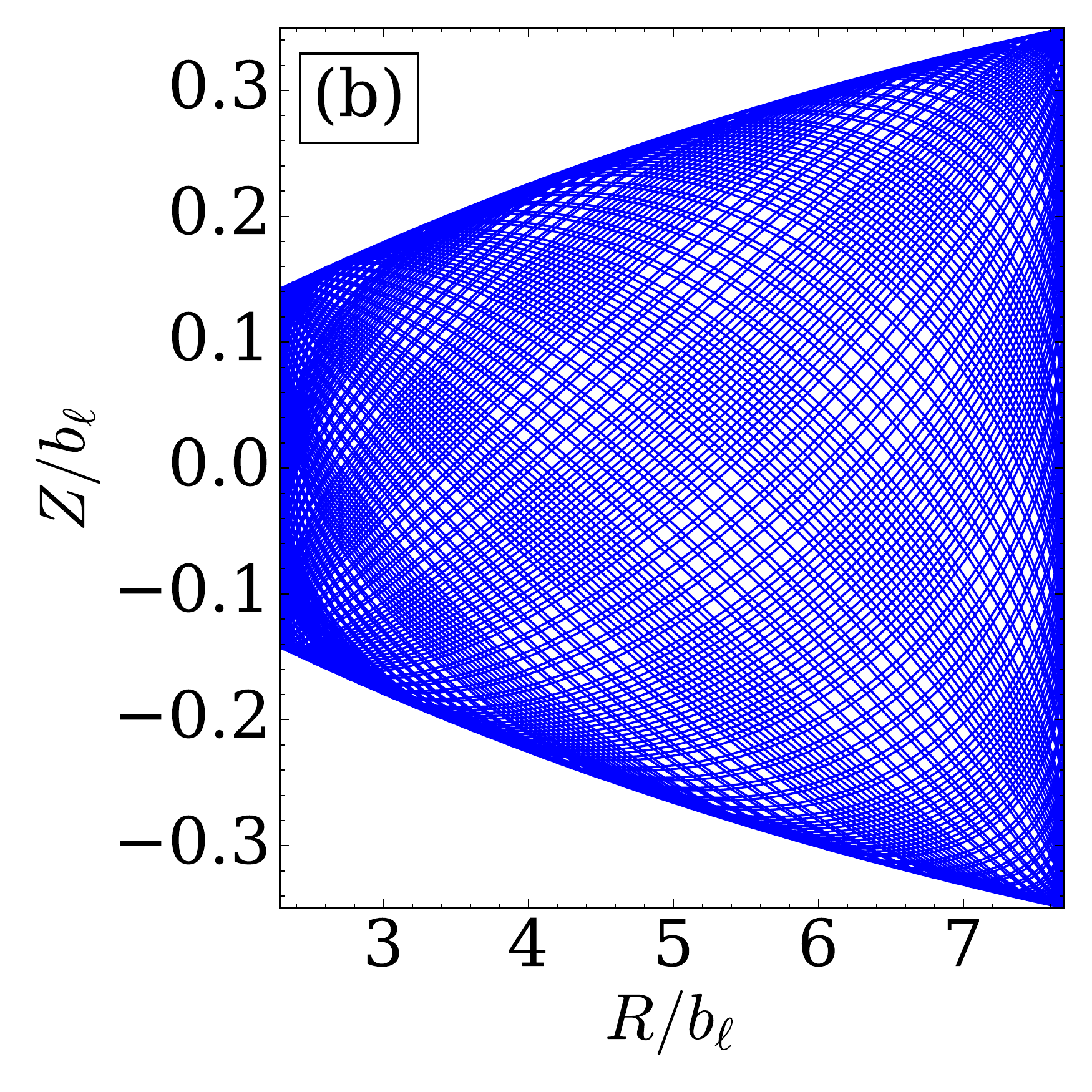}\quad
\includegraphics[width=0.34\linewidth, trim=0 0 25 0,clip]{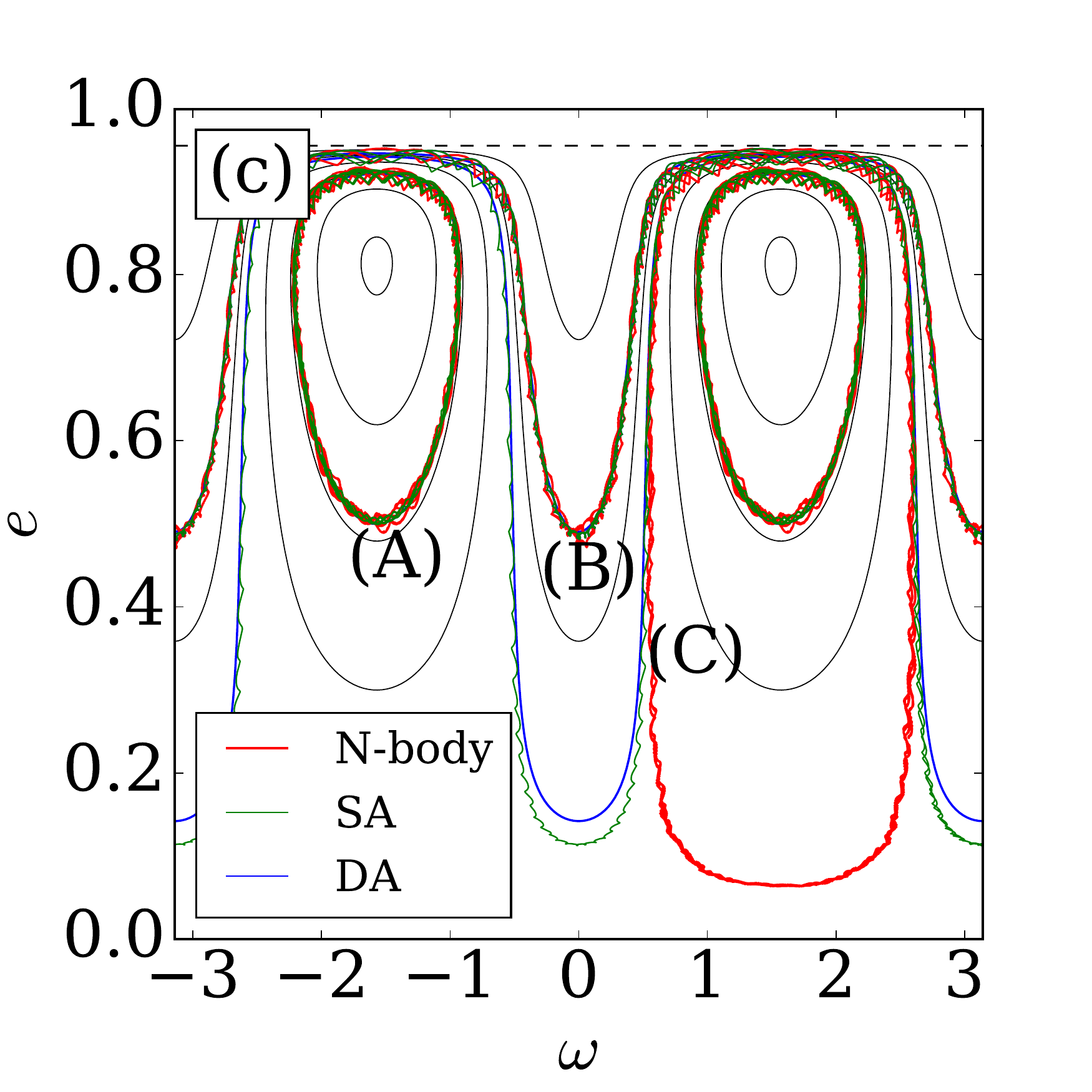}
\includegraphics[width=0.38\linewidth]{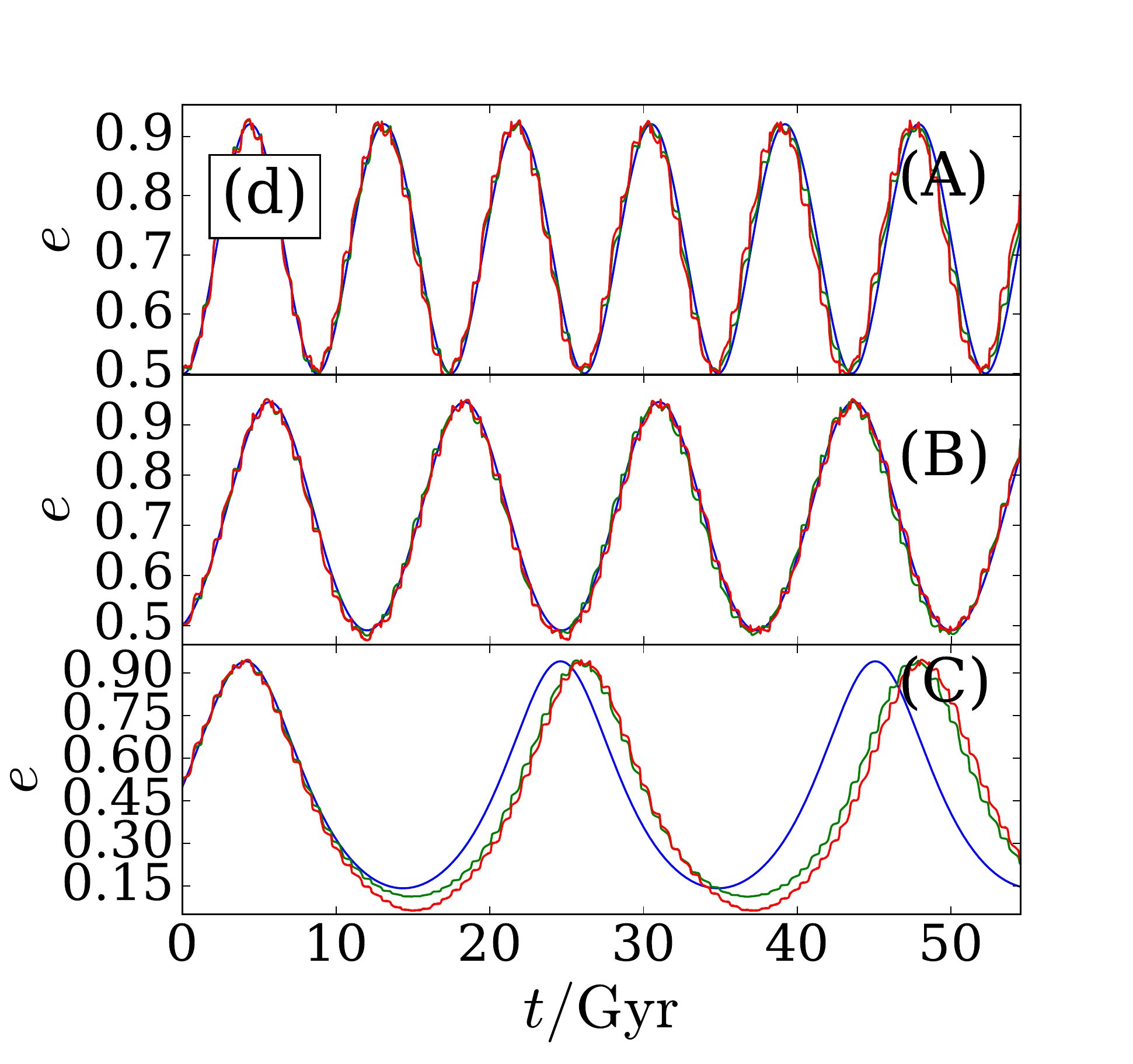}
\caption{Comparison of N-body, SA and DA integrations in a non-spherical cluster. The binary ($m_1=m_2=0.5M_\odot$) orbits a Miyamoto-Nagai potential \eqref{MNPot} with total mass $M=10^{11}M_\odot$ and $b_\ell=b_h=3.5$kpc.  The outer orbit has initial conditions $(R,v_\mathrm{R},Z,v_\mathrm{Z},\phi,v_\mathrm{\phi})=(2.29b_\ell,0,\,0.143 b_\ell,\,0,\,0,\,0.667\sqrt{GM/b_\ell})$, its projections onto the $(X,Y)$ (equatorial) and $(R,Z)$ planes are shown in panels (a) and (b); numerically we find $A_\mathrm{num} = 0.0149(GM/b_\ell^3)$ and $\Gamma_\mathrm{num}=0.370$. Panels (c) and (d) are similar to panels (b) and (c) of Figure \ref{Iso_GamPt4_Plots}. Initial orbital elements are $(a,e,i,\Omega,M)=(5\times 10^4\mathrm{AU},0.5,70^{\circ}, 17.188^{\circ},  161.36^{\circ})$, with initial $\omega$ taking the values (A) $91.67^{\circ}$, (B) $5.73^{\circ}$, (C) $30.65^{\circ}$. See \S\ref{sec_NumericalVerification1} for discussion.}
\label{MN_NumGamPt370_Plots}
\end{figure*}

In either case the binary's outer orbit $\Rg(t)$ is stipulated via its initial conditions $(R,v_\mathrm{R},Z,v_\mathrm{Z},\phi,v_\mathrm{\phi})$, where $\phi=\tan^{-1}({Y/X})$ is the azimuthal angle in cylindrical coordinates, $v_\mathrm{R}$ is the velocity of the binary in the direction of increasing $R$, etc. We integrate the orbit $\Rg(t)$ in this potential numerically using \texttt{galpy} (\citealt{Bovy2015}; see Appendix F of Paper I for details), and then feed the resulting time series into the SA and N-body equations\footnote{Note that in some examples where there is an extremely large separation between the secular timescale and inner orbital period it becomes prohibitively expensive to integrate the N-body equations of motion, so we just show the DA and SA results.}.  The DA equations \eqref{eom1}-\eqref{eom2} are integrated using the theoretical values of $A$ and $\Gamma$ when they are available (i.e. in the spherical isochrone case); otherwise we use the numerical prescription outlined in Appendix F of Paper I and denote them by\footnote{In spherical potentials, $A$ and $\Gamma$ can be calculated theoretically by stipulating the outer orbit's peri/apocentre $r_\mathrm{p/a}$. See Paper I for more information about calculating $A$, $A_\mathrm{num}, \Gamma$, $\Gamma_\mathrm{num}$ and possible small discrepancies between the theoretical and numerical values.} $A_\mathrm{num}, \Gamma_\mathrm{num}$. In all numerical examples the binary has constituent masses $m_1=m_2=0.5M_\odot$.

\begin{figure*}
\centering
\includegraphics[width=0.33\linewidth]{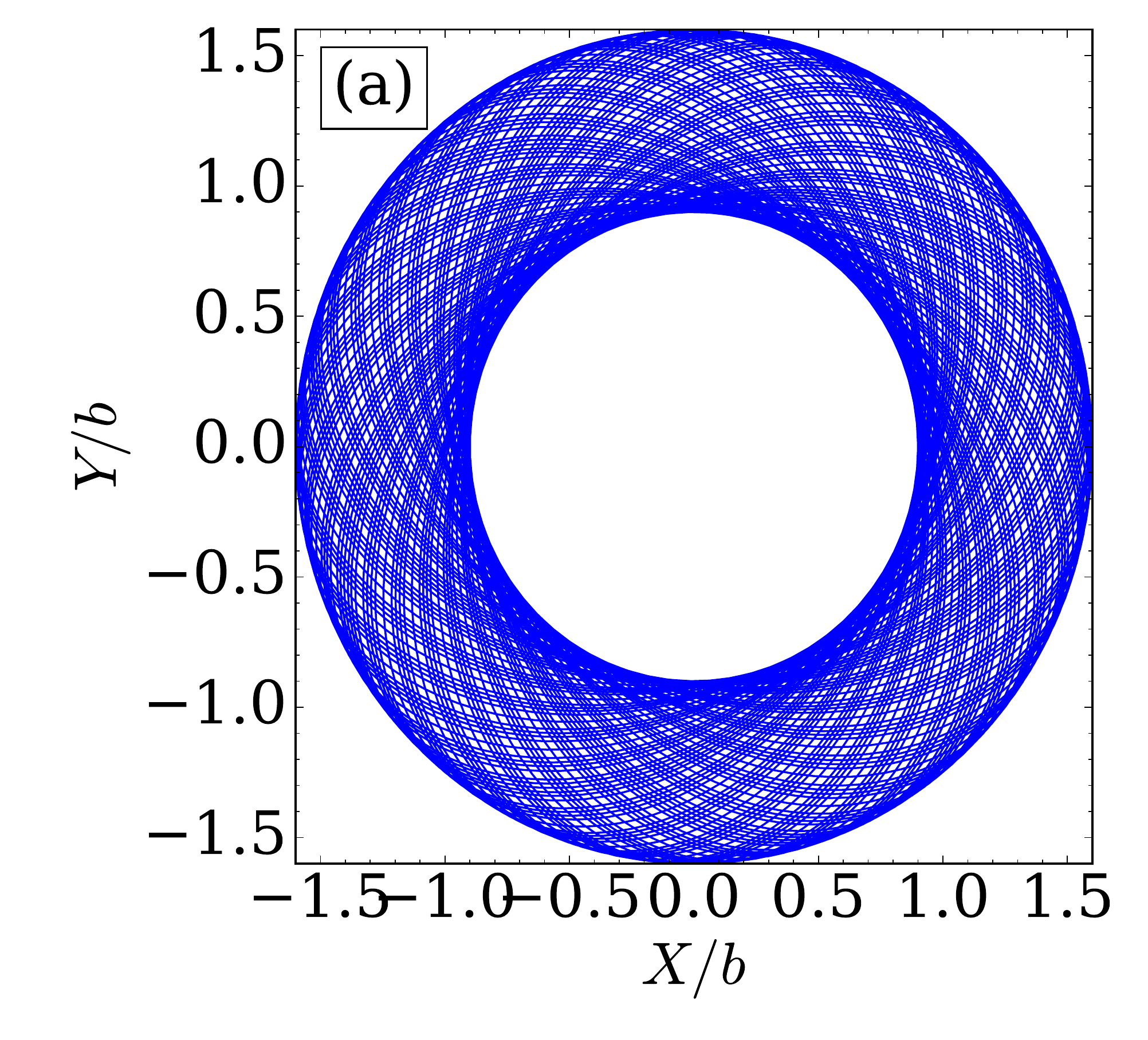}
\raisebox{0.05cm}{\includegraphics[width=0.305\linewidth]{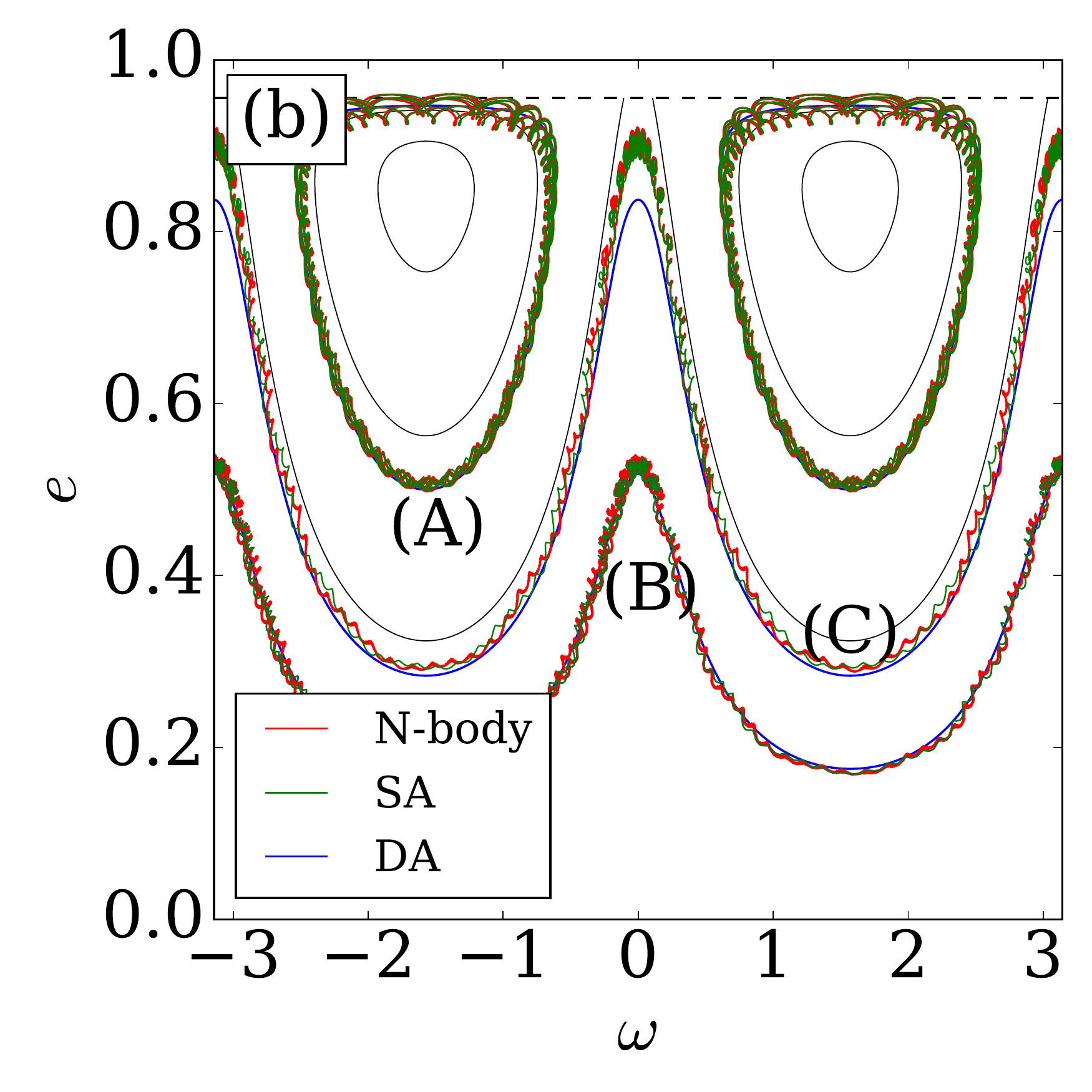}}
\raisebox{0.1cm}{\includegraphics[width=0.32\linewidth]{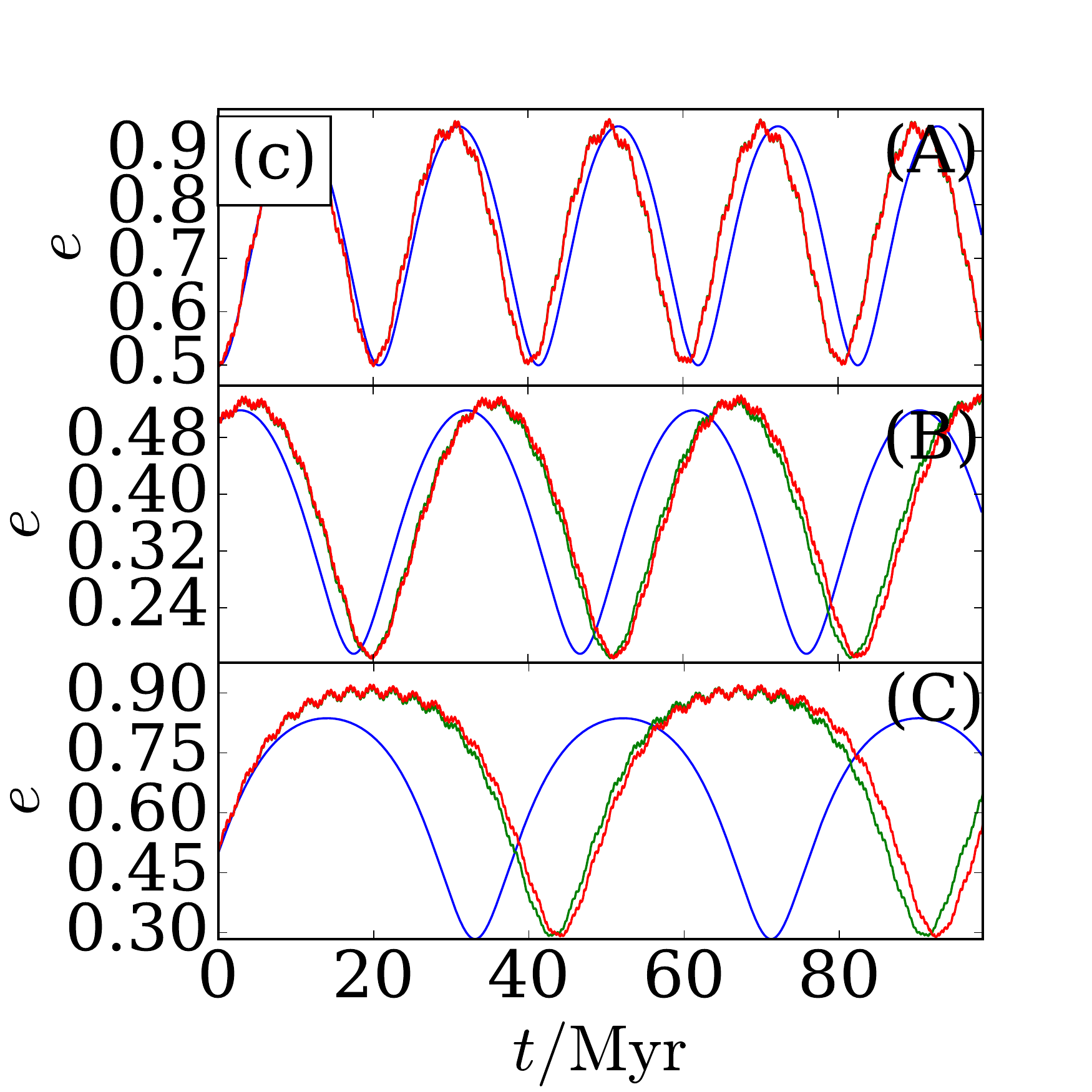}}
\includegraphics[width=0.32\linewidth]{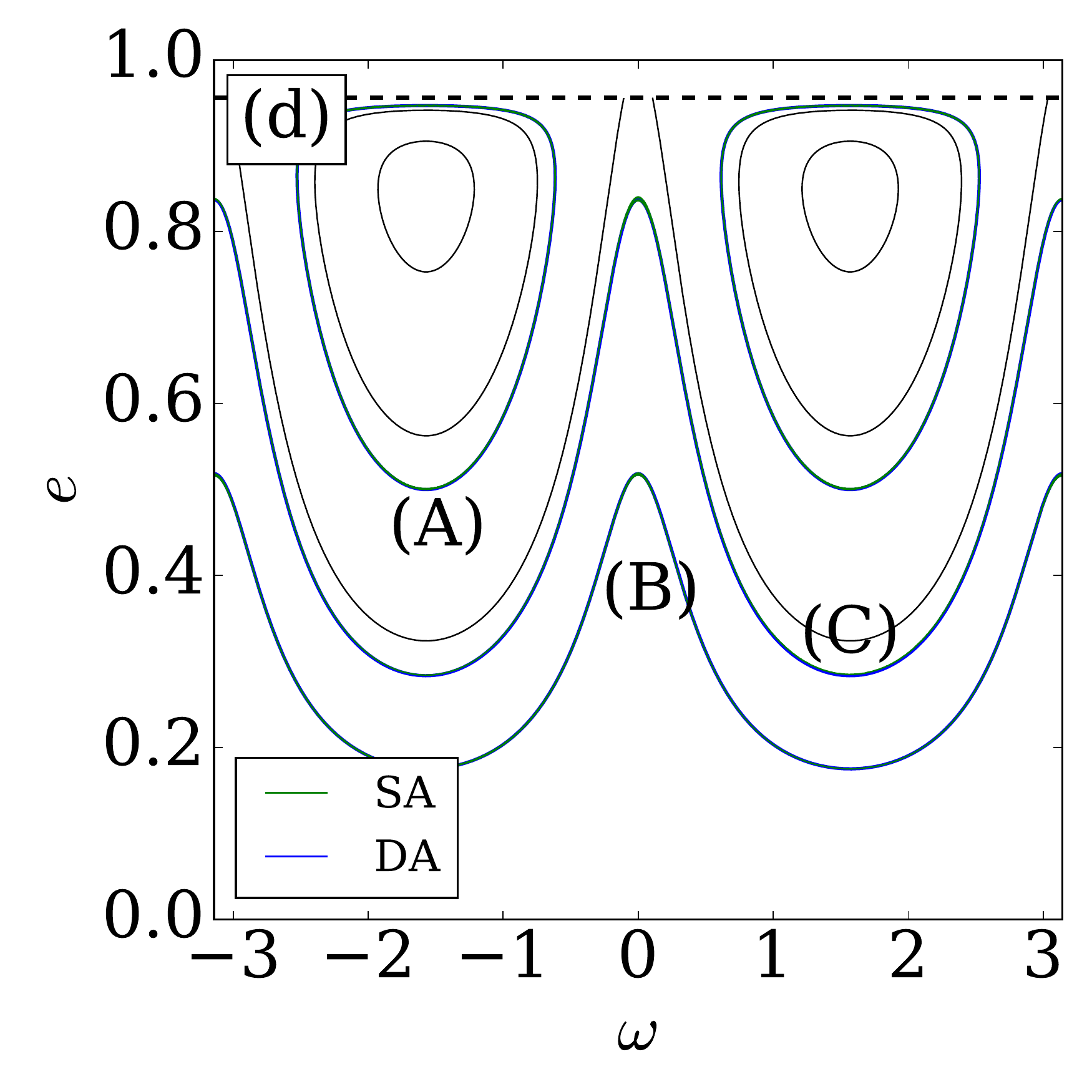}
\raisebox{0.03cm}{\includegraphics[width=0.335\linewidth]{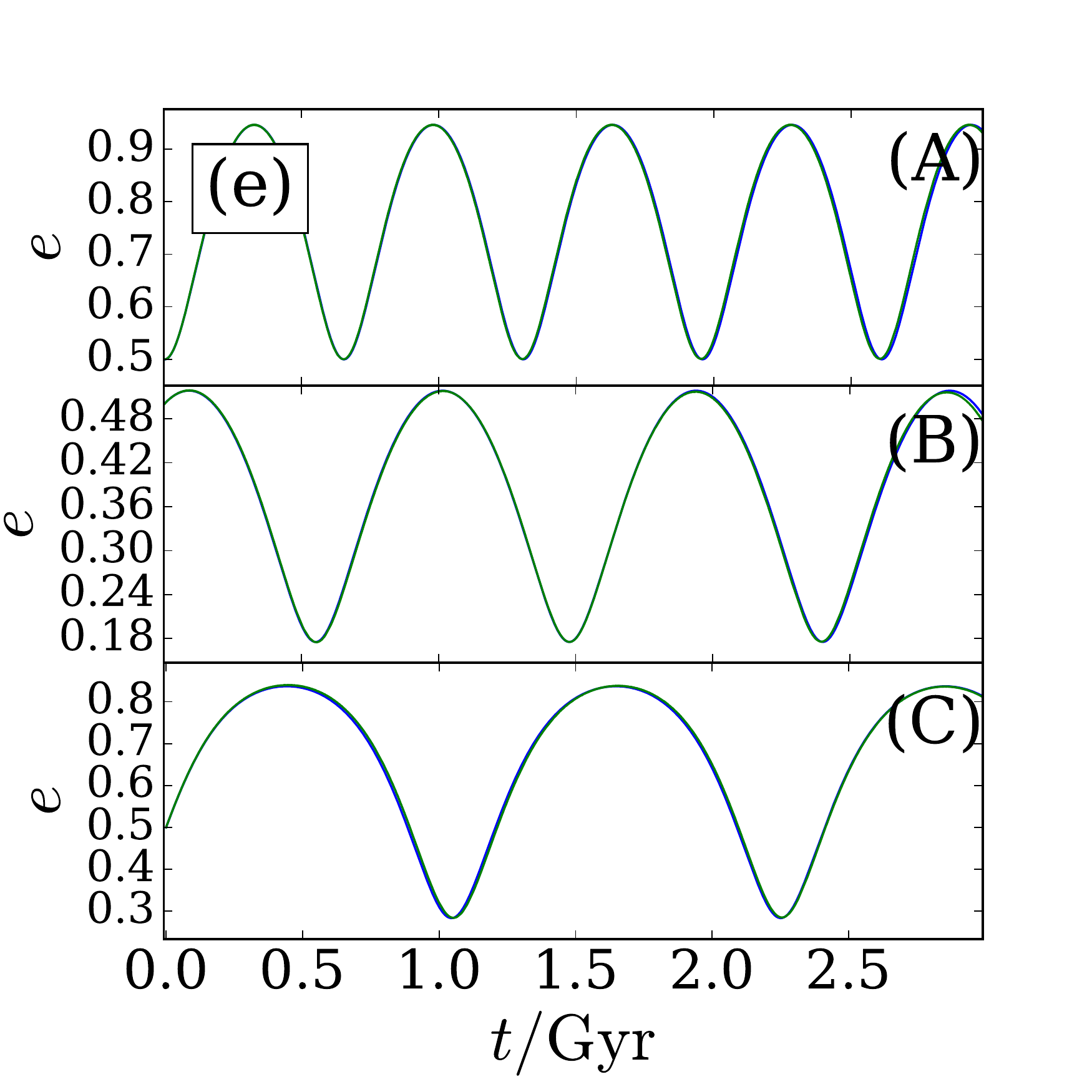}}
\caption{Same as Figure \ref{Iso_GamPt4_Plots}, including the form of the cluster potential, but now focusing on the regime $0<\Gamma\leq 1/5$. The binary's outer orbit (shown for $100T_\phi$ in panel (a)) has initial conditions $(R,v_\mathrm{R},Z,v_\mathrm{Z},\phi,v_\mathrm{\phi})=(1.6b,0,\,0,\,0,\,0,\,0.27\sqrt{GM/b})$, resulting in $A=0.124 (GM/b^3)$ and $\Gamma=0.162$. In panels (b) and (c) the initial binary semi-major axis is  $a=10^3\mathrm{AU}$ and we integrate the equations of motion for $100T_\phi$, while in panels (d) and (e) it is $a=100 \mathrm{AU}$ and we integrate for $3000T_\phi$.  In each case, the other initial orbital elements are $(e,i,\Omega,M)=(0.5,70^{\circ}, 17.188^{\circ},  161.36^{\circ})$, with initial $\omega$ taking the values (A) $91.67^{\circ}$, (B) $5.73^{\circ}$, (C) $34.14^{\circ}$. See \S \ref{sec_NumericalVerification2} for discussion.}
\label{Iso_GamPt162_Plots}
\end{figure*}


\subsection{Accuracy of the doubly-averaged approximation for \texorpdfstring{$\Gamma>1/5$}{}.}
\label{sec_NumericalVerification1}


Here we give two examples where the N-body, SA and DA integrations are in very good agreement, both in the regime $\Gamma > 1/5$ (explored in \S \ref{sec_GammaRegime1}).  In the first the binary orbits the spherical isochrone potential, and in the second it orbits the Miyamoto-Nagai potential.  The details of each example are given in the following paragraphs, and the results are shown in Figures \ref{Iso_GamPt4_Plots} and \ref{MN_NumGamPt370_Plots} respectively.

\paragraph*{[Figure \ref{Iso_GamPt4_Plots}: Isochrone potential, $\Gamma=0.4$.]} We consider a binary orbiting a spherical isochrone cluster \eqref{IsoPot} with scale radius $b=1 \mathrm{pc}$ and total mass $M=10^5 M_\odot$. The initial conditions for the outer orbit $\Rg$ are as follows: \begin{align}
(R,v_\mathrm{R},Z,v_\mathrm{Z},\phi,v_\mathrm{\phi})=(3.03b,\, 0,\,0,\,0,\,0,\,0.363\sqrt{GM/b}).
\end{align} It is easy to show that this choice of initial conditions is equivalent to a choice of peri/apocentre $(r_\mathrm{p}/b, r_\mathrm{a}/b)= (2.40, 3.03)$. The theoretical $A, \Gamma$ values that result are $A=0.0206(GM/b^3)$ and $\Gamma=0.4$.  The outer orbit is shown in Figure \ref{Iso_GamPt4_Plots}a.  In Figure \ref{Iso_GamPt4_Plots} all panels show $100 T_\phi$ worth of data.

\begin{figure*}
\centering
\raisebox{-0.2cm}{\includegraphics[width=0.32\linewidth]{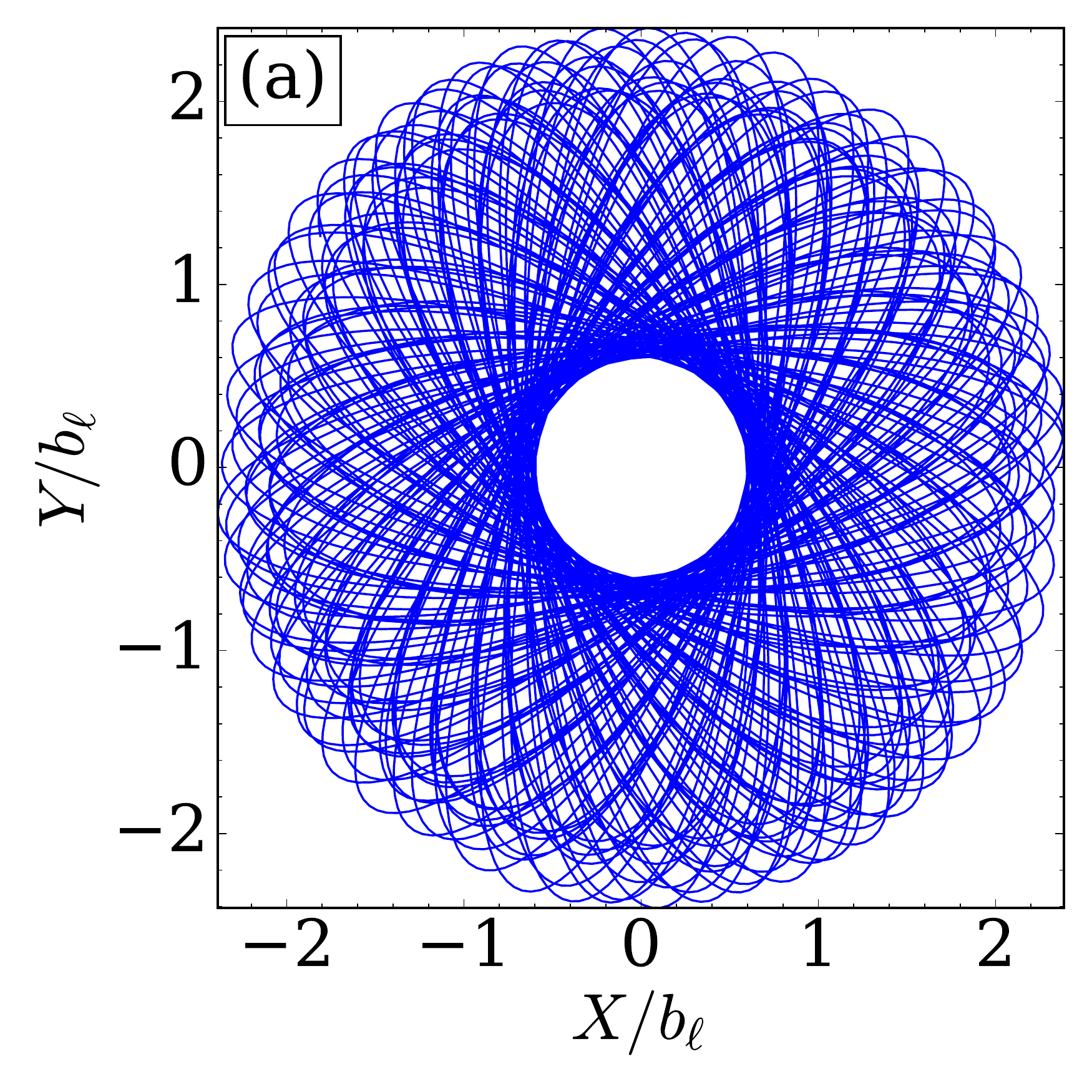}}
\includegraphics[width=0.32\linewidth, trim=0 0 25 0,clip]{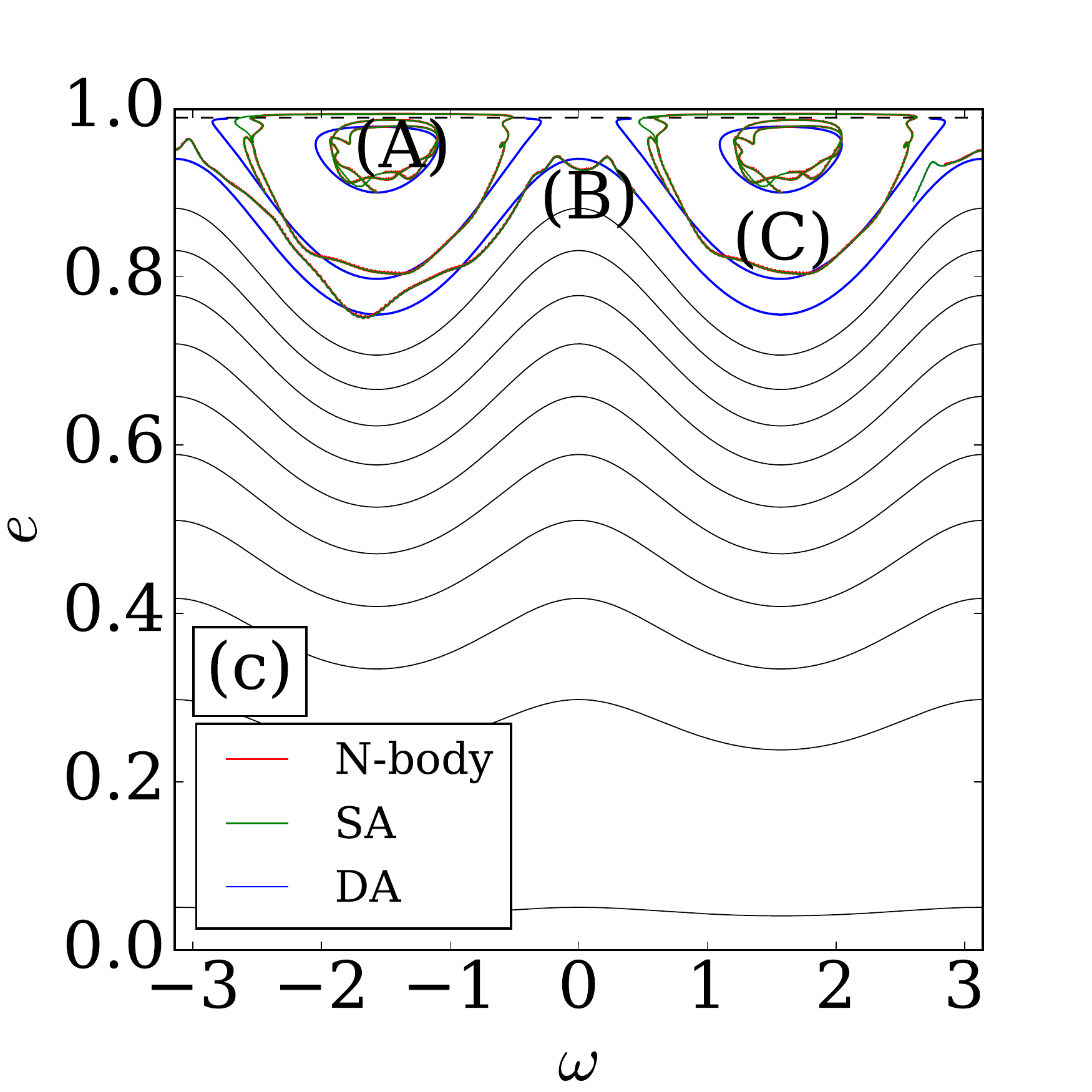}
\includegraphics[width=0.32\linewidth, trim=0 0 25 0,clip]{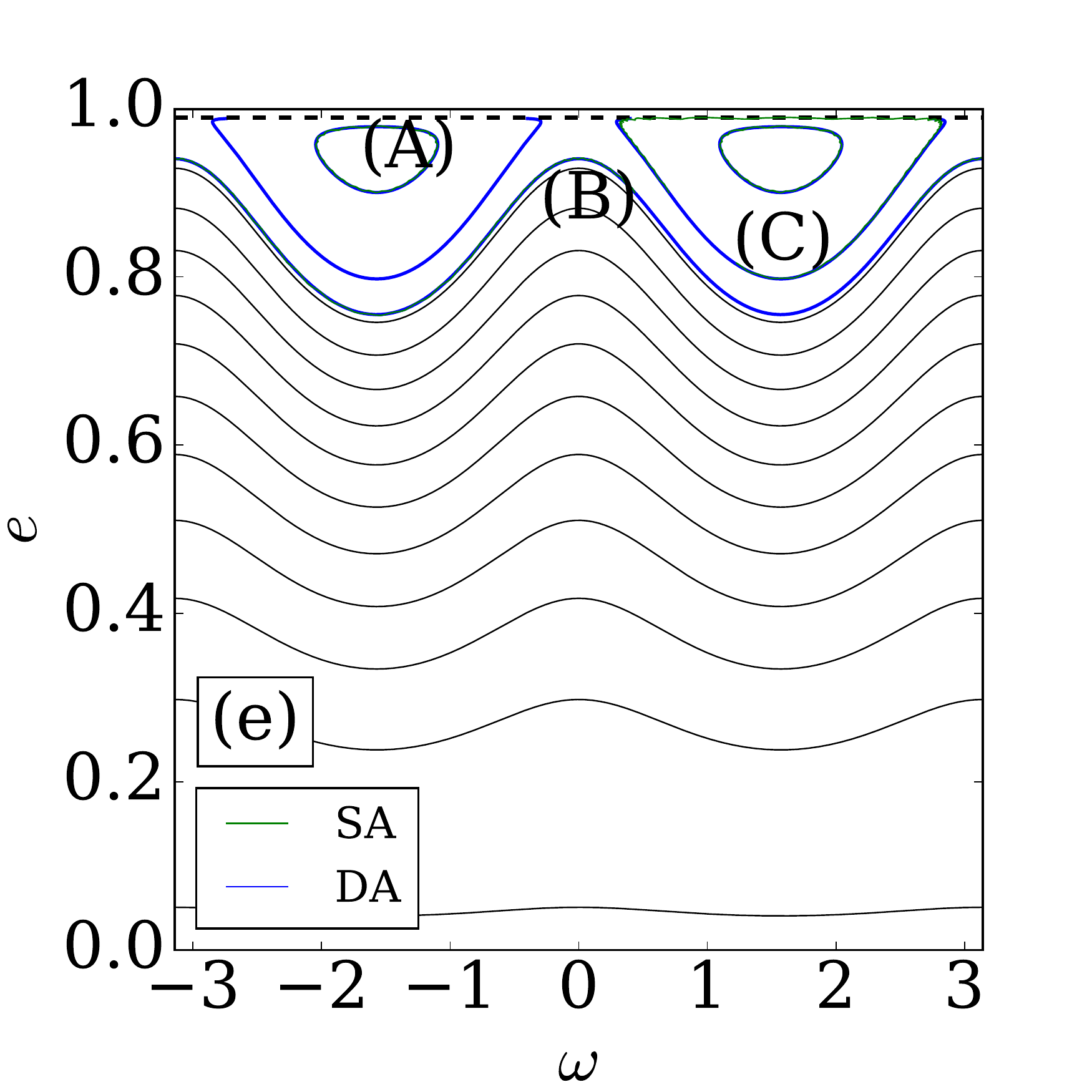}
\raisebox{-0.2cm}{\includegraphics[width=0.323\linewidth]{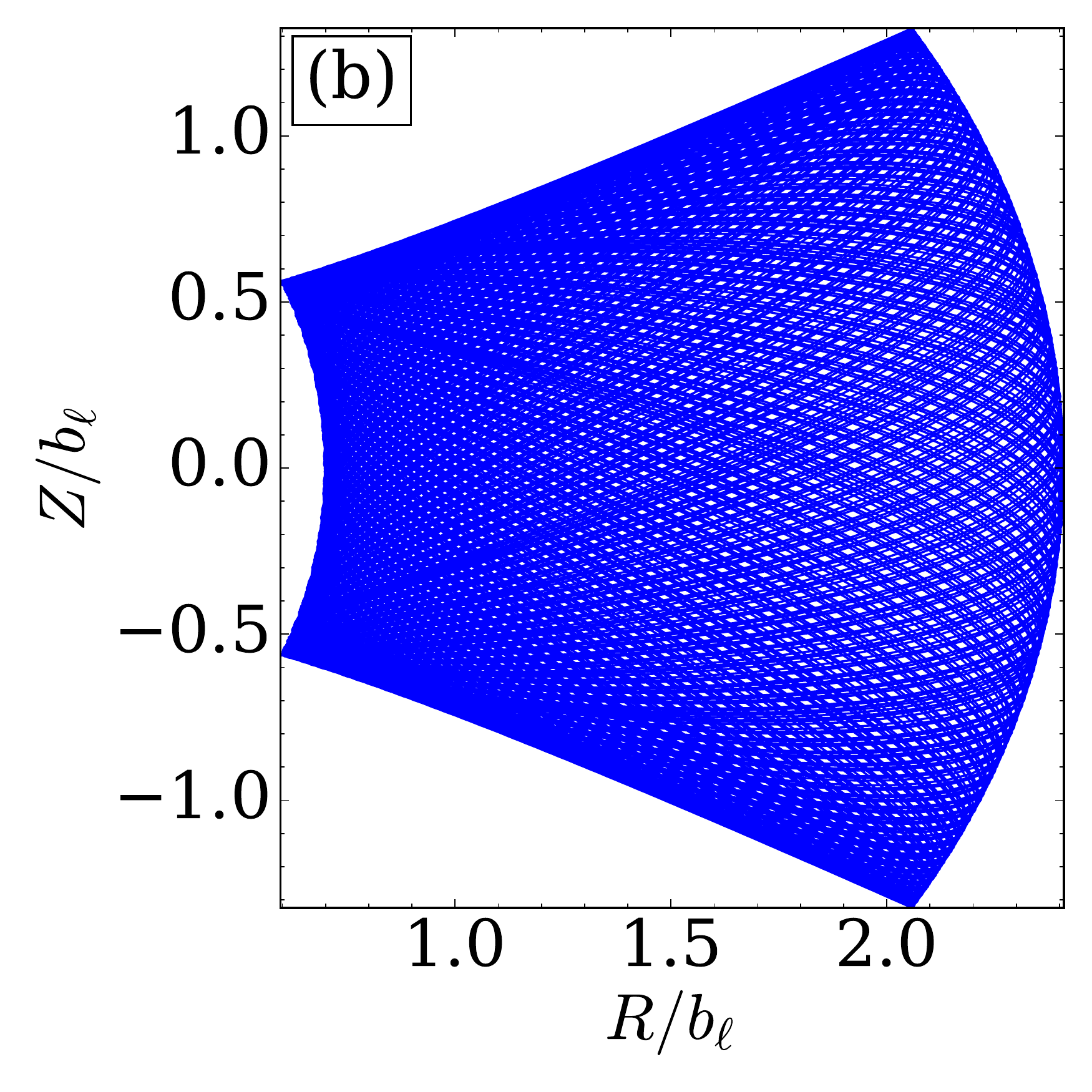}}
\includegraphics[width=0.335\linewidth]{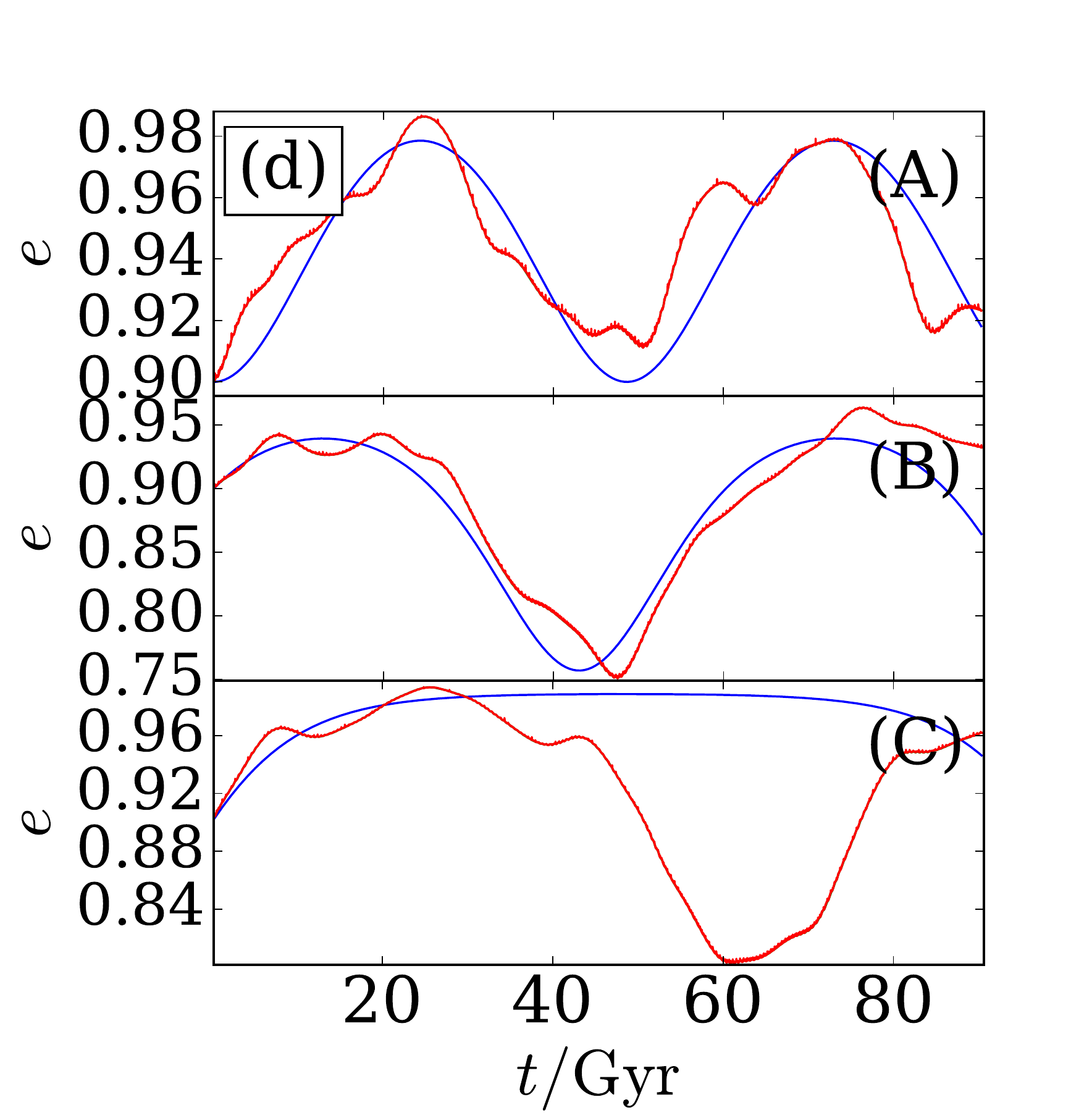}
\includegraphics[width=0.32\linewidth, trim=0 0 25 0,clip]{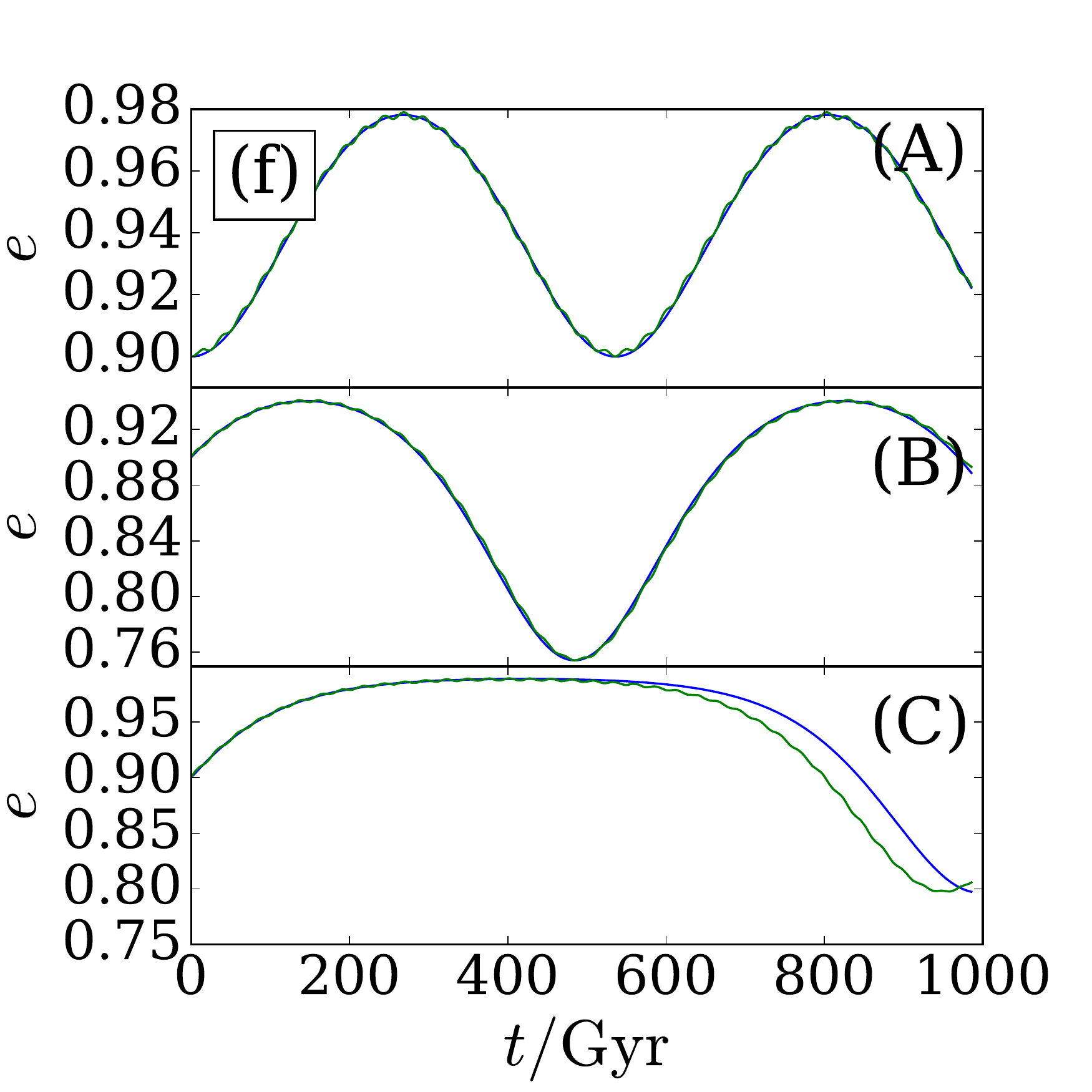}
\caption{Validity of the doubly-averaged secular approximation for a substantially inclined outer orbit of the binary in a non-spherical cluster.  The binary ($m_1=m_2=0.5M_\odot$) orbits a Miyamoto-Nagai potential \eqref{MNPot} with total mass $M=10^{11}M_\odot$ and $b_\ell=3.5$kpc and $b_h/b_\ell = 5$ (less flattened than in Figure \ref{MN_NumGamPt370_Plots}).  The outer orbit (shown for $100T_\phi$ in panels (a) and (b)) has initial conditions $(R,v_\mathrm{R},Z,v_\mathrm{Z},\phi,v_\mathrm{\phi}) = (2.29b_\ell,0.05\sqrt{GM/b_l},\,0.143b_\ell,\,0.05\sqrt{GM/b_l},\,0,\,0.05\sqrt{GM/b_l})$; numerically we find $A_\mathrm{num}=0.00826(GM/b_\ell^3)$ and $\Gamma_\mathrm{num}=0.045$. In panels (c) and (e) the initial binary semi-major axis is  $a=5\times 10^4\mathrm{AU}$ and we integrate the equations of motion for $100T_\phi$, while in panels (d) and (f) it is $a=10^4 \mathrm{AU}$ and we integrate for $1000T_\phi$. Initial orbital elements are $(e,i,\Omega,M)=(0.9,70^{\circ}, 17.188^{\circ},  161.36^{\circ})$, with initial $\omega$ taking the values (A) $90^{\circ}$, (B) $24.9^{\circ}$, (C) $40.1^{\circ}$. See \S\ref{sec_NumericalVerification2} for discussion.}
\label{MN_NumGamPt045_Plots}
\end{figure*}

The (rather soft) binary has initial orbital elements $(a,e,i,\Omega,M)=(10^3\mathrm{AU},0.5,70^{\circ}, 17.188^{\circ},  161.36^{\circ})$ with three different initial values of $\omega$, namely (A) $91.67^{\circ}$ (librating phase-space orbit), (B) $5.73^{\circ}$ (circulating orbit), and (C) $34.14^{\circ}$ (an orbit very close to the separatrix). From Figures \ref{Iso_GamPt4_Plots}b,c we see that the agreement is extremely good between N-body, SA and DA calculations for phase-space trajectories (A) and (B), even though their secular timescales are longer than $T_\phi$ by just several tens. Although initially the agreement in trajectory (C) is also excellent, there is a divergence between N-body, SA and DA calculations when we reach low eccentricity because (C) was chosen to be so close to the separatrix (where the secular timescale formally is infinite, see Figure \ref{fig_Timescales}). The DA result circulates while the others librate.

\paragraph*{[Figure \ref{MN_NumGamPt370_Plots}: Miyamoto-Nagai potential, $\Gamma_\mathrm{num}=0.370$.]} This time the binary orbits the Miyamoto-Nagai potential \eqref{MNPot} with $b_\ell=b_h=3.5\mathrm{kpc}$ and total mass $M=10^{11} M_\odot$.  The initial conditions of the outer orbit are \begin{align}
&(R,v_\mathrm{R},Z,v_\mathrm{Z},\phi,v_\mathrm{\phi}) \nn \\  &=(2.29b_\ell,0,\,0.143b_\ell,\,0,\,0,\,0.667\sqrt{GM/b_\ell}).  
\end{align} 
The projections of the outer orbit onto the $(X,Y)$ and $(R,Z)$ planes shown in Figures \ref{MN_NumGamPt370_Plots}a,b. Again all panels show the first $100T_\phi$ of integration time.  Since the outer orbit is not planar we do not have theoretical $A, \Gamma$ values; numerically we find $A_\mathrm{num}=0.0149(GM/b_\ell^3)$ and $\Gamma_\mathrm{num}=0.370$ (not too different from $\Gamma=1/3$ corresponding to binaries near the midplane of a thin disc, see Paper I).

The binary has initial orbital elements $(a,e,i,\Omega,M)=(5\times 10^4 \,\mathrm{AU},0.5,70^{\circ}, 17.188^{\circ},  161.36^{\circ})$, and we consider three initial values of $\omega$, namely (A) $91.67^{\circ}$ (librating orbit), (B) $5.73^{\circ}$ (circulating orbit), and (C) $30.65^{\circ}$ (an orbit very close to the separatrix). It is again evident (Figures \ref{MN_NumGamPt370_Plots}c,d) that the agreement is extremely good between N-body, SA and DA calculations for trajectories (A) and (B). The agreement in the eccentricity timeseries for the separatrix trajectory (C) is also good over the first half-oscillation, but there is then once again a discrepancy in the phase portrait as to whether the orbit ought to librate or circulate. Since the semi-major axis used in this example is typical for Oort Cloud comets, we conclude that the DA approximation should work well for characterising the secular evolution of the long-period comets \citep{Heisler1986}.
\\

In both of these examples the DA theory is very accurate, except for describing phase-space trajectories that lie extremely close to the separatrix between librating and circulating orbits. Note that in each case, $100$ outer orbital periods was enough time for at least two secular cycles to take place, so $t_\mathrm{sec}/T_\phi$ was at most $\sim 50$ and usually smaller.  We will see in the upcoming sections that in some circumstances a much greater timescale separation $t_\mathrm{sec}/T_\phi$ is required for the secular approximation to be valid.


\subsection{Accuracy of the doubly-averaged approximation for \texorpdfstring{$0<\Gamma\leq 1/5$}{}.}
\label{sec_NumericalVerification2}


We now consider the case $0<\Gamma \leq 1/5$ studied in \S \ref{sec_GammaRegime2}. We again provide one example in the spherical isochrone potential and one in the non-spherical Miyamoto-Nagai potential.  

This time, in each instance we use two different initial semi-major axes to show how the secular approximation improves for smaller $a$ (shorter $T_\mathrm{b}$).

\paragraph*{[Figure \ref{Iso_GamPt162_Plots}: Isochrone potential, $\Gamma=0.162$.]} The potential is exactly as in the $\Gamma=0.4$ case discussed in \S\ref{sec_NumericalVerification1} (Figure \ref{Iso_GamPt4_Plots}), but now we choose different initial conditions for the outer orbit, namely 
\begin{align}
(R,v_\mathrm{R},Z,v_\mathrm{Z},\phi,v_\mathrm{\phi})=(1.6b,0,\,0,\,0,\,0,\,0.27\sqrt{GM/b}). 
\end{align} 
This corresponds to $(r_\mathrm{p}/b, r_\mathrm{a}/b)= (0.9,1.6)$, giving the theoretical values $\Gamma=0.162$ and $A=0.124(GM/b^3)$.  Figure \ref{Iso_GamPt162_Plots}a shows the outer orbit for the first $100T_\phi$ of integration time. 

\begin{figure*}
\centering
\hspace{-0.2cm}
\includegraphics[width=0.36\linewidth]{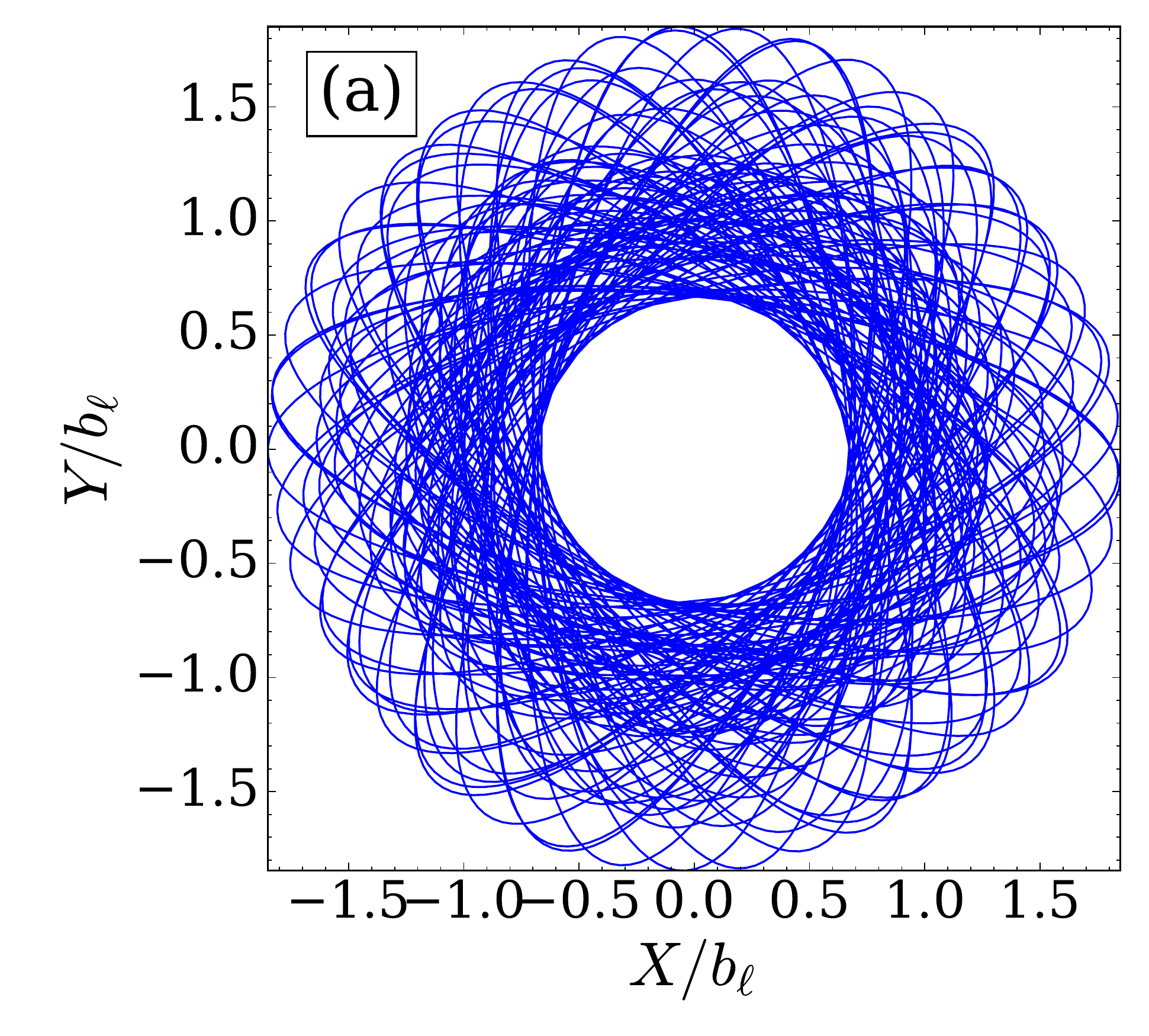}
\raisebox{0.0cm}{\includegraphics[width=0.34\linewidth]{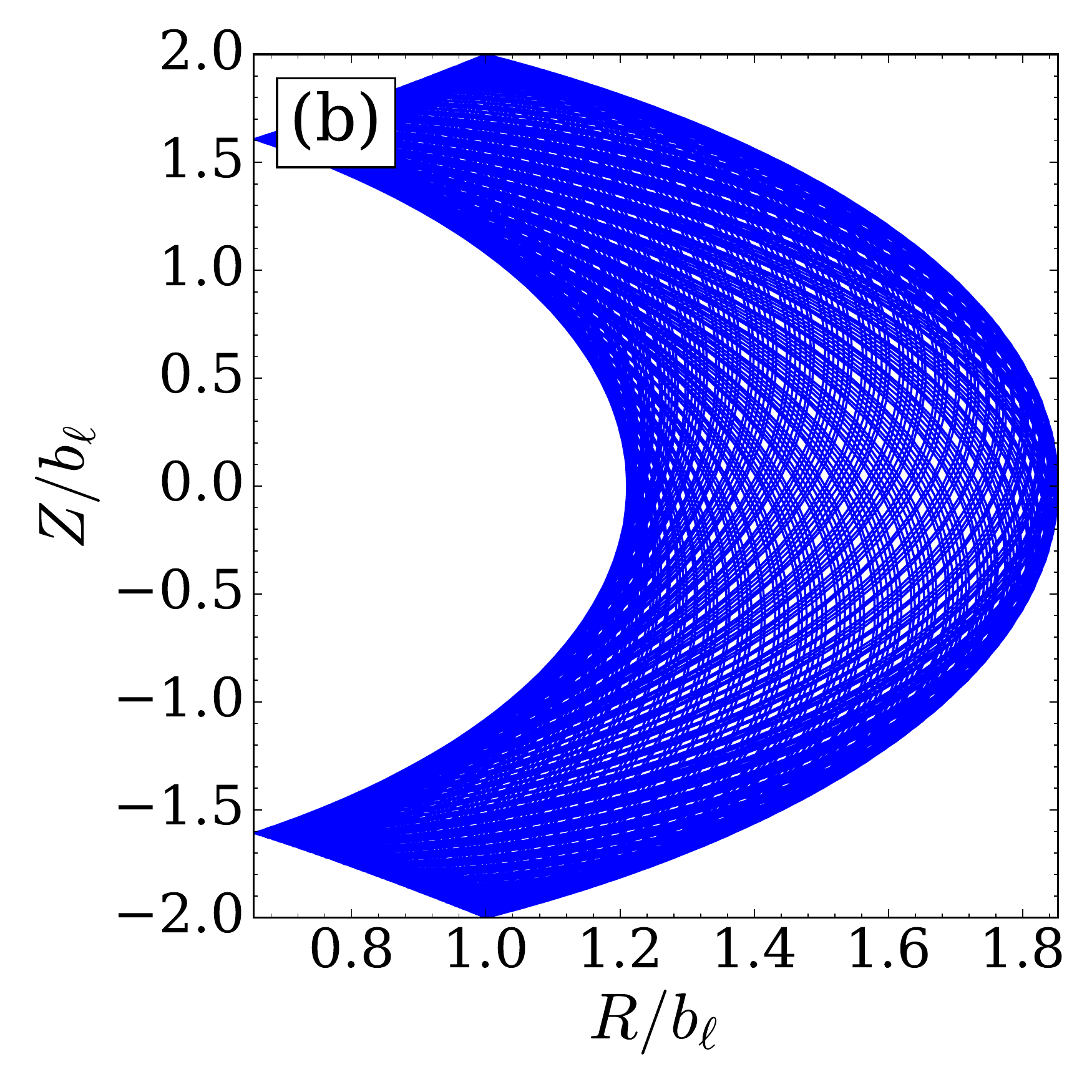}}
\includegraphics[width=0.34\linewidth]{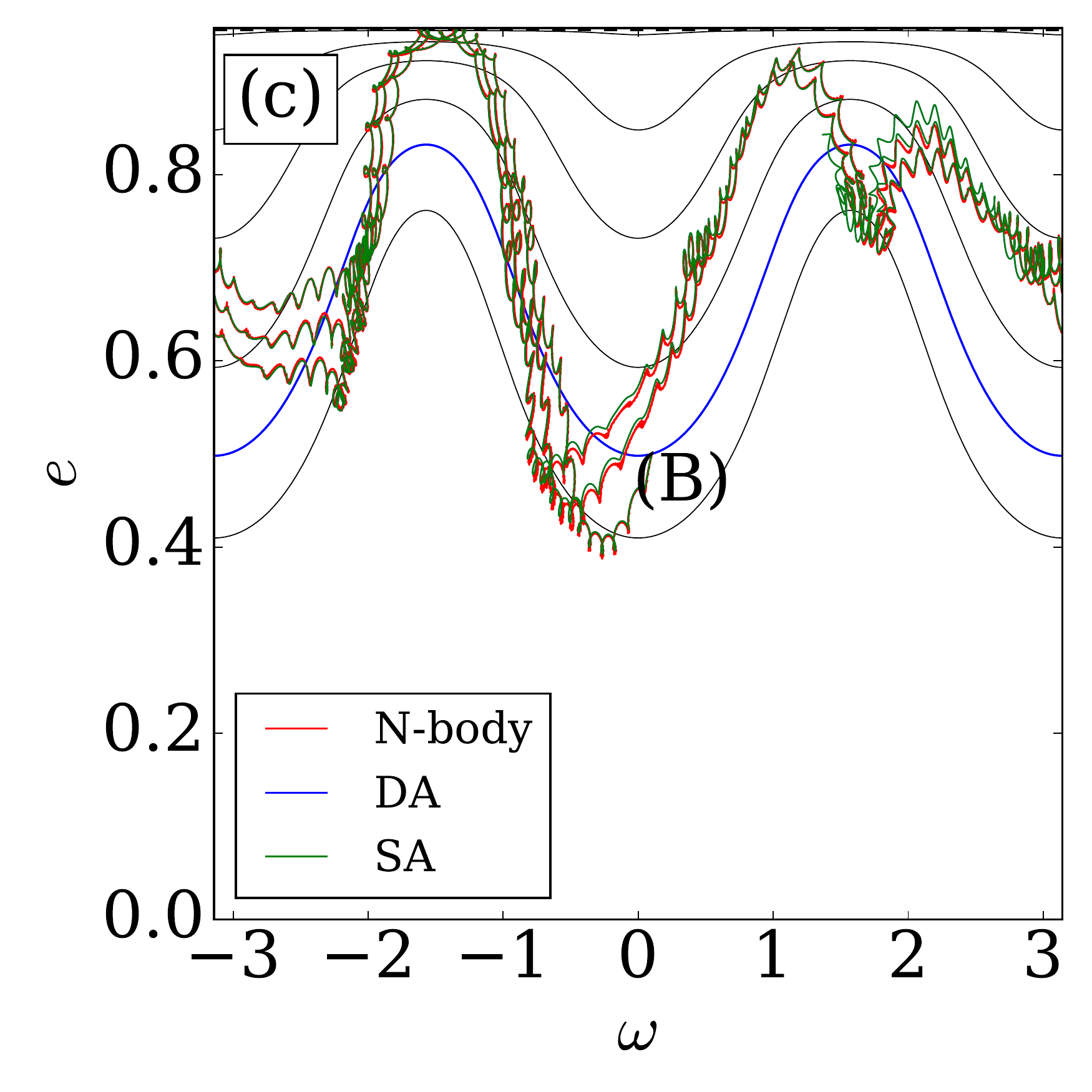}
\includegraphics[width=0.348\linewidth]{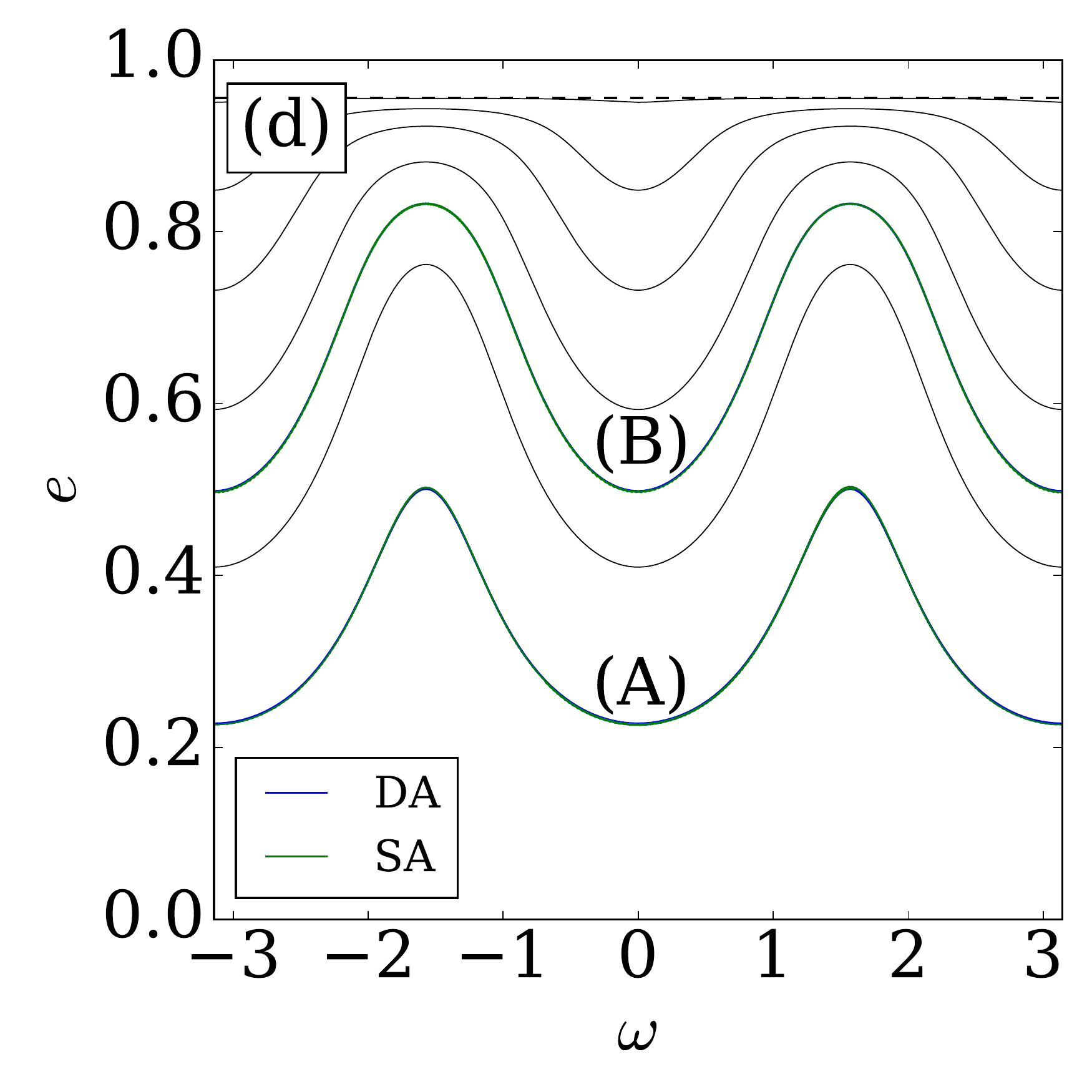}
\caption{Same as Figure \ref{MN_NumGamPt370_Plots} except that the initial vertical coordinate for the outer orbit is now $Z=2b_\ell$ (otherwise, the potential and initial conditions of the outer orbit are the same). This gives a negative $\Gamma$ value, namely $\Gamma_\mathrm{num}=-0.163$ and $A_\mathrm{num}=0.0392(GM/b_\ell^3$). The outer orbit is shown for the first $100T_\phi$ in panels (a) and (b). In panel (c) we integrate the equations of motion for a single phase-space trajectory (`B'), with $a=5\times 10^4\mathrm{AU}$, for $100T_\phi$. In panel (d) we show this trajectory and another trajectory (`A') for an $a=5\times 10^3\mathrm{AU}$ binary over $2000 T_\phi$. In all cases the other inner orbital elements of trajectories (A) and (B) are exactly the same as in Figure \ref{MN_NumGamPt370_Plots}, namely $ (e,i,\Omega,M) = (0.5,70^\circ,17.188^\circ,161.36^\circ)$, with initial $\omega$ taking the values (A) $91.67^\circ$, (B) $5.73^\circ$. See \S\ref{sec_NumericalVerification3} for discussion.}
\label{MN_NumGamMinusPt163_Plots}
\end{figure*}

In Figures \ref{Iso_GamPt162_Plots}b,c the binary's initial orbital elements are also unchanged from those in Figure 8 (in particular, we still use $a=10^3 \mathrm{AU}$). We again integrate for $100T_\phi$.  We see that the agreement between N-body, SA and DA calculations is good for trajectory (A) and reasonable for (B).  Trajectory (C), which is very near the separatrix, does not show very good agreement between the DA and other approximations. The reason that the DA theory is so much less accurate in this example than for $\Gamma=0.4$ (Figure \ref{Iso_GamPt4_Plots}) --- despite having similar ratios of $t_\mathrm{sec}/T_\phi$ --- is that this time the binary does not fill its annulus fast enough for the time-averaged potential coefficients $\overline{\Phi}_{\alpha\beta}$ to converge rapidly.  

To show that the secular approximation can be improved, we rerun integrations of the same three trajectories with exactly the same initial conditions except we use a smaller semi-major axis, $a=100 \mathrm{AU}$ (lowering $T_b$ by $\sim 30$ and, correspondingly, increasing $t_{\rm sec}$ by the same factor).  We integrate for $3000T_\phi$.  The results are shown in Figures \ref{Iso_GamPt162_Plots}d,e.  The secular timescales are much longer now (from $t_\mathrm{sec}/T_\phi \sim 650$ for trajectory (A), to $\sim 1300$ for (C)). As a result, the binary fills its annulus many times per secular period and the agreement between DA and SA integrations is almost perfect.

\begin{figure*}
\centering
\includegraphics[width=0.34\linewidth]{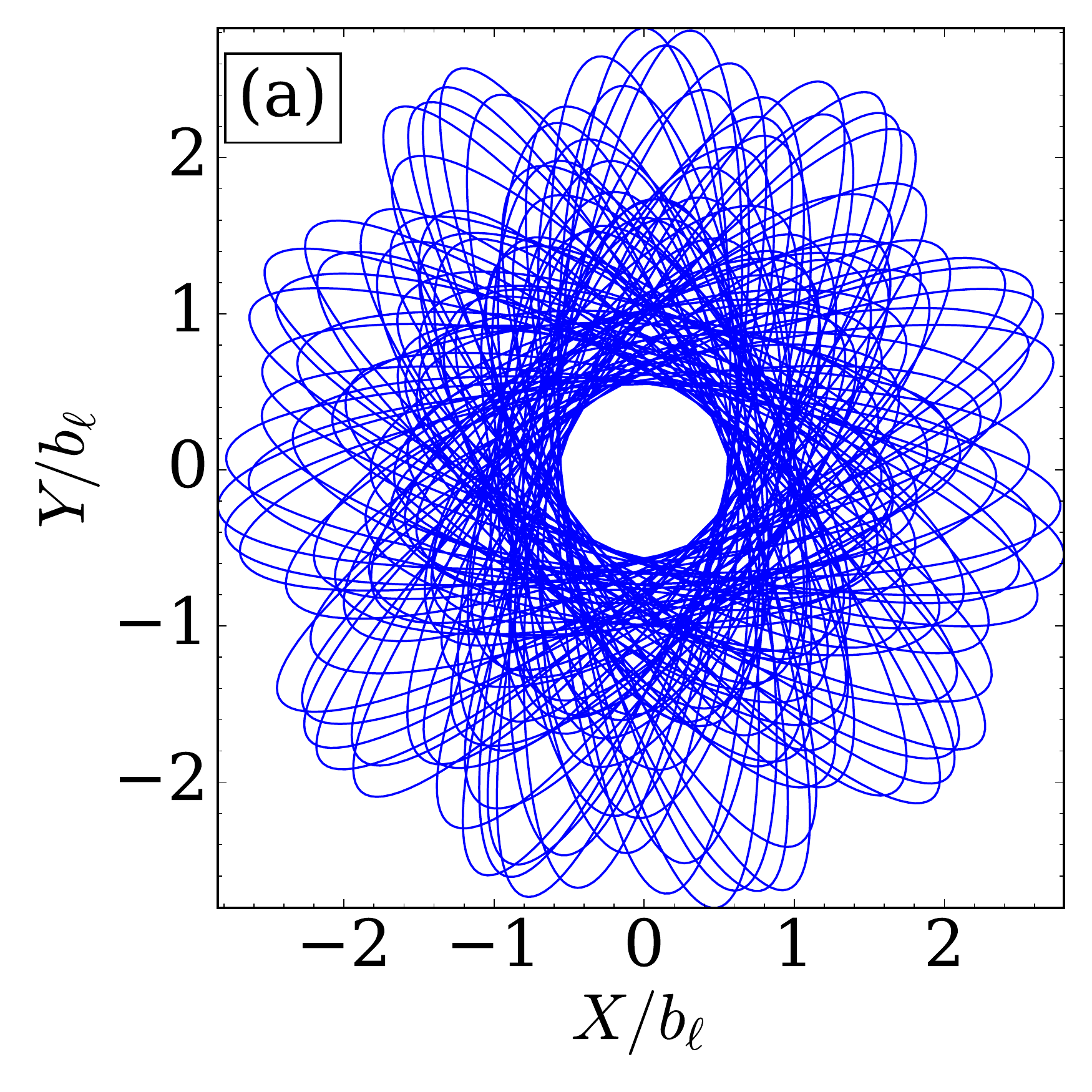}
\raisebox{0.05cm}{\includegraphics[width=0.345\linewidth]{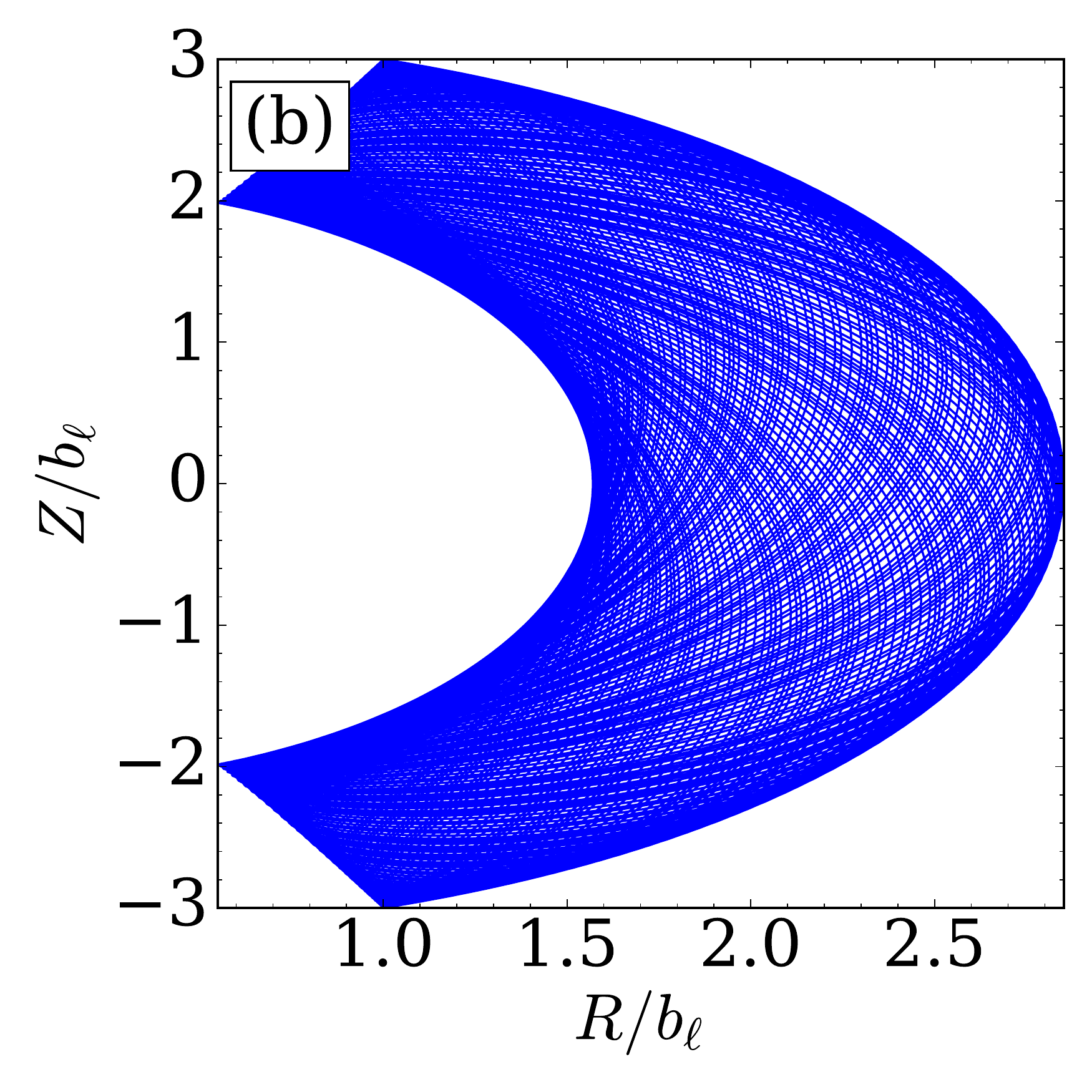}}
\raisebox{0.1cm}{\includegraphics[width=0.34\linewidth]{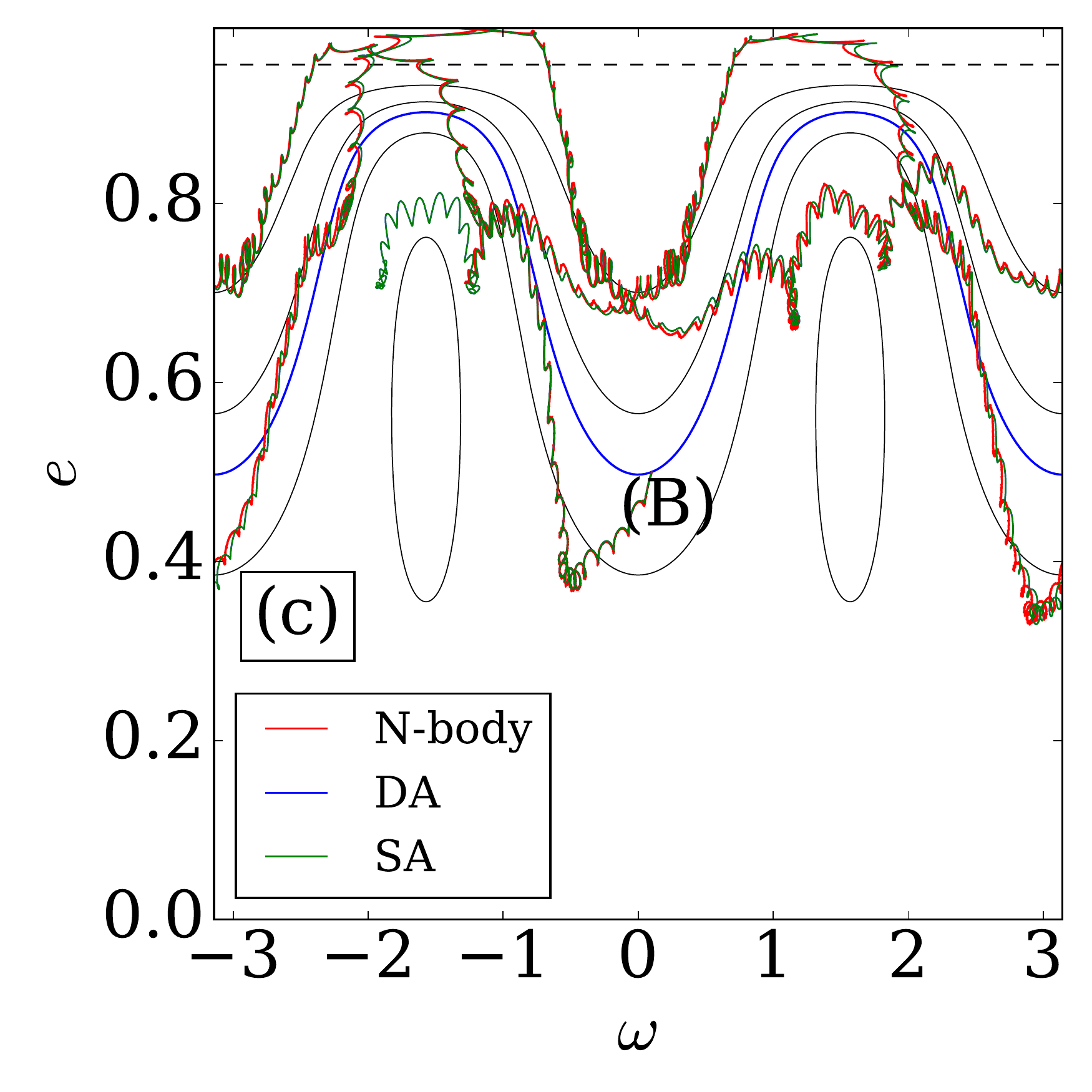}}
\raisebox{0.13cm}{\includegraphics[width=0.348\linewidth]{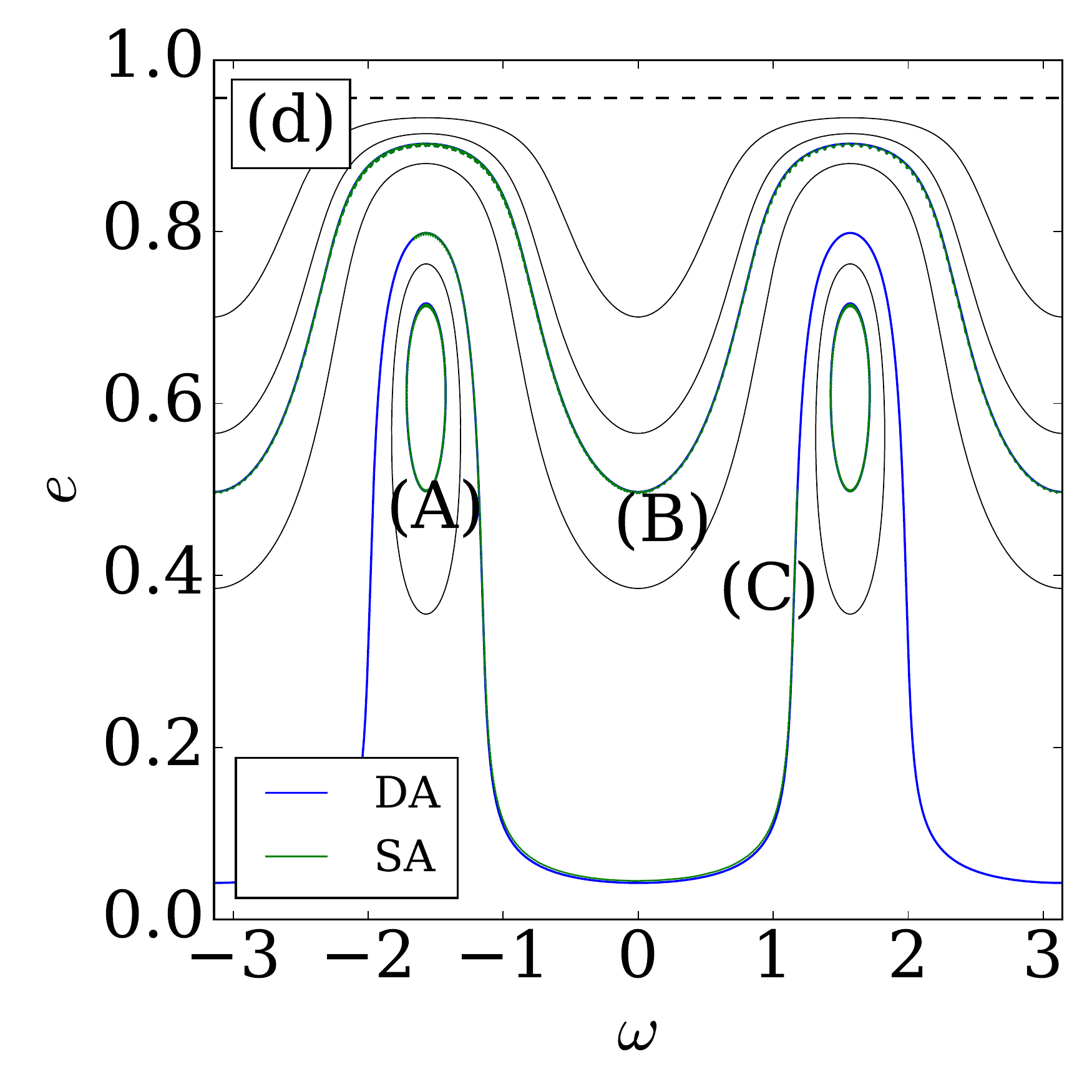}}
\caption{Same as Figure \ref{MN_NumGamPt370_Plots} except that the initial vertical coordinate for the outer orbit is now $Z=3b_\ell$ (other initial conditions and the potential are the same).  This gives a highly inclined outer orbit illustrated in panels (a) and (b) with $\Gamma_\mathrm{num}=-0.384$ and $A_\mathrm{num}=0.0142(GM/b_\ell^3)$. In panel (c) we integrate the equations of motion for a single (circulating) phase-space trajectory, with $a=5\times 10^4\mathrm{AU}$, for $100T_\phi$. In panel (d) we show three trajectories for an $a=5\times 10^3\mathrm{AU}$ binary over $2000 T_\phi$. In each case the other orbital elements are $ (e,i,\Omega,M) = (0.5,70^\circ,17.188^\circ,161.36^\circ)$, with initial $\omega$ taking the values (A) $91.67^\circ$ (librating), (B) $5.73^\circ$ (circulating), (C) $67.6^\circ$ (close to the separatrix). See \S\ref{sec_NumericalVerification3} for discussion.}
\label{MN_NumGamMinusPt384_Plots}
\end{figure*}

\paragraph*{[Figure \ref{MN_NumGamPt045_Plots}: Miyamoto-Nagai potential, $\Gamma_\mathrm{num}=0.045$.]} The binary orbits the Miyamoto-Nagai potential \eqref{MNPot} with $b_\ell=3.5\mathrm{kpc}$ and $b_h/b_\ell=5$, which is a less flattened potential than the $b_h/b_\ell=1$ example from Figure \ref{MN_NumGamPt370_Plots}.  The total mass is again $M=10^{11} M_\odot$. The initial conditions of the outer orbit are \begin{align}
(R,v_\mathrm{R},Z,v_\mathrm{Z},\phi,v_\mathrm{\phi}) \nn= (&2.29b_\ell,0.05\sqrt{GM/b_l},\,0.143b_\ell,\, \\ &0.05\sqrt{GM/b_l},\,0,\,0.05\sqrt{GM/b_l}).\end{align} 
From Figures \ref{MN_NumGamPt045_Plots}a,b (which both show the first $100T_\phi$ of integration time) we see that $\Rg$ makes large excursions in the $Z$ direction: the binary's barycentric orbit is about as thick vertically as it is radially. Numerically we find $A_\mathrm{num}=0.00826(GM/b_\ell^3)$ and $\Gamma_\mathrm{num}=0.045$. 

In Figures \ref{MN_NumGamPt045_Plots}c,d the binary has initial orbital elements $(a,e,i,\Omega,M)=(5 \times 10^4\mathrm{AU},0.9,70^{\circ}, 17.188^{\circ},  161.36^{\circ})$ and three initial values of $\omega$, namely (A) $90^{\circ}$ (librating orbit), (B) $24.9^{\circ}$ (circulating orbit), and (C) $40.1^{\circ}$ (separatrix).  We integrate the equations of motion for $100T_\phi$. Although the N-body and SA integrations agree extremely well (the red and green lines in Figures \ref{MN_NumGamPt045_Plots}c,d are almost indistinguishable), the agreement with the DA integration is only reasonable for trajectories (A) and (B) and is poor for trajectory (C).  Moreover, the smooth periodicity of the DA solution has disappeared in the SA and N-body integrations, which show chaotic small-scale oscillations. This is unsurprising --- despite the fact that $t_\mathrm{sec}/T_\phi \gtrsim 50$ in each case, the binary does not fill its torus densely enough over sufficiently few $T_\phi$ to render the axisymmetric secular approximation valid.

The DA theory fares much better in Figures \ref{MN_NumGamPt045_Plots}e,f, in which we rerun the same three trajectories except with a smaller semi-major axis, $a=10^4 \mathrm{AU}$ (i.e. lowering $T_b$ by $\sim 10$ and increasing $t_{\rm sec}$ by the same amount).  We integrate for $1000T_\phi$, so that $t_\mathrm{sec}/T_\phi \gtrsim 500$.  Of course, the timescales involved here are much longer than the age of the universe and so are not relevant in practice, but this example demonstrates how the secular approximation becomes more accurate when the binary's outer orbit has a better chance to fill its axisymmetric torus.


\subsection{Accuracy of the doubly-averaged approximation in the cases \texorpdfstring{$-1/5 < \Gamma \leq 0$ and $\Gamma \leq -1/5$}{}} \label{sec_NumericalVerification3}


In this section we present one numerical example in each of the regimes $-1/5 < \Gamma \leq 0$ and $\Gamma \leq -1/5$ (explored in \S\ref{sec_GammaRegime3} and \S\ref{sec_GammaRegime4} respectively).  In both cases we use a Miyamoto-Nagai potential with $b_h/b_\ell=1$; each time we give an example in which the DA theory works very poorly, and one in which it works well.

\paragraph*{[Figure \ref{MN_NumGamMinusPt163_Plots}: Miyamoto-Nagai potential, $\Gamma_\mathrm{num}=-0.163$.]} The potential and outer orbit initial conditions are exactly as in Figure \ref{MN_NumGamPt370_Plots}, except that the initial vertical coordinate is $Z=2b_\ell$.  The resulting (vertically extended) outer orbit (shown in Figures \ref{MN_NumGamMinusPt163_Plots}a,b for the first $100T_\phi$ of integration time) results in a negative $\Gamma$ value: we find $\Gamma_\mathrm{num}=-0.163$ and $A_\mathrm{num}=0.0392(GM/b_\ell^3)$.

For clarity we only show a single phase-space orbit (`B') in Figure \ref{MN_NumGamMinusPt163_Plots}c (integrated for $100T_\phi$), which obviously circulates since there are no fixed points in this $\Gamma$ regime (\S\ref{sec_GammaRegime3}). The initial orbital elements for trajectory (B) are identical to those in Figure \ref{MN_NumGamPt370_Plots} --- in particular, we again use $a=5\times 10^4\mathrm{AU}$. From the $(\omega,e)$ phase-space portrait (Figure \ref{MN_NumGamMinusPt163_Plots}c) we see that the N-body and SA integrations agree nicely, but the agreement with the DA theory is very poor.  Comparing with Figure \ref{MN_NumGamPt370_Plots}, we note that we have caused the DA theory to fail simply by changing a single parameter, the initial $Z$ coordinate of the outer orbit.  This is true because the outer orbit is now much more vertically extended and so takes many more outer orbital periods to fill its torus, delaying the convergence of the time averages of the potential derivatives $\Phi_{\alpha\beta}$ that enter the DA Hamiltonian.

Much better agreement is found in Figure \ref{MN_NumGamMinusPt163_Plots}d, in which we use a semi-major axis of $a=5\times 10^3\mathrm{AU}$ (increasing the secular timescale by a factor $\sim 30$) and integrate for $2000T_\phi$.  Other than semi-major axis, trajectory (B) in this figure has the same initial conditions as trajectory (B) in Figure \ref{MN_NumGamMinusPt163_Plots}c. Trajectory (A) differs from (B) only in that it has initial $\omega=91.67^\circ.$

\paragraph*{[Figure \ref{MN_NumGamMinusPt384_Plots}: Miyamoto-Nagai potential, $\Gamma_\mathrm{num}=-0.384$.]} The potential and the initial conditions of the outer orbit are exactly the same as in Figure \ref{MN_NumGamMinusPt163_Plots}, except the initial $Z$ coordinate is now even larger, $Z=3b_\ell$, making the orbit very highly inclined. This thickens the outer orbit's torus further (the first $100T_\phi$ are shown in Figures \ref{MN_NumGamMinusPt384_Plots}a,b) and results in $\Gamma_\mathrm{num} = -0.384$ and $A_\mathrm{num}=0.0142(GM/b_\ell^3)$. From panel (a) it is clear that the outer orbit has not come close to filling its torus after $100T_\phi$.

Figure \ref{MN_NumGamMinusPt384_Plots}c shows a trajectory with exactly the same initial conditions as trajectory (B) in Figure \ref{MN_NumGamMinusPt163_Plots}c (i.e. $a=5\times 10^4$ AU), integrated for $100T_\phi$.  The DA theory is seen to be very inaccurate here.  For a better example, in Figure \ref{MN_NumGamMinusPt384_Plots}d we again lower the semi-major axis to $a=5\times 10^3 \mathrm{AU}$ and integrate for $2000T_\phi$. We see that this time (B) is captured almost perfectly by the DA theory, as are two other trajectories, namely (A) (identical initial conditions except $\omega=91.67^\circ$) and (C) (with $\omega=67.6^\circ$).


\subsection{Discussion: validity of the doubly-averaged secular theory} 
\label{discussnum}


The examples presented in \S \ref{sec_NumericalVerification1}-\ref{sec_NumericalVerification3}  illustrate two possible ways in which the DA theory can be in error:\\ \\
(i) The secular approximation is only a good one if the timescale for evolution of the inner orbital elements is much longer than the time for the time-averages of $\Phi_{\alpha\beta}$ to converge (i.e. for the binary to `fill its torus').  Otherwise, the secular approximation can break down and the DA equations can fail completely to describe the evolution.\\ \\
(ii) The torque experienced by the binary fluctuates on the timescale of its outer orbital period, leading to small fluctuations in the orbital elements on this timescale.  Even when the DA equations provide a good description of the dynamics on average, they will always fail to resolve these short-timescale fluctuations (which are fully captured at the level of the SA approximation).
\\ \\
So far in this section we primarily explored the validity of the secular approximation in the sense of $\Phi_{\alpha\beta}$ convergence (effect (i)). However, in practice both effects (i) and (ii) are typically present in our calculations and distinguishing them is important.

The best illustration of the two effects operating simultaneously is provided by Figures \ref{MN_NumGamMinusPt163_Plots}c and \ref{MN_NumGamMinusPt384_Plots}c, which show that the secular (DA) approximation completely fails to follow the dynamics quantitatively over long timescales. First, the overall shape of each trajectory (the `carrier signal') computed with both the N-body and SA theory does not line up with the DA prediction. This mismatch is entirely the consequence of effect (i). Additionally, the N-body and SA curves also exhibit short-timescale `wiggles' on top of the smoother `carrier' phase curve. These wiggles are caused by effect (ii), i.e. torque variations along the outer orbit.

Regardless of how good is the secular approximation by the measure of $\Phi_{\alpha\beta}$ convergence, there are always small short-timescale fluctuations due to effect (ii) that the DA theory cannot capture (e.g. see Figures \ref{MN_NumGamPt370_Plots}c,d, \ref{Iso_GamPt162_Plots}b,c, \& \ref{MN_NumGamPt045_Plots}c,d). In many applications of the (quadrupole) LK theory effect (i) is largely absent because there the timescale for a binary to effectively `fill its torus' is simply equal to its outer orbital period. The primary deviations from the DA prediction in the LK case are then due to short-timescale fluctuations (effect (ii)), which has been studied widely in this setup (\citealt{Ivanov2005, Katz2012,Antonini2012,Bode2014,Antonini2014,Antognini2014,Luo2016,Grishin2018}). While we largely pass over them from now on, the short-period SA fluctuations can become vitally important when eccentricities get very close to $1$ (see LK references above).  Their magnitude depends strongly on the shape of the outer orbit. Accounting for such fluctuations in a systematic way in the general $\Gamma$ case is a topic for future work.

Focusing now on the issue of the $\Phi_{\alpha\beta}$ convergence, the two examples presented in \S\ref{sec_NumericalVerification1} (Figures \ref{Iso_GamPt4_Plots} and \ref{MN_NumGamPt370_Plots} --- both in the regime $\Gamma>1/5$) showed very good agreement between DA theory and direct numerical integration, even for binaries whose secular evolution timescales were significantly shorter than $100 T_\phi$ (for example, trajectory (A) in Figure \ref{Iso_GamPt4_Plots} had $t_\mathrm{sec}/T_\phi \sim 25$).  However, all other examples required a much larger ratio of $t_\mathrm{sec}/T_\phi$ for the DA theory to be rendered accurate (typically $\sim$ a few $\times 100$).  This is because the secular approximation is valid only when the timescale for secular evolution is much longer than the time for the binary to fill its torus densely. The number of outer orbital periods required to fill the torus densely depends strongly on the form of the potential and the choice of outer orbit.

The DA approximation is often most easily satisfied (i.e. it works for relatively small values of $t_\mathrm{sec}/T_\phi$) in spherical potentials, because then the `torus' reduces to a two-dimensional annulus (e.g. Figure \ref{Iso_GamPt4_Plots}).  Not only does this decrease the volume that must be filled, but also the derivatives $\Phi_{xz}, \Phi_{yz}$ automatically vanish and so pose no problem to the convergence. Circular outer orbits and orbits that avoid any central core tend to fill their annuli particularly efficiently (and typically correspond to $\Gamma > 1/5$; see Paper I). 

However, many spherical cluster potentials are cored (such as the isochrone and Plummer models), and so binaries that spend significant time near the centre of these clusters (i.e. those with small $\rp$) experience an almost-harmonic potential.  Since orbits in a harmonic potential are closed ellipses, the apsidal precession of such outer orbits can be very slow; this frequently leads to unfilled gaps being left in the annulus even after  $\sim 100T_\phi$ (see Paper I for an example).  As a result, the secular approximation may require relatively large values of $t_\mathrm{sec}/T_\phi$ to be valid, as was the case in Figure \ref{Iso_GamPt162_Plots}.  In spherical potentials, small $\rp$ tends to correspond to small (but always positive) $\Gamma$, so this issue will arise most often for binaries in the regime $0 < \Gamma \leq 1/5$.

In non-spherical potentials the situation is often worse simply because there is a third dimension of the torus for the outer orbit to fill.  In addition, the derivatives $\Phi_{xz}, \Phi_{yz}$ are no longer identically zero in general, so we must wait for them to converge, and this typically takes longer than for the other $\Phi_{\alpha\beta}$ (see \S7.2 of Paper I). In these potentials the secular approximation is most easily satisfied by binaries on outer orbits that are coplanar or nearly coplanar --- then the torus is small in volume, and, if the potential is strongly flattened, the vertical oscillations tend to be very rapid, so the torus is filled efficiently (a good example is Figure \ref{MN_NumGamPt370_Plots}). Orbits confined near the midplane of a strongly flattened potential have $\Gamma \approx 1/3$. 

On the other hand, when the outer orbit has a large vertical extent, filling a torus takes many more outer orbital periods and hence very large values of $t_\mathrm{sec}/T_\phi$ are required for the secular theory to be accurate (e.g. Figures \ref{MN_NumGamMinusPt163_Plots} \& \ref{MN_NumGamMinusPt384_Plots}). This in turn implies that for $\Gamma<0$, the DA theory may be of limited practical use in certain cases (when $t_{\rm sec}/T_\phi$ is not large enough) because achieving negative $\Gamma$ typically requires outer orbits that make large excursions in the $Z$ direction.

	
\section{Effect of general relativistic precession on the cluster-tide driven evolution} 
\label{GReffect}


So far our secular theory considered only the gravitational tidal effect of a stellar cluster on binary evolution.  However in a realistic astrophysical situation there could be other, short-range forces which must be taken into account, particularly at high eccentricity when the pericentre distance becomes small (see \S \ref{sec_highe}).  Depending on the type of binary (i.e the masses and sizes of its components) and its semi-major axis these could include (i) prograde precession of the argument of pericentre $\omega$ due to general relativity (GR), (ii) precession due to the oblateness or tidal distortions of the binary components, (iii) loss of energy and angular momentum due to gravitational wave emission, (iv) tidal dissipation within the components of the binary, etc. The first two effects are conservative in that they do not change the energy of the system and preserve binary semi-major axis, while the latter two lead to energy losses and tend to shrink the binary orbit. 

In this section we will only consider (i), namely GR pericentre precession. We can include GR precession in our doubly-averaged theory by adding the following term to the Hamiltonian \eqref{H1Mt} \citep{Fabrycky2007}: 
\begin{align} 
\label{HGR} 
\langle H_\mathrm{GR} \rangle_M = -\frac{3G^2(m_1+m_2)^2}{c^2a^2\sqrt{1-e^2}} = - \frac{3G^4(m_1+m_2)^4}{c^2L^3J}. 
\end{align} 
The angle brackets remind us that this term is derived by averaging over the binary's mean anomaly. The Hamiltonian $\langle H_{\mathrm{GR}} \rangle_M$ is independent of the longitude of the ascending node $\Omega$, so the $z$-component of angular momentum $J_z$ is conserved; hence, $\Theta = (1-e^2)\cos^2 i$ remains an integral of motion. Another integral of motion is the full perturbation energy $\overline{\langle H_1 \rangle}_{M} + \langle H_{\mathrm{GR}} \rangle_M$ --- the sum of the cluster tide and GR Hamiltonians, equations \eqref{H1Mt} and \eqref{HGR}.

We put \eqref{HGR} into dimensionless form by dividing by $C=Aa^2/8$ (see equation \eqref{H1star}).  Then we must compare the perturbation $H_1^*$ due to the potential of the cluster to the corresponding dimensionless perturbation due to GR:  
\begin{align} 
H^*_\mathrm{GR} &= -\frac{24G^2(m_1+m_2)^2}{c^2Aa^4\sqrt{1-e^2}} = -\frac{24G^6(m_1+m_2)^6}{c^2AL^8j} \nn \\ &=-\frac{\epsilon_\mathrm{GR}}{\sqrt{1-e^2}}=-\frac{\epsilon_\mathrm{GR}}{j}. 
\label{eq:HGR}
\end{align} 
where the relative strength of GR precession compared to the cluster tide is measured by the (not necessarily small) parameter
\begin{align} 
\epsilon_\mathrm{GR} &\equiv \frac{24G^2(m_1+m_2)^2}{c^2Aa^4} \\ \nn &=  0.258 \times \left( \frac{A^*}{0.5}\right)^{-1}\left( \frac{M}{10^5M_\odot}\right)^{-1}\left( \frac{b}{\mathrm{pc}}\right)^{3}  \\ &\times  \left( \frac{m_1+m_2}{M_\odot}\right)^{2}  \left( \frac{a}{20 \, \mathrm{AU}}\right)^{-4}. 
\end{align} 
In the numerical estimate we have again assumed that the binary is orbiting a spherical cluster with scale radius $b$ and total mass $M$, and typical values of $A^*$ are given in Paper I. The parameter $\epsilon_\mathrm{GR}$ represents, up to a constant factor, the ratio of the GR apsidal precession rate for a {\it circular} binary to the binary precession rate due to the cluster tide.

The prograde pericentre precession rate induced by GR is \begin{align} 
\label{omegadotgr} 
\dot{\omega}_\mathrm{GR} &\equiv \frac{C}{L}\frac{\partial H_\mathrm{GR}^*}{\partial j} = \frac{3G^{3/2}(m_1+m_2)^{3/2}}{a^{5/2}c^2(1-e^2)}.
\end{align} 
With this we can evaluate the ratio of $\dot{\omega}_\mathrm{GR}$ to the precession rate $\dot{\omega}_1$ due to the background cluster tide alone (equation \eqref{eom1}): 
\begin{align} 
\frac{\dot{\omega}_\mathrm{GR}}{\dot{\omega}_1} \nn &= \frac{\epsilon_\mathrm{GR}j}{6}\left(5\Gamma\Theta - j^4 - 5\Gamma\cos 2 \omega(\Theta-j^4) \right)^{-1} \\ &= \frac{\epsilon_\mathrm{GR}j}{6\Theta}\left(10\Gamma \sin^2 \omega + \frac{j^4}{\Theta}(5\Gamma-1) \right)^{-1}. 
\label{omegaratio} 
\end{align} 
We expect GR effects to be most important at high eccentricity, when $j^2\sim\Theta \ll 1$. Also, in most cases of interest $j_\mathrm{min}$ occurs at $\omega=\pi/2$. Plugging these relations into \eqref{omegaratio} we can evaluate 
\begin{align} 
\frac{\dot{\omega}_\mathrm{GR}}{\dot{\omega}_1} \approx \frac{\epsilon_\mathrm{GR}}{60\Gamma}\frac{j}{\Theta}. 
\end{align}  
We can ignore GR precession only when $\vert \dot{\omega}_\mathrm{GR}/\dot{\omega}_1 \vert \lesssim 1$, which requires rather small $\epsilon_\mathrm{GR}\lesssim 60\Gamma \Theta^{1/2}(j/\Theta^{1/2})^{-1}$.

Since $H^*_{\mathrm{GR}}$ is independent of $\omega$, the equation for $\md j/\md t$ is unchanged from \eqref{eom2}. As a result the maximum eccentricity reached by a binary perturbed by the Hamiltonian $H^* \equiv H_1^* + H_{\mathrm{GR}}^*$ will be found at the same value of $\omega$ as for a binary perturbed only by $H_1^*$. Following the derivation of equation \eqref{djsquareddt}, we can express $\omega$ through $j$ and $H^*$ using equations (\ref{H1star}) and (\ref{eq:HGR}) and plug the result into equation \eqref{eom2}. After some manipulations we find that
\begin{align}
\frac{\md j^2}{\md t} =&
\pm \frac{12C\sqrt{25\Gamma^2-1}}{Lj}
\left[j^2(j_0^2-j^2)+\frac{\epsilon_\mathrm{GR}}{3(5\Gamma-1)}j\right]^{1/2}
\nonumber\\
&\times \left[(j_+^2-j^2)(j^2-j_-^2)-\frac{\epsilon_\mathrm{GR}}{3(1+5\Gamma)}j\right]^{1/2},
\label{eq:djdtGR}
\end{align}
where the definitions of $j_\pm, j_0$ now involve $H^*$ replacing $H_1^*$ everywhere. In the limit $\epsilon_\mathrm{GR}\to 0$ this equation reduces to \eqref{djsquareddt}.  We have used equation \eqref{eq:djdtGR} to find maximum eccentricities when calculating merger rates of compact object binaries in globular and nuclear star clusters \citep{Hamilton2019c}.

In principle, the minima and maxima of $j$ (and hence $e$) in the presence of GR precession can be calculated analytically. Indeed, the 8th order polynomial inside the radical on the right-hand side of equation (\ref{eq:djdtGR}) can be factorized into the product of $j$, a depressed cubic polynomial, and a quartic polynomial; the latter two have analytic roots, which correspond to the extrema of $j$ (since $\md j^2/\md t=0$ when $j$ equals one of these roots). However, the symbolic expressions for these roots are too complicated to be worth presenting here. 

One can draw qualitative conclusions regarding the role of GR precession in our problem. Typically its effect is to increase the rate at which $\omega$ precesses, i.e. the rate at which the binary's apsidal line changes its orientation with respect to the fixed cluster axes. Then the cluster has less time to coherently torque the binary per secular cycle, resulting in a reduced maximum eccentricity (see e.g. \citealt{Fabrycky2007,Liu2015}).  However, there are subtleties in this problem. In the LK case the maximum eccentricity can actually be \textit{increased} by GR precession if one includes octupole terms in the perturbing Hamiltonian (\citealt{Ford2000,Naoz2013b}).  In our (strictly quadrupole) case it turns out that increasing the strength of GR precession can increase $e_\mathrm{max}$ for some binaries if $\Gamma < 1/5$, because of the way that GR precession modifies the phase space morphology. We defer a more careful exploration of the secular problem with the GR precession in all $\Gamma$ regimes to a future study.


\section{Discussion} 
\label{discussion}


The main result of this work is the unveiling of a variety of new dynamical regimes that characterise the orbital evolution of a binary system subject to an external gravitational tidal field (`cluster'). While the results of \S\ref{sec_General} are completely general, we found that we need to investigate dynamics in four separate regimes, corresponding to certain ranges of the parameter $\Gamma$ characterising the external tidal field of the cluster and the binary orbit in it. For $\Gamma>1/5$ (\S\ref{sec_GammaRegime1}), the results were found to be qualitatively very similar to those previously derived in the test particle quadrupole Lidov-Kozai problem, which is recovered exactly by taking $\Gamma = 1$  \citep{Vash1999,Kinoshita2007,Antognini2015}.  

However, when leaving the regime $\Gamma > 1/5$ several qualitative differences emerge. The condition for the existence of fixed points changes, as do locations of minimum and maximum eccentricities in the $(\omega,e)$ phase-space and the morphology of the $(D,\Theta)$ plane; even the very existence of the fixed points and orbits librating around them changes with $\Gamma$.  The first bifurcation in the qualitative dynamics happens at $\Gamma=1/5$ but there are others at $\Gamma=0$ and $\Gamma=-1/5$, so we separately treated the regimes $0 < \Gamma \leq 1/5$ (described in \S\ref{sec_GammaRegime2}), $-1/5 < \Gamma \leq 0$ (in \S\ref{sec_GammaRegime3}), and $\Gamma \leq -1/5$ (in \S\ref{sec_GammaRegime4}).

In Tables \ref{table_Ranges} \& \ref{table_Hamiltonian} we collect the results of \S\S\ref{sec_GammaRegime1}-\ref{sec_GammaRegime4}. In Table \ref{table_Ranges} we provide the locations and values of minimum/maximum $j$, the values of $\Delta$ which enter the secular timescale \eqref{timescaleeqn}, and the allowed ranges of the constants of motion $D$ and $\Theta$, for each family of phase-space orbit and in each $\Gamma$ regime. Table \ref{table_Hamiltonian} collects the locations and values of the extrema of the dimensionless Hamiltonian $H_1^*$, depending on the $\Gamma$ regime and whether or not fixed points exist.

In this work we never explicitly considered the possibility of $\Gamma > 1$ --- all our examples were given for $\Gamma\leq 1$. Situations in which $\Gamma$ exceeds unity are possible. However, we found in Paper I that this regime is realised only for rather extreme binary orbits inside the cluster, e.g. close to polar, which justifies our overall neglect of the $\Gamma>1$ possibility. Also, we should note that none of the results obtained in the $\Gamma>1/5$ regime (\S\ref{sec_GammaRegime1}) explicitly assumed $\Gamma\leq 1$; they also hold when $\Gamma>1$. The only substantial difference in this case would be that $A$ becomes negative for $\Gamma>1$. As a result, maxima and minima of the (dimensional) perturbing Hamiltonian would swap their locations in the $(\omega,e)$ phase-space, and the phase-space trajectories would be traversed in the opposite direction compared to the $1/5<\Gamma\leq 1$ case.    

For our doubly-averaged theory to properly characterise binary orbital evolution certain conditions should be met. We already saw in \S \ref{sec_NumericalVerification} that the description based on the doubly-averaged Hamiltonian (\ref{H1Mt}) may fail when the secular timescale $t_{\rm sec}$ is not much longer than the period of the outer orbit of the binary $T_\phi$. In such cases one should resort to using the singly-averaged framework described in Paper I, which always works very well. Other phenomena that may affect the orbital evolution of binaries in clusters (on top of the smooth cluster tide-driven secular evolution) are discussed in \S \ref{scattering}.

\begin{table*}
\setlength{\tabcolsep}{10pt}
\caption{Summary of $(\omega,j)$ locations and values of minima and maxima of $j$, values of $\Delta$, and $\Theta$ and $D$ ranges, for different types of orbit in each $\Gamma$ regime.  These determine minimum and maximum eccentricities as well as timescales of secular eccentricity oscillations via equation \eqref{timescaleeqn}.}\label{table_Ranges}
\begin{tabular}{c c c c c c c c c c c c c c c} 
 \hline
Type of orbit & Regime & $(\omega,j_\mathrm{min})$& $(\omega,j_\mathrm{max})$  & $\Delta$ & $\Theta \in$ & $D \in$ \\ 
  \hline \hline
   Librating, $\Gamma>1/5$ & $j_-^2 < j^2 < j_+^2 < j^2_0$ & $(\pm \pi/2,j_-)$ & $(\pm \pi/2, j_+)$ & $j_0^2-j_-^2$ & $(0,\frac{1+5\Gamma}{10\Gamma}) $ & $(D_-,0)$\\ 
   \hline
Circulating, $\Gamma>1/5$ & $j^2_- < j^2 < j^2_0 < j^2_+ $ & $(\pm \pi/2, j_-)$ & $(0, j_0)$& $j_+^2-j_-^2$& $(0,1) $ & $(0,1-\Theta)$\\ 
\hline \hline
   Librating, $0< \Gamma \leq 1/5$ & $j^2_0 < j^2_- < j^2 < j^2_+$ & $(\pm \pi/2, j_-)$ & $(\pm \pi/2, j_+)$& $j_+^2-j_0^2$& $(0,\frac{10\Gamma}{1+5\Gamma}) $ &$(1-\Theta,D_-)$\\ 
 \hline
 Circulating, $0< \Gamma\leq 1/5$ & $j^2_- < j_0^2 < j^2 < j^2_+$ & $(0, j_0)$&$(\pm \pi/2, j_+)$& $j_+^2-j_-^2$& $(0,1) $ &$(0,1-\Theta)$\\ 
 \hline \hline
$-1/5 < \Gamma \leq 0$ & $j^2_- < j^2 _+ < j^2 < j^2_0$ & $(\pm\pi/2,j_+)$ & $(0, j_0)$& $j_0^2-j_-^2$& $(0,1) $ &$(0,1-\Theta)$\\ 
\hline \hline
   Librating, $\Gamma\leq -1/5$ & $j^2_+ < j^2 < j_-^2 < j^2_0$ & $(\pm\pi/2,j_+)$ & $(\pm\pi/2, j_-)$& $j_0^2-j_+^2$& $(0,\frac{1+5\Gamma}{10\Gamma}) $ &$(D_+,0)$\\ 
\hline
 Circulating, $\Gamma\leq -1/5$ & $j^2_+ < j^2  < j_0^2 < j^2_-$ & $(\pm\pi/2,j_+)$ & $(0, j_0)$& $j_-^2-j_+^2$&$(0,1) $ & $(0,1-\Theta)$\\ 
[1ex] 
 \hline
\end{tabular}
\end{table*}

\begin{table*}
\setlength{\tabcolsep}{10pt}
\caption{Summary of $(\omega,j)$ locations and values of minima and maxima of $H_1^*$, depending on $\Gamma$ and whether fixed points exist.}\label{table_Hamiltonian}
\begin{tabular}{c c c c c c c c c c c c c c c} 
 \hline
Category & $H_1^*$ min location & $H^*_{1,\mathrm{min}}$ & $H_1^*$ max location & $H^*_{1,\mathrm{max}}$ \\ 
  \hline   \hline
   $\Gamma>1/5$, fixed points exist & $j=\sqrt{\Theta}$ & $(5-3\Theta)(1-3\Gamma)$ & $(\pm \pi/2, j_\mathrm{f})$ & $H_-$\\
   \hline
   $\Gamma>1/5$, fixed points do not exist & $j=\sqrt{\Theta}$ & $(5-3\Theta)(1-3\Gamma)$ & $j=1$ & $2(1-3\Theta\Gamma)$\\ 
   \hline \hline
   $0< \Gamma \leq 1/5$, fixed points exist & $j=1$ & $2(1-3\Theta\Gamma)$ & $(\pm \pi/2, j_\mathrm{f})$ & $H_-$\\
   \hline
   $0<  \Gamma \leq 1/5$, fixed points do not exist & $j=1$ & $2(1-3\Theta\Gamma)$ & $j=\sqrt{\Theta}$ & $(5-3\Theta)(1-3\Gamma)$\\  
      \hline \hline
   $-1/5 < \Gamma \leq 0 $ & $j=1$ & $2(1-3\Theta\Gamma)$ & $j=\sqrt{\Theta}$ & $(5-3\Theta)(1-3\Gamma)$\\ 
      \hline \hline
   $\Gamma \leq -1/5$, fixed points exist & $(\pm\pi/2,j_\mathrm{f})$ & $H_+$ & $j=\sqrt{\Theta}$ & $(5-3\Theta)(1-3\Gamma)$\\
   \hline
   $\Gamma \leq -1/5$, fixed points do not exist & $j=1$ & $2(1-3\Theta\Gamma)$ & $j=\sqrt{\Theta}$ & $(5-3\Theta)(1-3\Gamma)$\\   
[1ex] 
 \hline
\end{tabular}
\end{table*}


\subsection{Critical inclination for the existence of fixed points}
\label{sec:crit_inc_circular}


A classic result of LK theory is the value of the critical initial inclination $i_0=i_\mathrm{c}$, above which fixed points appear in the $(\omega,e)$ phase-space. This critical angle marks the onset of large eccentricity oscillations, and a qualitative departure from classical Laplace-Lagrange dynamics. It provides a constraint on which binary orientations can lead to large eccentricity excursions. 

Assuming an initial binary eccentricity of zero, we can calculate $i_\mathrm{c}$ for general $\Gamma$ using the conservation of $\Theta=\cos^2 i_0$.  The upper bounds on $\Theta$ for fixed points to exist ($=3/5$ in the test particle quadrupole LK case) are given by equations \eqref{thetarange}, \eqref{thetarange2} and \eqref{thetarange4}; it then easily follows that the existence of fixed points requires $i_0 > i_\mathrm{c}$ where \begin{align}
i_\mathrm{c} = \begin{cases} \cos^{-1}(\Lambda^{1/2}), \,\,\,\,\,\,\,\,\,\,\,\,\,\, \vert \Gamma \vert > 1/5,\\ \cos^{-1}(\Lambda^{-1/2}), \,\,\,\,\,\,\,\,\, 0 < \Gamma \leq 1/5, \end{cases} 
\label{eq:critinc} 
\end{align} 
with $\Lambda(\Gamma)$ given in equation \eqref{jfeqn}. There are no fixed points for any initial inclination if $-1/5 <\Gamma \leq 0$, see \S\ref{sec_GammaRegime3}. If the initial eccentricity of the binary $e_0$ is non-zero, the argument of the $\cos^{-1}$ needs to be additionally divided by $\sqrt{1-e_0^2}$, further lowering $i_{\rm c}$. 

\begin{figure}
\centering
\includegraphics[width=0.95\linewidth]{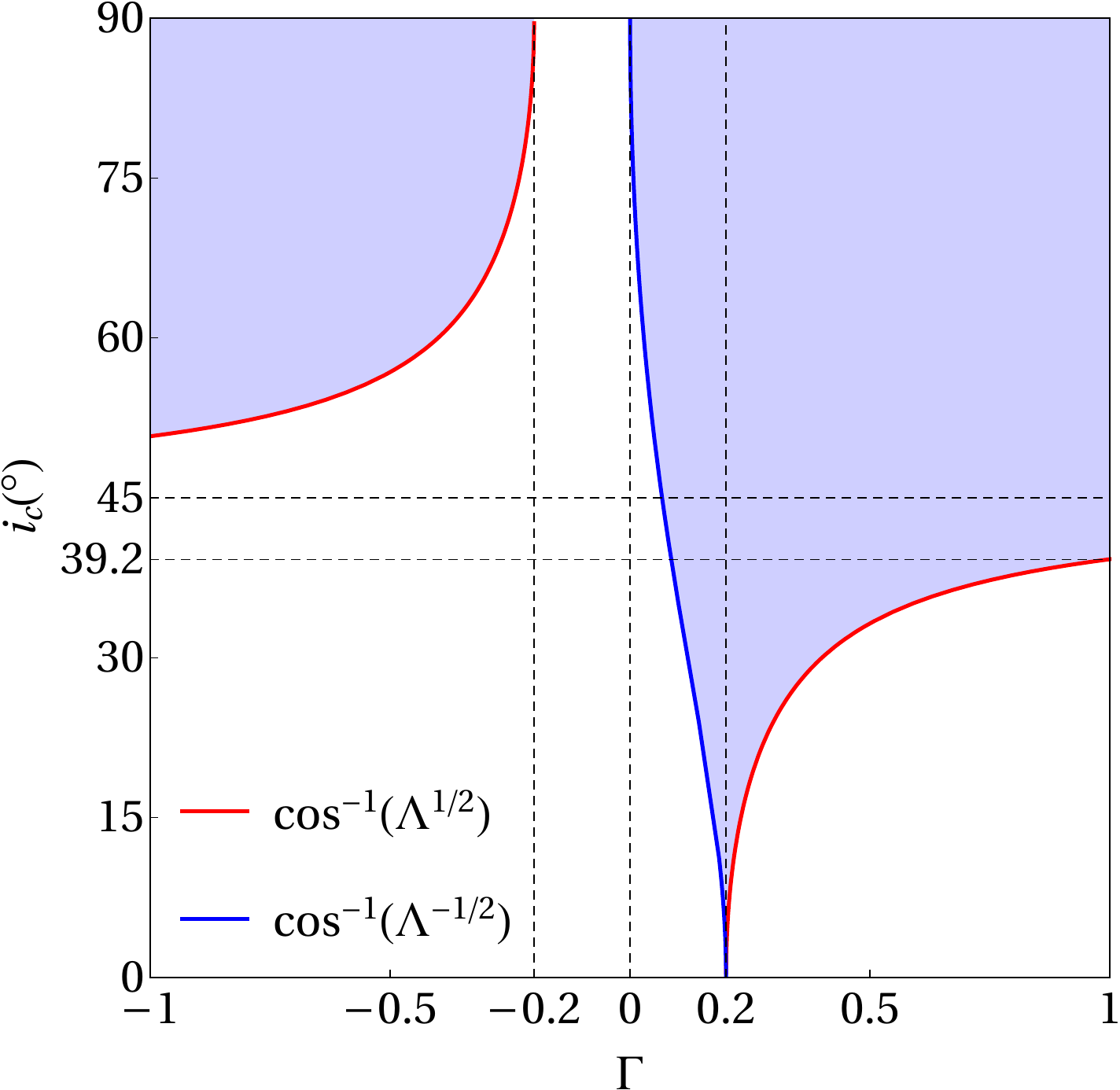}
\caption{Critical inclination $i_\mathrm{c}$ assuming zero initial eccentricity (equation \eqref{eq:critinc}), as a function of $\Gamma$. For initial inclinations greater than $i_\mathrm{c}$ (shaded regions), fixed points exist in the $(\omega,e)$ phase-space.  There are no fixed points in the range $-1/5 < \Gamma \leq 0$.  The classic LK result $i_\mathrm{c}=39.2^\circ$ is recovered when $\Gamma=1$.}
\label{Critical_Inclination_Plot}
\end{figure}

We plot $i_\mathrm{c}$ as a function of $\Gamma$ in Figure \ref{Critical_Inclination_Plot}. In the LK limit $\Gamma=1$ we recover the classic result $i_\mathrm{c} = \cos^{-1}\sqrt{3/5} \approx 39.2^\circ$. As we decrease $\Gamma$ from $1$, fixed points exist for ever smaller initial inclinations, until $i_\mathrm{c}$ reaches zero at $\Gamma=1/5$; note however that the secular timescale diverges as $\vert \Gamma \vert \to 1/5$. Asymptotes at $\Gamma=-1/5$ and $\Gamma=0$ ensure that fixed points never exist between those values.  As $\Gamma \to -\infty$ we find $\Lambda\to 1/2$ and so $i_\mathrm{c}\to 45^\circ$.

\subsection{High eccentricity behaviour} 
\label{sec_highe}


One is often interested in the high-eccentricity behaviour of binaries undergoing LK-like cycles, because it is at small pericentre distances that exotic effects like GR precession (\ref{GReffect}), mass transfer, gravitational wave emission and tidal circularisation become important. To explore these possibilities we consider orbits that are capable of reaching $e\to 1$ or $j\to 0$, which necessarily requires 
\begin{align}
\Theta\ll 1,
\label{eq:Theta}
\end{align}
(see definitions \ref{eq:AMdefs}) and study their behavior as they evolve through highest eccentricity. We will focus on orbits that start at eccentricities that are not too close to unity, and ignore the effects of GR precession (\S\ref{GReffect}). 

We base our discussion on equation \eqref{djsquareddt}. One of the roots ($j_\pm,j_0$) corresponds to the smallest value of $j$ satisfying the constraint (\ref{eq:AMconstr}), which we have called $j_{\rm min}$. Normally, when \eqref{eq:Theta} is true, the other two roots are much larger in magnitude. With this in mind equation \eqref{djsquareddt} can be approximately integrated in the vicinity of this root $j_{\rm min}$ as 
\begin{align}
j(t)=\sqrt{1-e^2(t)} \approx j_\mathrm{min} \left[ 1+ \left( \frac{t}{t_\mathrm{min}}\right)^2 \right]^{1/2},
\label{eq:time_ev}
\end{align}
where we defined 
\begin{align}
t_{\rm min}=\frac{L}{6C\sqrt{25\Gamma^2-1}}\frac{j_{\rm min}}{j_1 j_2}
=\frac{t_1\Theta^{1/2}}{2j_1 j_2\sqrt{25\Gamma^2-1}}\frac{j_{\rm min}}{\Theta^{1/2}},
\label{eq:t_min}
\end{align}
(see equation (\ref{eq:t1})) and $j_1$ and $j_2$ are the other two roots of equation \eqref{djsquareddt}, i.e not $j_{\rm min}$ (which we normalized by its smallest possible value $\Theta^{1/2}$); $|j_1|$,$|j_2|\gg j_{\rm min}$. Time $t$ is counted from the point of reaching highest eccentricity, i.e. $j(t=0)=j_{\rm min}$, while $t_{\rm min}$ is the characteristic evolution timescale in the vicinity of $j_{\rm min}$ --- the time it takes for $j$ to change from $j_{\rm min}$ (at $t=0$) to $2^{1/2}j_{\rm min}$. Note also that $j(t)\propto t$ when $j\gtrsim j_{\rm min}$.

The fact that the time spent near maximum eccentricity is of order $t_\mathrm{min} \propto j_\mathrm{min} = \sqrt{1-e_\mathrm{max}^2}$ is a familiar result from the Lidov-Kozai case (e.g. \citealt{Miller2002}).  It has been used to characterise the timescale for gravitational-wave induced mergers of binaries driven to high eccentricity through the LK effect (e.g. \citealt{Thompson2011,Antonini2012,Bode2014,Liu2018,Grishin2018,Randall2018}).  We will employ the more general result (\ref{eq:time_ev})-(\ref{eq:t_min}) applicable for arbitrary axisymmetric perturbation (not just that of a point mass companion) when exploring the rate of compact-object binary mergers in stellar clusters predicted by our theory in future work. 

Finally, phase space morphology (as determined by the value of $\Gamma$) is a crucial factor in determining how many phase space orbits are able to reach $e\to 1$.  For example, provided \eqref{eq:Theta} is satisfied, all orbits in the regime $\Gamma > 1/5$ will reach very high eccentricities, whereas this ceases to be true for $0 < \Gamma < 1/5$ (compare Figures \ref{EccOmegaPlots}a,d,g with Figures \ref{EccOmegaPlots2}d,g). This effect is important when calculating merger rates of compact object binaries in stellar clusters \citep{Hamilton2019c}. 


\subsection{Stellar scattering and other non-ideal effects} 
\label{scattering}


In our work we have assumed that the gravitational field of the cluster can be adequately approximated as time-independent. This is of course not true in general. For example, globular clusters in the Milky Way can be shocked and tidally stripped as they move through the Galactic disk.  Also, globular clusters undergo secular evolution on $\gtrsim \mathrm{Gyr}$ timescales which eventually leads to core collapse.  Both of these effects would directly modify the mean field potential $\Phi$ and could also alter a binary's outer orbit dramatically.

Additionally, we have assumed that the cluster's potential is perfectly smooth. However, one must remember that the cluster's true potential is in fact the sum of the potentials of the many individual stars comprising it. As a result the true potential felt by the binary is both granular in space and stochastic in time. In practice, these effects can be explored by considering the impact of individual stellar passages in the vicinity of a binary on its orbital elements.  
The issue of binaries undergoing flyby encounters has been studied widely. \citet{Heggie1996} first considered the case of `secular encounters', where the scattering event takes much longer than the orbital period of the inner binary \citep{Hamers2018}. This regime is appropriate if one is studying perturbations to the orbits of relatively tight systems (hard binaries), such as millisecond pulsars or hot Jupiters.  
On the other hand, \citet{Collins2008} (see also \citealt{Collins2010}) considered the opposite regime in which the timescale for the flyby interaction is much shorter than the inner binary period, so that the encounter can be treated in the impulse approximation. This is the correct description when studying the dynamics of the Oort Cloud comets in the Galaxy or very soft binaries in clusters.  
Finally, when the approach distance and velocity of the external perturber are comparable to the semi-major axis and the orbital speeds of the binary components, the binary changes its orbital elements in a dramatic fashion on a short (non-secular) timescale, with a high chance of being disrupted \citep{Heggie1975,Goodman93}. This would completely reset the course of the smooth secular evolution of the binary orbit explored in this work. Thus, the prescription needed for estimating the effects of stellar scattering depends on the physical problem one wishes to address. 

We defer a careful study of the coupling between the effects of stochastic stellar encounters and the smooth cluster tide-driven evolution of binaries to a future work. Here we simply estimate the characteristic time between close encounters of a binary with field stars.  Assuming that all perturbers have mass $m$ and can be drawn from an homogeneous, isotropic Maxwellian distribution with number density $n$ and velocity disperision $\sigma$, we can estimate the typical time elapsed before the binary experiences a collision with impact parameter $q_\mathrm{coll} = a/2$ as \citep{Binney2008}:
\begin{align}
t_\mathrm{coll} &= \frac{4}{\pi n \sigma a^2}\left( 1+\frac{4G(m_1+m_2+m)}{3\sigma^2a} \right)^{-1}
\nonumber\\
& \approx 5 \,\, \mathrm{Gyr} \times \frac{1}{1+\xi_\mathrm{GF}} \nonumber\\
&\times \left( \frac{n}{10^{4} \, \mathrm{pc}^{-3}}\right)^{-1}  \left( \frac{\sigma}{10 \, \mathrm{km s}^{-1}}\right)^{-1}  \left( \frac{a}{10 \mathrm{AU}}\right)^{-2}, 
\label{scatest}
\end{align}
where
\begin{align}
\xi_\mathrm{GF} \nn &\equiv  \frac{4G(m_1+m_2+m)}{3\sigma^2 a} \\ &= 1.2 \times \left(\frac{m_1+m_2+m}{M_\odot}\right)
\left(\frac{\sigma}{10\mathrm{km s}^{-1}}\right)^{-2}\left(\frac{a}{10\mathrm{AU}}\right)^{-1}.
\label{eq:Safronov}
\end{align} 
is a measure of gravitational focusing, and we have used typical values of $n$ and $\sigma$ for a globular cluster (although as the binary moves through the cluster, the velocity dispersion and number density of field stars it experiences may change dramatically). 
One can see that depending on cluster mass, number density, binary semi-major axis, etc., $t_\mathrm{coll}$ can be larger or smaller than the secular timescale due to cluster tides (equations \eqref{timescaleeqn} and (\ref{eq:t1})). Moreover, weaker (secular) encounters \citep{Heggie1996} which cause slow random walk of the binary orbital elements would occur more frequently. 

Thus, it is usually very important to take into account the effect of stellar flybys.  However, we do not believe that this diminishes the astrophysical relevance of cluster tides, for several reasons.  First, tidal effects can be important even in the outskirts of clusters where $n$ is low and stellar encounters are rare. In fact, tides can drive compact-object binaries to merge out to cluster-centric distances of several parsecs \citep{Hamilton2019c}.  Second, in massive centrally cusped clusters (such as nuclear clusters with or without a central massive black hole), secular timescales due to tides can be as short as $\sim 10^5 \mathrm{yr}$, potentially leading to interesting effects before close encounters occur. Third, while dense stellar environments can lead to frequent disruption of binaries they can also result in efficient binary formation, i.e. they can act as a {\it source} of new binaries that can then undergo tidal evolution.


\subsection{Relation to previous work}


The secular dynamics of binaries presented in this paper have been investigated thoroughly by other authors in the LK ($\Gamma=1$) limit (the `test particle quadrupole' LK problem).  In particular, \citet{Vash1999} and \citet{Kinoshita2007} derived analytically the maximum and minimum eccentricities and the timescale of LK oscillations.  \citet{Antognini2015} rederived the same results and provided an approximate fitting formula for the timescale.

A study of the phase-space portrait of binaries perturbed by the Galactic tide --- a problem investigated by \citet{Heisler1986} and many others, see Paper I --- has been performed by \citet{Brasser2001}.  Keeping only the $\partial^2\Phi/\partial Z^2$ contribution in the tidal expansion of the potential (equivalent to $\Gamma=1/3$), they derived the fixed points, secular timescale, and criteria for circulation and libration in $(\omega,e)$ space.

\citet{Petrovich2017} considered an extension to the LK problem in which a binary orbits a supermassive black hole (SMBH), and its (outer) orbit is perturbed by a non-spherical nuclear cluster potential (the inner orbit is assumed to be unperturbed by the cluster).  Unlike our study, \citet{Petrovich2017} only looked at the effect of the cluster potential on the outer orbit of the binary and completely ignored the direct effect of the cluster potential on the secular dynamics of the inner orbital elements.  Relevant for this work, part of their paper involves an investigation of the $(\omega,e)$ phase portrait of the inner binary in the quadrupole approximation, assuming (a) the outer orbit is almost circular and (b) the cluster potential is only weakly flattened. However, our doubly-averaged formalism does not cover this part of \citet{Petrovich2017}'s paper, because in this particular limit the outer orbit's nodal precession timescale is long compared to the secular evolution time, so the perturbing potential cannot be considered axisymmetric (the situation here is similar to that described in \S \ref{sec_NumericalVerification3}).


\section{Summary} 
\label{conclusions}


We considered the secular dynamics of binaries arising from the general doubly-averaged tidal Hamiltonian derived in Paper I. Our study focused on exploring the phase portraits describing the evolution of binaries perturbed by the tidal field of a host cluster. We unravelled a number of new dynamical regimes previously not accounted for in studies of binary evolution problem, and provided their full classification. Our results can be briefly summarized as follows.

\begin{itemize}
\item We find that that under a wide range of initial conditions, a generic axisymmetric potential (including spherical potentials) can generate a sufficient tidal torque on a binary to allow it to perform large-amplitude secular eccentricity oscillations reminiscent of the LK mechanism.  
 
\item The morphology of the binary evolution in the phase-space of its orbital elements (e.g. $\omega$ and $e$) is uniquely set by the value of a single dimensionless parameter $\Gamma$, which encodes all information about the shape of the cluster potential and the binary orbit within it. We mapped out different dynamical behaviours of the binary as a function of $\Gamma$. 
 
\item Although the dynamics are qualitatively similar to the LK mechanism for $\Gamma > 1/5$, there are bifurcations in the phase-space portrait when $\Gamma = \pm 1/5$ and $0$ such that the dynamics become drastically different from LK case. We provide detailed descriptions of the binary evolution in each of the corresponding dynamical regimes.
 
\item We numerically verify our theoretical predictions and find that they work well when the timescale for secular evolution is much longer than the time for the binary's outer orbit to fill an axisymmetric torus inside the cluster.  Such circumstances may be rare when $\Gamma<0$, because this regime typically requires strongly non-coplanar outer orbits that may take large number (several hundred) of orbital periods to fill a torus.
 
\item General relativistic pericentre precession typically acts to quench secular eccentricity oscillations. Its effect can be easily included in our general doubly-averaged formalism.
 
\item While the LK mechanism is efficient at driving high eccentricity oscillations, it requires the presence of a long-term distant companion to a binary. In contrast, every binary in a cluster feels the cluster potential, just as every comet feels the Galactic tide.  As a result, the effect considered in this work, while possibly weaker than in the standard LK scenario, should be more ubiquitous in nature since it is available to any binary bound to an axisymmetric host system.

 \end{itemize}

The theory we have developed in Papers I and II could be of importance to various astrophysical problems, such as formation of blue stragglers, X-ray binaries, hot Jupiters, and compact-object mergers (i.e. gravitational wave sources) in globular and nuclear star clusters. The possibility of the cluster tide-driven evolution explored here presents an interesting alternative to the well-explored hierarchical triple (LK) scenario.

In a third paper in this series we apply our formalism to the problem of binary compact-object mergers in globular and nuclear star clusters \citep{Hamilton2019c}.  In subsequent work we will investigate the role of stochastic orbital element perturbations due to stellar flybys, among other things.

\section*{Acknowledgements}

We thank Henrik Latter, Eugene Vasiliev and Antranik Sefilian for useful conversations.  The orbit integration in \S \ref{sec_NumericalVerification} was done using \texttt{galpy} (http://github.com/jobovy/galpy, \citealt{Bovy2015}). The N-body calculations were done with \texttt{MERCURY} \citep{Chambers1999}. CH is funded by a Science and Technology Facilities Council (STFC) studentship.




\bibliographystyle{mnras}
\bibliography{Bibliography} 




\appendix
	

\section{Detailed characteristics of the $-1/5 < \Gamma \leq 0$ regime.} 
\label{ap:regimeIII}


Here we provide more details about the properties of the dynamical regime $-1/5 < \Gamma \leq 0$, see \S \ref{sec_GammaRegime3}.


\subsection{Fixed points and orbit families}


When $\Gamma=0$ the dimensionless Hamiltonian is simply $H_1^* = 2+3e^2$, so the phase portrait consists of straight horizontal lines: all orbits circulate with $e(t)=e(0)$. 

When $-1/5<\Gamma< 0$, we use the constraint (\ref{eq:Theta_constr}) to explore the possibility of fixed points. Since both $\Lambda$ and $\Lambda^{-1}$ are negative (see Figure \ref{fig_LambdaOfGamma}), while $\Theta > 0$, we conclude there are no fixed points in this $\Gamma$ regime. As a result, all orbits circulate, in agreement with Figure \ref{EccOmegaPlots3}.


\subsection{Range of parameter values} 
\label{rangeof3}


In the absence of fixed points in this $\Gamma$ regime, the only possible bounds on $D$ are $D=0$ and $D=1-\Theta$.  Hence the $(D,\Theta)$ plane consists simply of a triangle of circulating orbits (see the third row of Figure \ref{fig_Timescales}): 
\begin{align} 
D \in \left(0,  1-\Theta\right) ,\,\,\,\,\,\,\,\,\,\, &-1/5< \Gamma \leq 0.  
\label{Drange3} 
\end{align}

It is easy to show that the Hamiltonian is maximised at $j^2=\Theta$ (i.e. at the upper limit on eccentricity in Figure \ref{EccOmegaPlots3}), so $H_{1,\mathrm{max}}^*$ obeys equation \eqref{eq:HjTheta}.  Similarly it is minimised along the line of zero eccentricity ($j^2=1$), so $H_{1,\mathrm{min}}^*$ is given by equation \eqref{eq:Hj1}.


\subsection{Maximum and minimum eccentricities} 


Figure \ref{EccOmegaPlots3} shows that circulating orbits' maximum and minimum eccentricities are back at $\omega=\pm\pi/2$ and $\omega=0$ respectively, as they were in the $\Gamma>1/5$ case (\S\ref{sec_GammaRegime1}). 

To understand the ordering of $j_\pm^2,j_0^2$ we only need to consider panel (a) of Figure \ref{jpmplots}, because we have only circulating orbits in this $\Gamma$ regime. For $-1/5 < \Gamma \leq 0$ the ordering of $j_0^2$ and $j_+^2$ has flipped compared to $0 < \Gamma \leq 1/5$, while $j_-^2$ still lies outside of the physical region (i.e. it is not bounded by the horizontal dotted lines $j^2=\Theta,1$). Hence we must have $j_-^2<j_+^2<j^2<j_0^2$, so that $j_\mathrm{min} = j_+$, $j_\mathrm{max}=j_0$, and $\Delta = j_0^2-j_-^2$.


\subsection{Timescales of eccentricity oscillations}


The timescale $\log_{10}(t_\mathrm{sec}/t_1)$ is plotted in the third row of Figure \ref{fig_Timescales} for $\Gamma=-0.01,-0.1,-0.18$. We have only a triangle of circulating orbits, and their secular timescale is rather well approximated by $t_1$.


\section{Detailed characteristics of the $\Gamma \le -1/5$ regime.} 
\label{ap:regimeIV}


Here we provide more details about the properties of the dynamical regime $\Gamma \le -1/5$, see \S \ref{sec_GammaRegime4}.


\subsection{Fixed points and orbit families}


When $\Gamma=-1/5$ there are still no librating orbits, because $\Lambda(-1/5)=0$. However, librating orbits emerge as we decrease $\Gamma$ further.  In terms of the constraint on $\Theta$ for fixed points and librating orbits to exist, the regime $\Gamma < -1/5$ mirrors the first $(\Gamma>1/5)$ regime in that we again require $\Theta<\Lambda$ (in Figure \ref{fig_LambdaOfGamma}, the shaded region is bounded by the red curve): \begin{align} 
\Theta \in \left(0, \Lambda \right),\,\,\,\,\,\,\,\, \Gamma < -1/5. 
\label{thetarange4} 
\end{align}

	
\subsection{Range of parameter values} 


It is perhaps unsurprising from the morphology of the phase portraits (compare Figures \ref{EccOmegaPlots} and \ref{EccOmegaPlots4}) that the $(D,\Theta)$ plane for $\Gamma \leq -1/5$ (bottom row of Figure \ref{fig_Timescales}) looks similar to the $\Gamma > 1/5$ case (top row of Figure \ref{fig_Timescales}).  However, in this case the librating orbits are bounded to the left by $D_+$ (see equation \eqref{eq:Djf}):
\begin{align} 
D\in \begin{cases} (D_+,0), \,\,\,\,\,\,\, &\Gamma<1/5, \mathrm{librating \,\, orbits,} \\ (0, 1-\Theta), \,\,\,\,\,\,\, &\Gamma<1/5, \mathrm{circulating \,\, orbits.} \end{cases} \label{Drange4} 
\end{align} 

As for the extrema of $H_1^*$, the only change from the case $-1/5<\Gamma\leq 0$ is that we now have fixed points available.  If they exist then the Hamiltonian is minimised at the fixed point and so $H_{1,\mathrm{min}}^*$ obeys equation \eqref{eq:Hjf}; if not, it is minimised at the zero eccentricity line $j^2=1$ (equation \eqref{eq:Hj1}).  The maximum $H_{1,\mathrm{max}}^*$ is always found at $j=\sqrt{\Theta}$ and is therefore given by equation \eqref{eq:HjTheta}.


\subsection{Maximum and minimum eccentricities}

For circulating orbits we may again inspect Figure \ref{jpmplots}a.  When $\Gamma \leq -1/5$ the physical solutions $j^2$ run from $j_+^2$ to $j_0^2$ as in the $-1/5 < \Gamma < 0$ case, but $j_-^2>1$ is suddenly larger than the others, so $j_+^2 <j^2< j_0^2 < j_-^2$.   Thus $j_\mathrm{min} = j_+$, $j_\mathrm{max} = j_0$ and $\Delta = j_-^2 - j_+^2$. 

For librating orbits we read off from Figure \ref{jpmplots}b that $j_+^2<j^2<j_-^2<j_0^2$ (with $j_0>1$), giving $j_\mathrm{min} = j_+$, $j_\mathrm{max} = j_-$ and $\Delta = j_0^2 - j_+^2$.


\subsection{Timescales of eccentricity oscillations} 
\label{toeo4}


We plot $\log_{10}(t_\mathrm{sec}/t_1)$ for $\Gamma=-0.25,-0.5,-0.8$ in the bottom row of Figure \ref{fig_Timescales}.  Bounds on $\Theta$ and $D$ are given by equations \eqref{thetarange4} and \eqref{Drange4} respectively.  The separatrix lies along $D=0$.  

Along the separatrix the timescale for secular oscillations once again diverges.  The timescale also diverges everywhere in $(D,\Theta)$ space for $\Gamma=-1/5$, see equation \eqref{timescaleeqn}. However, as $\Gamma$ is lowered, one can see that $t_{\rm sec}$ becomes substantially smaller than $t_1$, just as in the case of $\Gamma\to 1$ considered in \S\ref{toeo_a}.
\\

\bsp	
\label{lastpage}
\end{document}